\documentclass[preprint]{aastex63}

\usepackage{amsmath,amssymb,amstext}

\usepackage{enumitem}

\usepackage{graphbox}

\begin{document}
\title{Variations in Finite Difference Potential Fields}

\correspondingauthor{Ronald M. Caplan}
\email{caplanr@predsci.com}

\author[0000-0002-2633-4290]{Ronald M. Caplan}
\affil{Predictive Science Inc., 9990 Mesa Rim Road, Suite 170, San Diego, CA 92121, USA}

\author[0000-0003-1759-4354]{Cooper~Downs}
\affil{Predictive Science Inc., 9990 Mesa Rim Road, Suite 170, San Diego, CA 92121, USA}

\author[0000-0003-1662-3328]{Jon A. Linker}
\affil{Predictive Science Inc., 9990 Mesa Rim Road, Suite 170, San Diego, CA 92121, USA}

\author[0000-0003-1662-3328]{Zoran Mikic}
\affil{Predictive Science Inc., 9990 Mesa Rim Road, Suite 170, San Diego, CA 92121, USA}
\affil{Retired}

\keywords{Solar corona (1483); Solar coronal holes(1484); Solar surface(1527); Astronomy software(1855); Astronomy data analysis(1858)}

\begin{abstract}
The potential field (PF) solution of the solar corona is a vital modeling tool for a wide range of applications, including minimum energy estimates, coronal magnetic field modeling, and empirical solar wind solutions. Given its popularity, it is important to understand how choices made in computing a PF may influence key properties of the solution. Here we study PF solutions for the global coronal magnetic field on 2012 June 13, computed with our high-performance finite difference code POT3D. Solutions are analyzed for their global properties and locally around NOAA AR 11504, using the net open flux, open field boundaries, total magnetic energy, and magnetic structure as metrics. We explore how PF solutions depend on 1) the data source, type, and processing of the inner boundary conditions, 2) the choice of the outer boundary condition height and type, and 3) the numerical resolution and spatial scale of information at the lower boundary. We discuss the various qualitative and quantitative differences that naturally arise by using different maps as input, and illustrate how coronal morphology and open flux depend most strongly on the outer boundary condition. We also show how large-scale morphologies and the open magnetic flux are remarkably insensitive to model resolution, while the surface mapping and embedded magnetic complexity vary considerably. This establishes important context for past, current, and future applications of the PF for coronal and solar wind modeling. 
\end{abstract}


\section{Introduction}
\label{sec_intro}
The simplest description of the coronal magnetic field, based on a photospheric boundary radial magnetic field, is a potential (current-free) field model.  
These solutions have been used for many years to understand and interpret a wide variety of solar and heliospheric phenomena, including  the behavior of the interplanetary magnetic field \citep[e.g.,][]{hoeksemaetal1983,wangsheeley1994}, the structure and evolution of coronal hole boundaries \citep[e.g.,][]{wangetal1996,wangsheeley2004}, the source locations of solar energetic particles \citep[e.g.,][]{nittaetal2006}, coronal heating and X-ray emission \citep{schrijveretal2004}, coronal magnetic field topology \citep[e.g.,][]{antiochosetal2007,titovetal2011}, and as initial conditions for coronal MHD simulations \citep[e.g.,][]{linker99a}.  Empirical solar wind models, derived from the properties of potential field solutions \citep[e.g.,][]{wangsheeley1990,argeetal2003,rileyetal2015}, are used to predict solar wind properties in interplanetary space.  In particular, the Wang-Sheeley-Arge (WSA) model \citep{argeetal2004} is an element of NOAA Space Weather Prediction Center's operational solar wind model \citep{pizzoetal2011}. 

The most common versions of PF models are the potential field source-surface \citep[PFSS,][]{altschulernewkirk1969,schattenetal1969} and potential field current-sheet (PFCS) models \citep{schatten1971}.  In the PFSS model, the field at the upper radial boundary (the source-surface) is assumed to be fully radial.  The PFCS model is used to extend the field from the source surface to 20-30~$\text{R}_{\odot}$ and reproduce the latitudinally independent behavior of the magnitude of the radial component of magnetic field ($|B_r|$) observed by Ulysses \citep{smith_balogh1995,smith_balogh2008}.  

PF solutions have been frequently calculated using low resolution input data and spherical harmonic expansion algorithms \citep[e.g,][]{schrijver2003photospheric}.  The PFSS package in SolarSoftWare \citep{SSW} is frequently used.  These solutions are adequate for many purposes. However, with the availability of higher resolution data and modern computing techniques, much higher resolution solutions can now be computed rapidly.  Thus far, relatively little attention has been paid to how commonly inferred physical properties of the solutions change with increasing resolution.  The nature of the boundary data (i.e. the source map) and model parameters (e.g. source-surface radius) can also affect the solutions.  

In this paper, we explore these differences using our finite difference PF solver code POT3D.  A finite difference approach can be preferable to the harmonic approach, because mismatches between the order of the harmonics and the resolution of the data can lead to ringing of harmonic solutions \citep{toth2011obtaining}.  It also allows localized higher resolution to be implemented.  POT3D 
utilizes a non-uniform logically-rectangular spherical grid and is designed for high performance parallelization using the Message Passing Interface (MPI) with a three-dimensional decomposition, and GPU-acceleration using OpenACC (see Sec.~\ref{sec_pot3d} for more details).  POT3D can compute PFSS, PFCS, and the closely related Open Field  \citep{barnes_sturrock1972} solutions for an input magnetic flux distribution.

\begin{figure}[htbp]
\centering
\includegraphics[align=m,width=\textwidth]{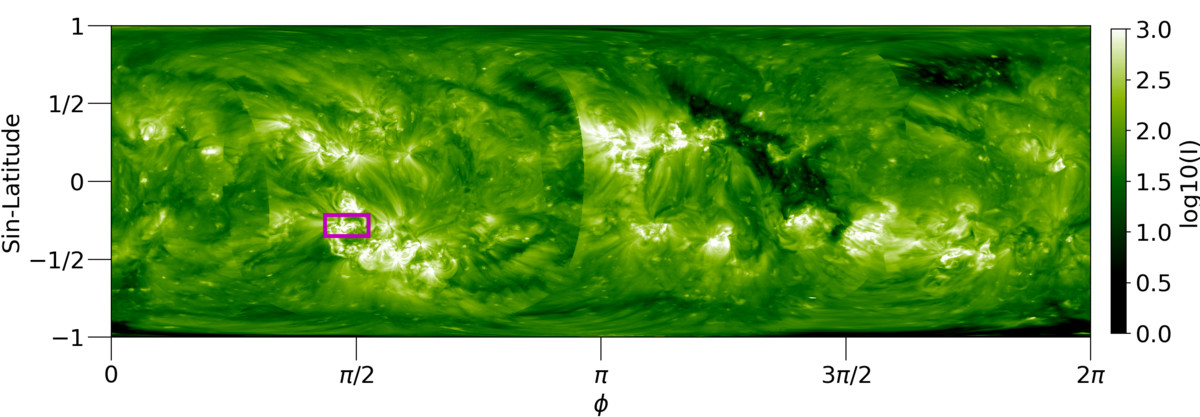}
\caption{Synchronic EUV composite map for 2012 June 13 12:00 UT \citep[from][\url{www.predsci.com/chd}]{caplan16}.  The map is constructed from STEREO A/B EUVI 195 and SDO AIA 193 images, and the targeted region around AR 11504 is highlighted.\label{fig_euv}} 
\end{figure}

To demonstrate coronal solutions, we select a nominal date of 2012 June 13 near 13:11:36 TAI, which is part of  Carrington Rotation (CR) 2124.  We examine how global coronal properties change between each solution, as well as a local region centered on NOAA active region (AR) 11504. Fig.~\ref{fig_euv} shows a synchronic extreme ultraviolet (EUV) map using data near the targeted time. This map gives an instantaneous snapshot of the corona's morphological appearance over the full sun. For each solution we compute quantities such as the total magnetic energy and open magnetic flux, and others based on field line mappings, including maps of the open field and the generalized squashing factor, $Q$ \citep{titov2007generalized}. $Q$ maps are a useful tool for visualizing magnetic topology and complexity on large and small scales.

A comparison of finite-difference PF solutions for CR2144 (approximately a year and a half from our target date) was performed in \citet{hayashi2016comparison} using various observatory data, flux balancing methods, and smoothing filters, with a constant resolution and boundary condition.  Their focus was on comparisons to observations, such as EUV coronal holes to the open field, and OMNI magnetic data to the open flux.  In contrast, here we vary the resolution, boundary conditions, and data source, and focus on the inherent differences between solutions, as well as both global and localized comparisons.

We note that a number of PF finite difference solver codes have been developed over the years.  Two presently publicly available solvers are the FDIPS\footnote{\url{http://csem.engin.umich.edu/tools/FDIPS}} Fortran code \citep{toth2011obtaining} and the recently released PFSSPY\footnote{\url{https://pfsspy.readthedocs.io}} python package \citep{pfsspy}.  While both these packages work very well for their intended use cases, in this study we require higher performance and more grid flexibility than either tool currently offers.
  
The paper is outlined as follows:  
In Sec.~\ref{sec_pot3d} we describe how global coronal PF solutions are computed with the POT3D code.  Sec.~\ref{sec:mapprep} describes how we prepare the photospheric magnetic field data needed for the model.  Our standard setup and comparison metrics are described in Sec.~\ref{sec:compare_method}.  Variations of input data, boundary conditions, and resolutions are described in Secs.~\ref{sec:var_input}, \ref{sec:var_bc}, and \ref{sec:var_res} respectively.  We discuss these results and conclude in Sec.~\ref{sec:discussion}.  


\section{Potential Field Solutions with POT3D}
\label{sec_pot3d}
POT3D is a Fortran code that computes potential field solutions to approximate the solar coronal magnetic field using observed photospheric magnetic fields as a boundary condition.  It is also used for computing Open Field and PFCS models (see Appendix~\ref{sec_appx_pot3d_models} for details).  It has been (and continues to be) used for numerous studies of coronal structure and dynamics \citep[e.g.,][]{linker2016empirically,titov20122010} and is the potential field solver for the WSA model in the CORHEL software suite publicly hosted at NASA's Community Coordinated Modeling Center (CCMC)\footnote{\url{https://ccmc.gsfc.nasa.gov/}}.  The POT3D source code is available as part of the Standard Performance Evaluation Corporation's (SPEC) beta version of the SPEChpc\textsuperscript{TM} 2021 benchmark suites\footnote{\url{https://www.spec.org/hpc2021}} and, with publication of this paper, as an open-source release on GitHub\footnote{\url{https://github.com/predsci/POT3D}}.

\subsection{Model description}
\label{sec:pot3d_model}
\begin{figure}[htbp]
\centering
\includegraphics[align=m,width=0.35\textwidth]{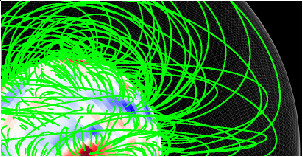}
\includegraphics[align=m,width=0.35\textwidth]{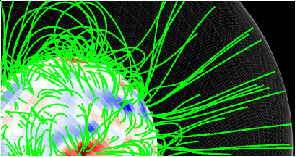}
\caption{Field lines of a potential field solution using `closed wall' (left) and 'source surface' (right) outer boundary conditions in POT3D. \label{fig_bc}} 
\end{figure}

A potential field (PF) is a force-free and current-free magnetic field.  Setting the current in Maxwell's equation ($\nabla \times {\bf B} =\mu_0\,{\bf J}$) to zero (${\bf J}=0$) leads to solutions of the form ${\bf B} = \nabla \Phi$, where $\Phi$ is the scalar potential.  Combined with the divergence-free condition ($\nabla\cdot{\bf B} = 0$), this yields a Laplace equation for $\Phi$:
\begin{equation}
\label{eq:laplace}
\nabla^2\Phi=0.
\end{equation}
We use spherical coordinates, and the lower boundary condition at $r=R_\odot$ (the solar surface) is set based on a surface-map of the radial magnetic field, $B_r$, as
\begin{equation}
\label{eq:laplace_bcr0}
\left.\frac{\partial \Phi}{\partial r}\right|_{R_{\odot}} = \left.B_r\right|_{R_\odot}.
\end{equation}
Depending on the needs of the application, the upper boundary ($r=r_1$) can be set to either a `closed wall' ($\left.B_r\right|_{Rr_1}=0$) or `source surface' ($\left.\Phi\right|_{r_1}=0$) condition as depicted in Fig.~\ref{fig_bc}.
The $\phi$ direction uses a periodic boundary ($\left.\Phi\right|_{\phi=0}=\left.\Phi\right|_{\phi=2\,\pi}$), while polar boundary conditions are set using the average
\begin{equation}
\label{eq:laplace_bcpole}
\left.\Phi\right|_{\theta_{0/\pi}}=\frac{1}{2\,\pi}\int_{\phi=0}^{\phi=2\,\pi}\left.\Phi\right|_{\theta_{0/\pi}\pm\epsilon}\,d\phi,
\end{equation}
where $\epsilon$ is a small distance to the pole.  Once Eq.~(\ref{eq:laplace}) is solved, we set ${\bf B} = \nabla \Phi$ to get the magnetic field.

\subsection{Numerical methods}
POT3D solves Eq.~\ref{eq:laplace} globally using finite differences on a non-uniform logically-rectangular spherical grid.  After solving for $\Phi$, each component of the resulting field ${\bf B}$ is saved on a staggered grid.  The finite-difference current derived from this field is identically zero-valued (${\bf J}=0$), while the field's divergence-free condition ($\nabla\cdot{\bf B}=0$) is satisfied to the level of the solver tolerance.  

POT3D uses a second-order finite-difference method \citep{NUG_PLAY_1992} in which Eq.~(\ref{eq:laplace}) takes the form 
\begin{alignat}{2}
\label{eq_diffuse_desc1}
\nabla^2 \Phi_{i,j,k} &\approx \frac{1}{\Delta r_i}\left[
\frac{\Phi_{i+1,j,k}-\Phi_{i,j,k}}{\Delta r_{i+\frac{1}{2}}}
-\frac{\Phi_{i,j,k}-\Phi_{i-1,j,k}}{\Delta r_{i-\frac{1}{2}}}
\right] \\
&+ \frac{1}{\sin\theta_j\,\Delta\theta_j}\left[
 \sin\theta_{i,j+\frac{1}{2}}\,\frac{\Phi_{i,j+1,k}-\Phi_{i,j,  k}}{\Delta\theta_{j+\frac{1}{2}}}
-\sin\theta_{i,j-\frac{1}{2}}\,\frac{\Phi_{i,j,  k}-\Phi_{i,j-1,k}}{\Delta\theta_{j-\frac{1}{2}}}
\right] \notag \\
&+\frac{1}{\sin^2\theta_j\,\Delta\phi_k}\left[     
 \frac{\Phi_{i,j,k+1}-\Phi_{i,j,k  }}{\Delta\phi_{k+\frac{1}{2}}}
-\frac{\Phi_{i,j,k  }-\Phi_{i,j,k-1}}{\Delta\phi_{k-\frac{1}{2}}}
\right]=0. \notag
\end{alignat}
Eq.~\ref{eq_diffuse_desc1} can be represented in matrix form as 
\begin{equation}
\label{eq:pot3d}
{\bf A}\,\Phi = 0.
\end{equation} 
In POT3D, the sparse symmetric matrix ${\bf A}$ is stored in a custom DIA sparse format \citep{DIACSR} for the inner grid points, while the boundary conditions are implemented matrix-free.  The system is solved using the Preconditioned Conjugate Gradient (PCG) method with two communication-free preconditioners: 1) A point-Jacobi/diagonal-scaling (PC1) which uses the inverse of the diagonal of ${\bf A}$, and 2) A non-overlapping domain decomposition with zero-fill incomplete LU factorization (PC2) \cite{IterativeMethods_SAAD_Book}.  PC1 has a computationally  inexpensive formulation and application, but it is limited in its effectiveness at reducing iterations. PC2 is more computationally expensive to formulate and apply, but is also much more effective at reducing iterations. In some cases, it is possible for PC1 to outperform PC2 (e.g. when solving problems requiring very few iterations, when using hardware that is very efficient for vectorizable algorithms, or when the ILU0 suffers `breakdown' \citep{ILU_breakdown}). For PC2, the LU matrix is stored in a memory-optimized CSR format \citep{CSRopt}, and solved with the standard backward-forward algorithm.  The PCG solver's convergence criteria is chosen to be when $|r_p|/|b_p| < \epsilon$, where $|r_p|$ is the norm of the preconditioned residual and $|b_p|$ is the norm of the preconditoned right-hand-side (whose values are set by the boundary conditions).   For the solutions performed in this work, we set $\epsilon=10^{-9}$.

\subsection{Parallelization}
POT3D is parallelized using MPI and can scale to thousands of CPU cores.  Each MPI rank takes one subsection of the grid (as cubed as possible) and treats it as its own local domain for all operations.  The only MPI communication needed are point-to-point messages for the local boundaries in the matrix-vector product, and collective operations for the inner products and polar boundary conditions.  Initial and final collectives are also used to decompose the domain and collect the solution for output.  

POT3D is GPU-accelerated using the OpenACC directive-based API \citep{OpenACCBook2}.  This allows the code to be compiled for either CPU or GPU systems using a single source code.  The code can scale to multiple GPU accelerators across multiple compute nodes. 

Performance results for a selection of solutions computed in this paper are shown in Appendix \ref{sec_appx_comp}.  

\section{Map Preparation}
\label{sec:mapprep}
For all input maps, several steps are taken to prepare them for use in POT3D.  The details of each step are beyond the current scope, but are summarized here.

We require full-Sun maps of the surface radial magnetic field ($B_r$).  These are commonly provided by observatories, often with $B_r$ inferred from the line-of-sight field, under the assumption that the field is radial where it is measured in the photosphere \citep{wang_sheeley1992}. Since all current magnetic imagers are situated on the Sun-Earth line, multiple images over time are combined in some way to form a full map.  Therefore, our first step is to  use either a publicly available full-Sun data product (such as a `Carrington' map), or manually combine data from these and various other sources. Sometimes the full-sun products are not as high-resolution as the original measurements, or have incomplete data at the poles. In these cases we `stitch' various data products together. For example, to resolve an AR at a high resolution closest to a desired time, we might insert high-resolution data into an otherwise prepared full map. The polar regions also present a unique challenge as they are always poorly observed and often need to be extrapolated.  Numerous pole-filled data products are now available, but if needed/desired, we utilize our own polar filling procedure.   

Since the magnetic field data for a map is constructed from data taken over time, the map is generally not flux-balanced (i.e. $\nabla\cdot{\bf B}\ne0$). We therefore calculate the total flux imbalance and modify the map through multiplication to obtain a flux-balanced input at the inner boundary.  The use of a multiplicative method instead of a simple additive approach preserves the location and structure of neutral lines (see \citep{hayashi2016comparison} where additive flux balancing resulted in very different PF solutions).  

Once a fully specified, flux-balanced, full-Sun map is constructed, we interpolate it to the desired resolution for the POT3D code.  This is done in a flux-preserving way by projecting the source pixel grid onto the destination grid and integrating the contained or fractionally overlapped portions cell-by-cell.

Finally, we smooth the map in order to allow the finite-difference representation to resolve any small structures.  In order to maintain flux balance and avoid aliasing issues, we use a time-dependent spherical-surface diffusion equation integrator.  The spatially-variant diffusivity is set based on the local grid size in order to only smooth the minimum amount needed for resolvablility.   

\begin{figure}[htbp]
\centering
\includegraphics[align=m,width=0.325\textwidth]{br_CompositeMap1_seam01_4ar_fb_w}
\includegraphics[align=m,width=0.325\textwidth]{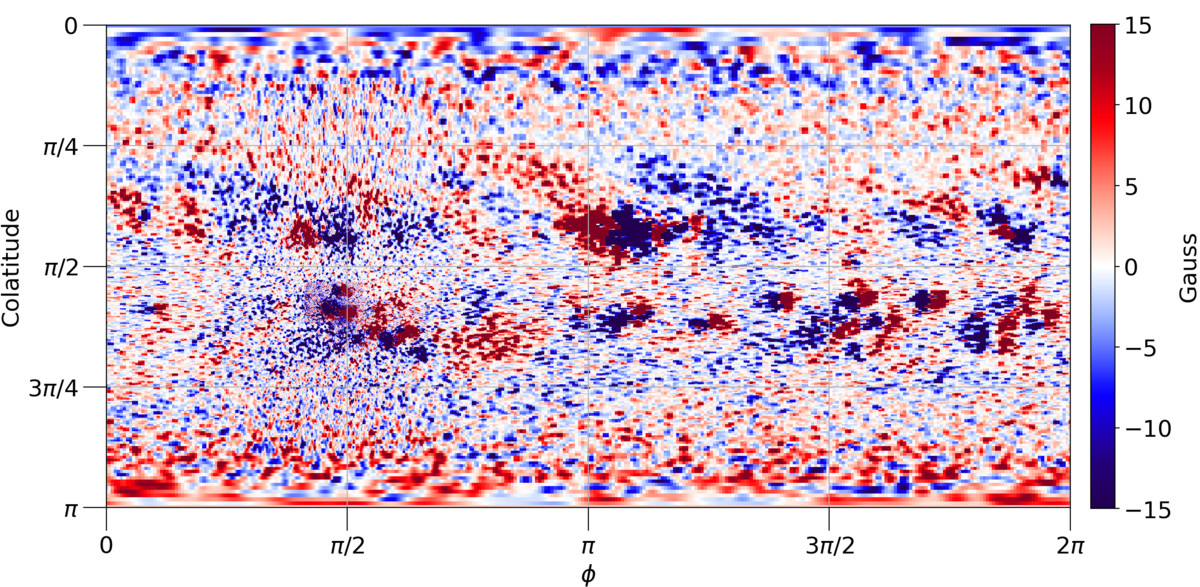}
\includegraphics[align=m,width=0.325\textwidth]{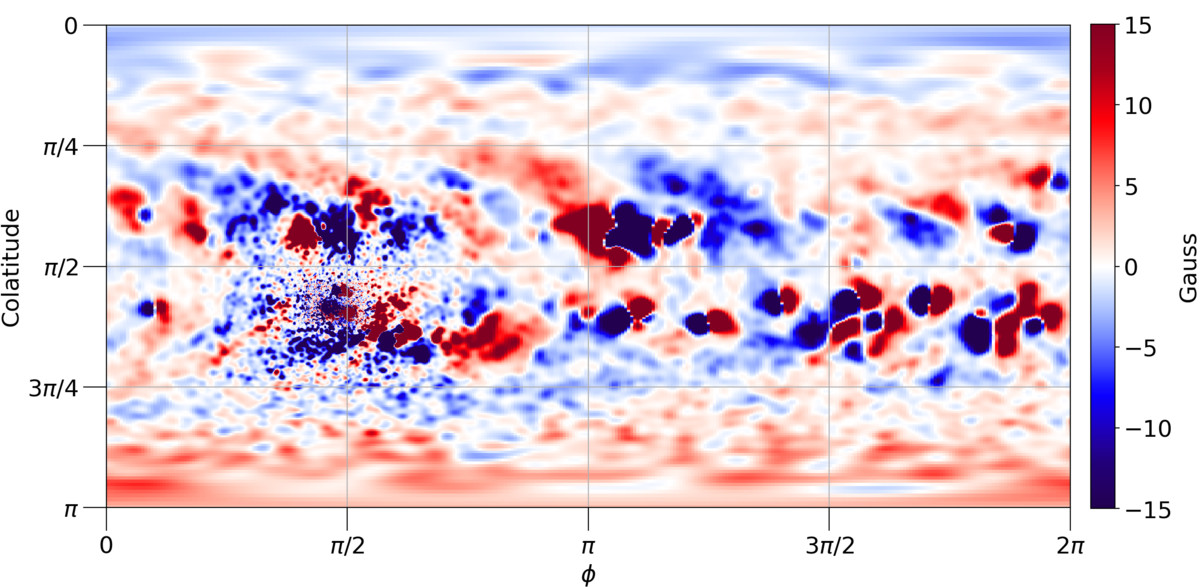}
\caption{Depiction of magnetic map processing.  A full-Sun, flux-balanced composite map is shown on the left.  The map interpolated to a non-uniform grid is shown in the center.  The final map after smoothing is shown on the right. \label{fig_map_proc}} 
\end{figure}

An example of applying the interpolation and smoothing steps are shown in Fig.~\ref{fig_map_proc}.

\section{Comparison Method and Default Parameters}
\label{sec:compare_method}
To compare the PF solutions, we compute several derived quantities.  By tracing field lines through the solutions, we generate open field and squashing factor, $Q$, maps \citep{titov2007generalized} both globally and locally near the AR.  We compute the total open flux, and display it as an extrapolated IMF value at 1AU as
\[
|B_{r1au}|=\frac{|\Phi_{open}|}{4\pi\,r_{1au}^2}=\frac{1}{4}\,\left(\frac{r_1}{r_{1au}}\right)^2\,\int_0^{4\pi}|B_r(r_1,\theta,\phi)|d\Omega,
\]
and the area of open field regions at the surface as
\[
A_{open}=\int_0^{4\pi}\mbox{OF}(r_0,\theta,\phi)d\Omega,
\]  
where at each pixel, $\mbox{OF}=1$ if the field is open, and $\mbox{OF}=0$ if it is closed.  We also compute the magnetic energy within both the AR region and full computational domain as
\begin{equation}
\label{eq_me}
W = \frac{1}{8\,\pi}\,\int_{\phi_0}^{\phi_1}\int_{\theta_0}^{\theta_1}\int_{r_{0}}^{r_{1}} |{\bf B}|^2\,r^2\,\sin\theta\,dr\,d\theta\,d\phi.
\end{equation}
In order to account for the truncated radial domain for global calculations, we add the average of analytic approximated upper and lower bounds of the magnetic energy out to infinity (see Appendix~\ref{sec:mage}).

The AR 11504 region is taken to be bounded by $r\in[1.0,2.0]\,R_{\odot}$, $\theta\in[1.79,1.93]$, and $\phi\in[1.37,1.65]$.  This region is depicted as the magenta contour in Fig.~\ref{fig_euv}.  

All solutions use a radial non-uniform grid which coarsens towards the outer radial boundary.  The non-uniformity is set such that the grid cells are as `square' as possible at the outer boundary.  To achieve this with our choice of radial non-uniformity, we set the number of $r$ grid points to be $1/6.67$ of the number of $\phi$ grid points.  The default $N_{\phi} \times N_{\theta}$ resolution is set to $360\times 180$.  The default radial domain is set to have an outer radii of $2.5\,R_{\odot}$ and the boundary condition is set to the source-surface condition described in Sec.~\ref{sec:pot3d_model}. The default input $B_r$ map is the one designated as map {\bf (PSI)} below in Sec.~\ref{sec:var_input}.  

In all comparisons, any parameters not explicitly mentioned conform to the defaults described in this section.

\section{Variations in Input Data}
\label{sec:var_input}
Although it is well known that different map sources/products will lead to different solutions \citep{hayashi2016comparison,linker2017open}, it is still illustrative to compare some of the more common data products in use today.  This serves to provide a baseline for the sections that follow, in order to assess how the level of variation  seen there compare to the variations seen here.  We also compare the standard map products with our custom map created from multiple data sources, as we will use the custom map in the rest of the paper as described in Sec.~\ref{sec:compare_method}.

For all maps (except the custom map) we limit ourselves to publicly available, pole-filled maps.  These maps come from data obtained from the National Solar Observatory's Global Oscillation Network Group (GONG) \citep{gong} and the Helioseismic Magnetic Imager (HMI) aboard the Solar Dynamics Observatory \citep{hmi}.  Our time of interest is 2012 June 13 (part of CR2124), and we use the following maps (filenames of the publicly available data from the Joint Science Operations Center \footnote{\url{http://jsoc.stanford.edu}} are indicated):
\begin{itemize}
\item[] {\bf (GS)} {--} GONG Pole-filled Synoptic CR2124 \\
({\tt mrmqs120608t0040c2124\_000.fits})
\item[] {\bf (GH)} {--} GONG Pole-filled Hourly 2012/06/13 11:54 \\
({\tt mrbqs120613t1154c2124\_047.fits})
\item[] {\bf (GJ)} {--} GONG Pole-filled Janis 2012/06/13 11:54 \\
({\tt mrbqj120613t1154c2124\_047.fits})
\item[] {\bf (HS)} {--} HMI Pole-filled Synoptic CR2124 \\
({\tt hmi.synoptic\_mr\_polfil\_720s.2124.Mr\_polfil.fits})
\item[] {\bf (HD)} {--} HMI Pole-filled Daily 2012/06/13 12:00 \\
({\tt hmi.mrdailysynframe\_polfil\_720s.20120613\_120000\_TAI.Mr\_polfil.fits})
\item[] {\bf (PSI)} {--} {\bf (HS)} + SHARP + Pole-filling (see text)
\end{itemize}

We note that maps {\bf (GH)}, {\bf (GJ)}, and {\bf (HD)} are not provided in Carrington coordinates.  In order to have all maps in equivalent coordinates for comparison, we apply a shift in $\phi$ ($-47^{\circ}$ for {\bf (GH)} and {\bf (GJ)}, and $-47.40796^{\circ}$ for {\bf (HD)} \citep{polarHMI}) to place these maps into Carrington coordinates.

Map {\bf (PSI)}, is our `best' map for the time period and AR of interest in this paper.  
To generate this map, we take vector $B_r$ data from the Spaceweather HMI Active Region Patch (SHARP) database \citep{sharp}.  We insert patches 118, 112, and 106 into the synoptic HMI map {\bf (HS)}.  Although the poles are already filled in the {\bf (HS)} data product, we refill the poles ourselves by inserting a random distribution of parasitic polarities whose net flux matches the pole-filled map in these regions \citep[similar to][]{mikic18}.

\begin{figure}[htbp]
\centering
$\begin{array}{rcc}
\rotatebox[origin=c]{90}{\mbox{{\bf G}ONG {\bf S}ynoptic}}
& 
\includegraphics[align=m,width=0.4\textwidth]{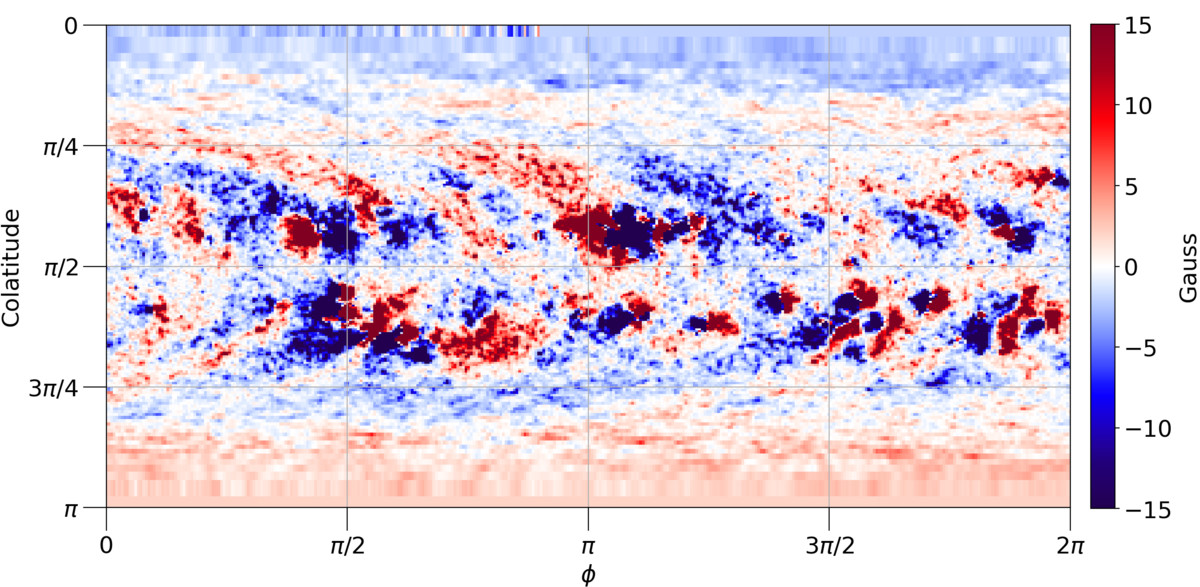}
&
\includegraphics[align=m,width=0.4\textwidth]{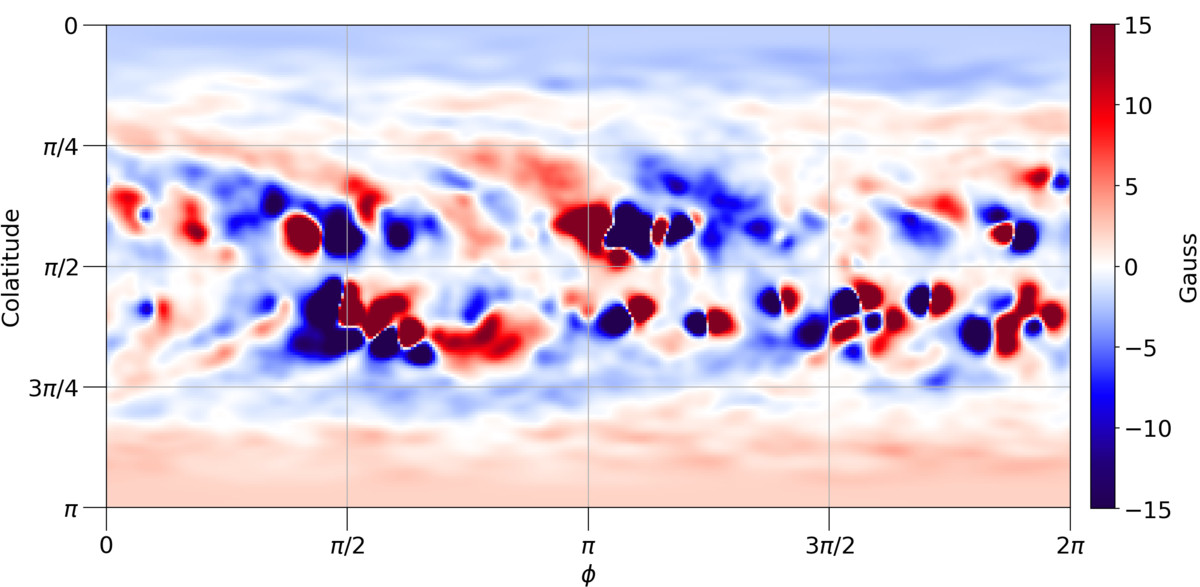}
\\
\rotatebox[origin=c]{90}{\mbox{{\bf G}ONG {\bf H}ourly}}
&
\includegraphics[align=m,width=0.4\textwidth]{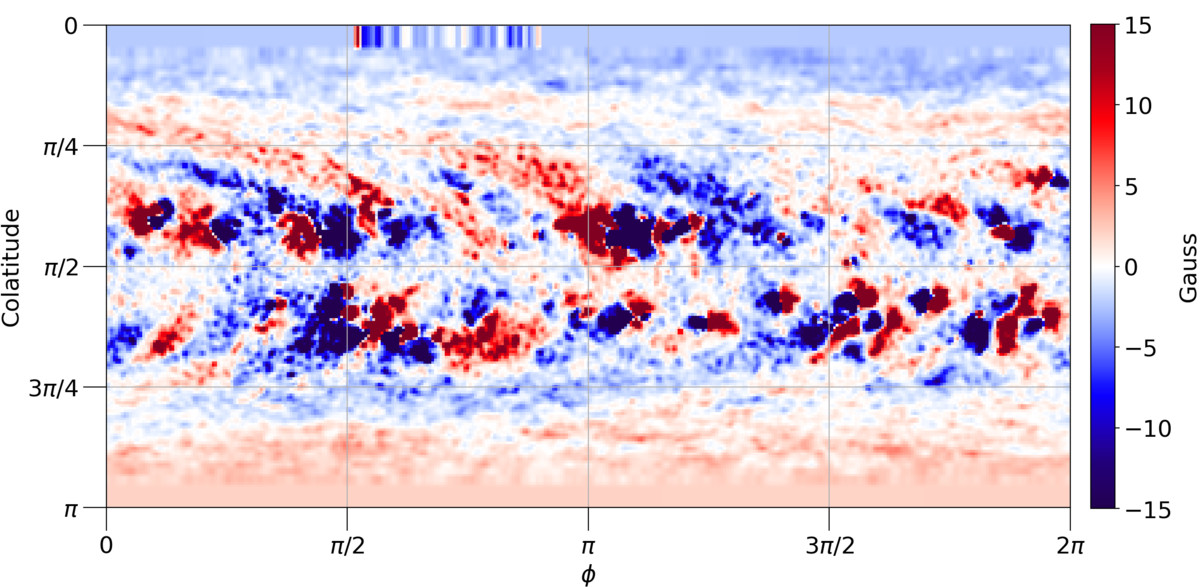}
&
\includegraphics[align=m,width=0.4\textwidth]{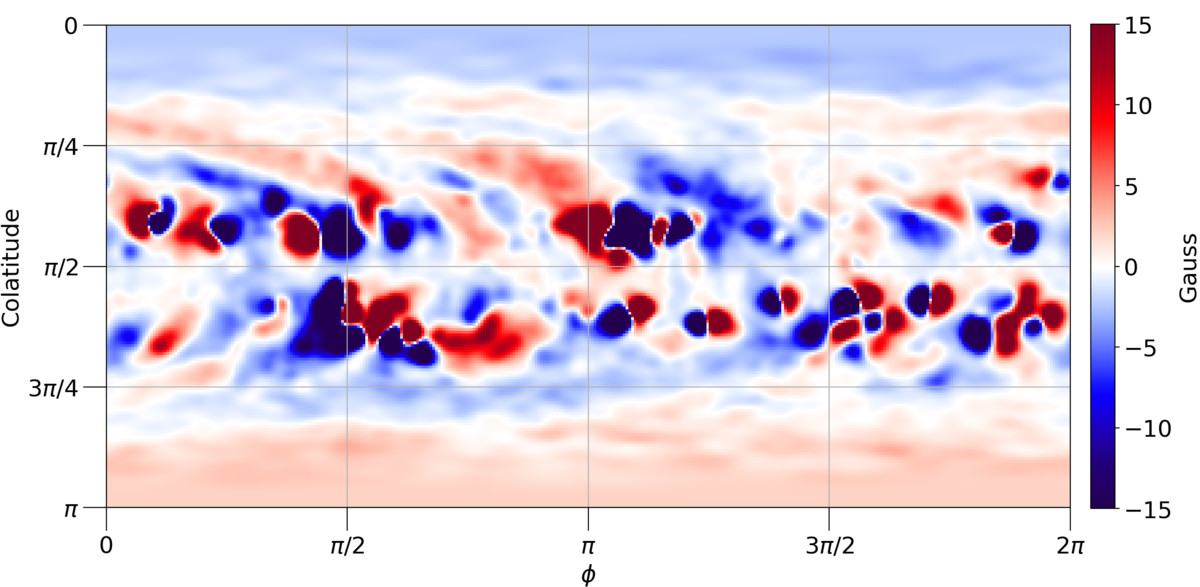}
\\
\rotatebox[origin=c]{90}{\mbox{{\bf G}ONG {\bf J}anis}}
& 
\includegraphics[align=m,width=0.4\textwidth]{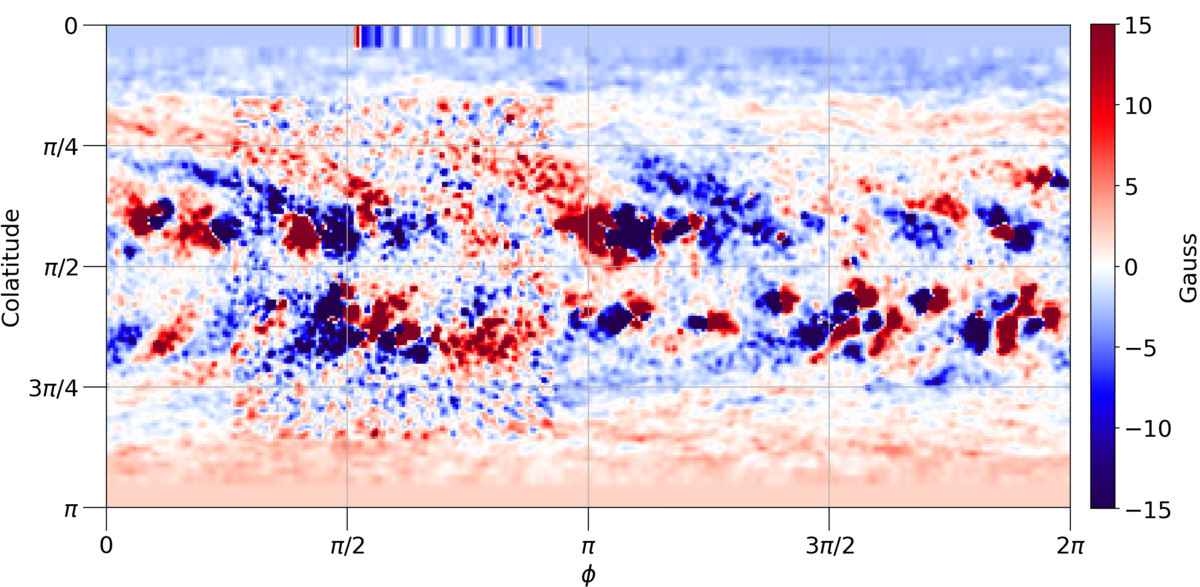}
&
\includegraphics[align=m,width=0.4\textwidth]{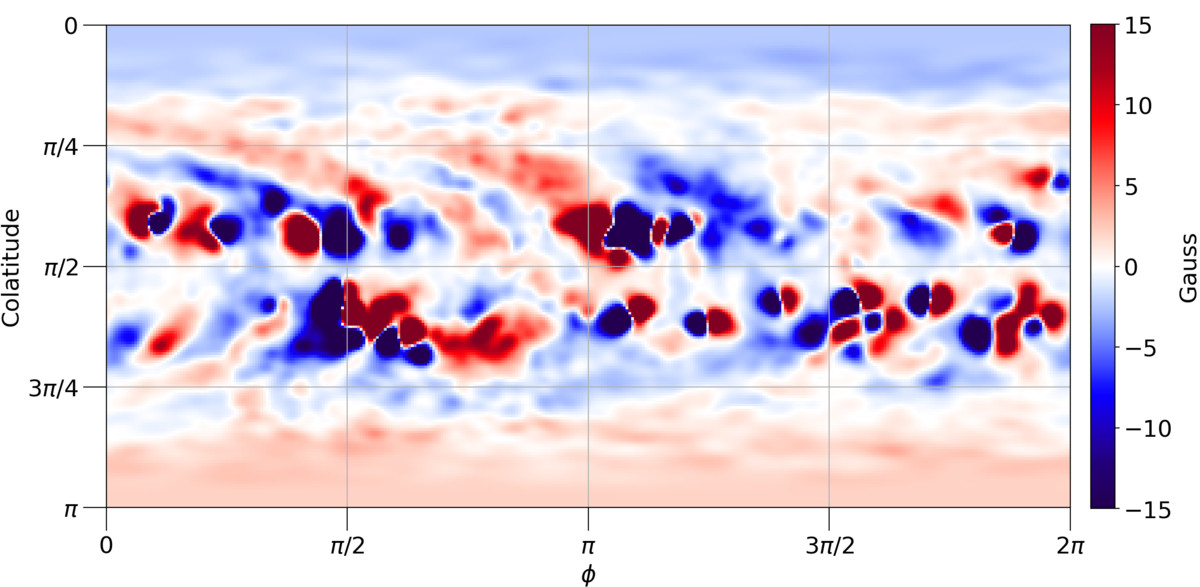}
\\
\rotatebox[origin=c]{90}{\mbox{{\bf H}MI {\bf S}ynoptic}}
& 
\includegraphics[align=m,width=0.4\textwidth]{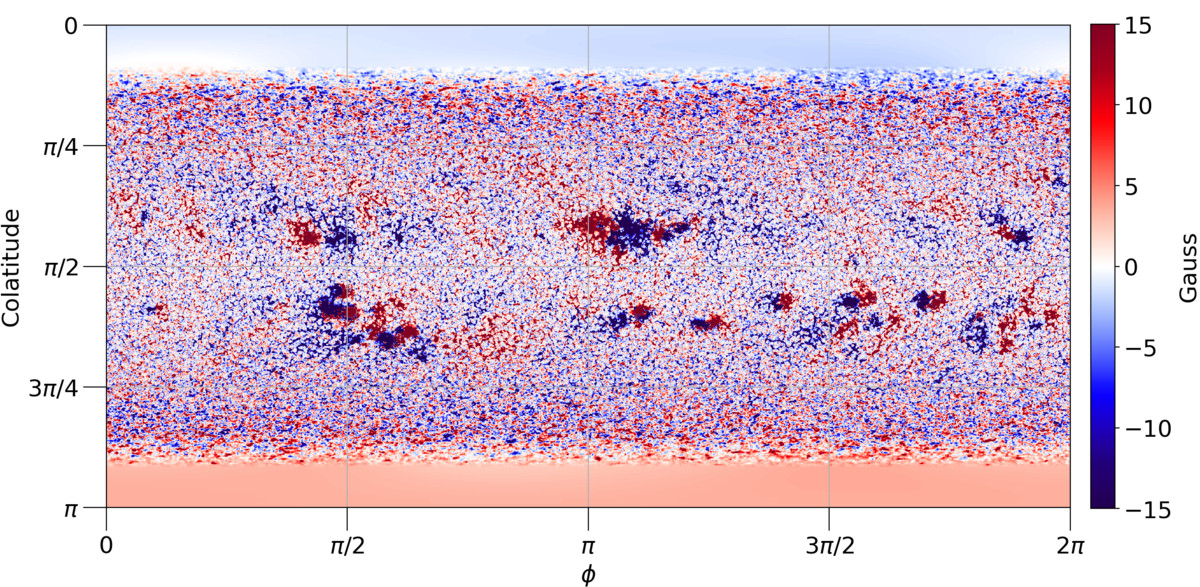}
&
\includegraphics[align=m,width=0.4\textwidth]{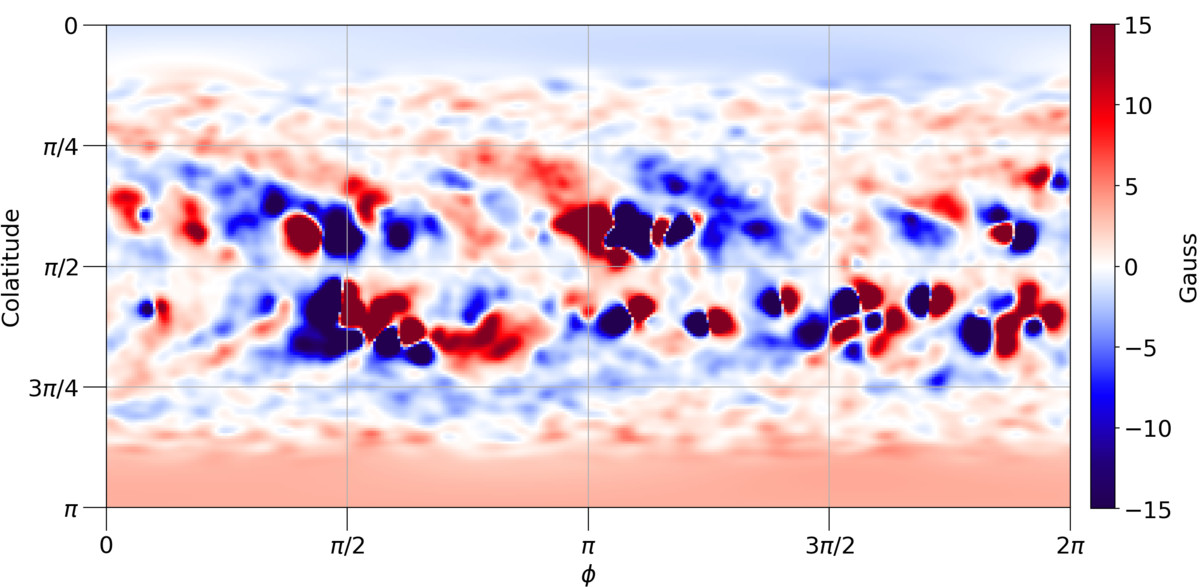}
\\
\rotatebox[origin=c]{90}{\mbox{{\bf H}MI {\bf D}aily}}
& 
\includegraphics[align=m,width=0.4\textwidth]{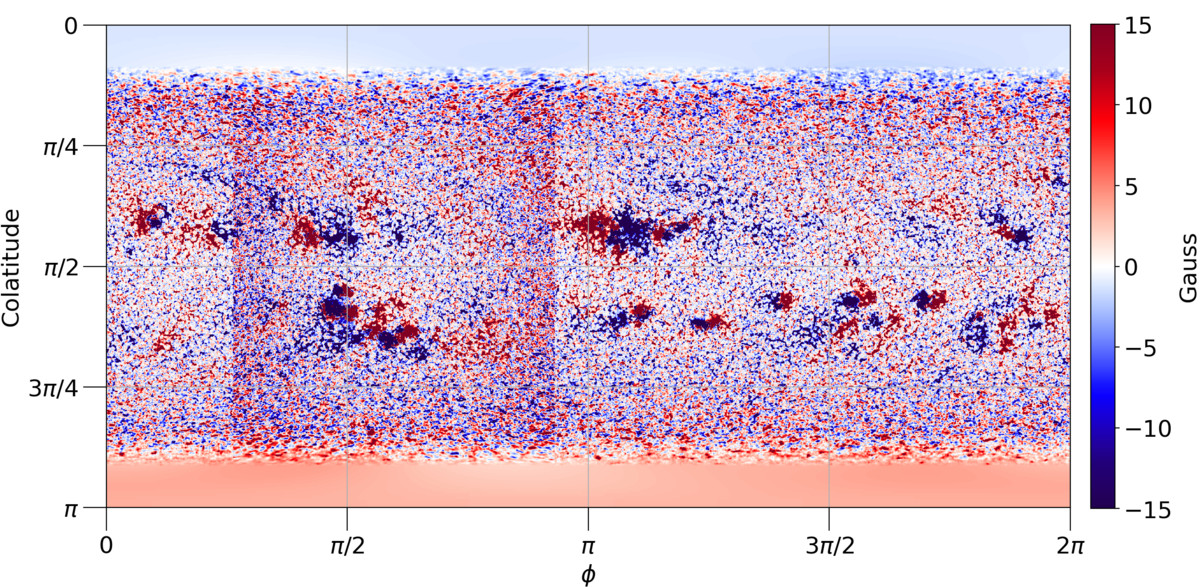}
&
\includegraphics[align=m,width=0.4\textwidth]{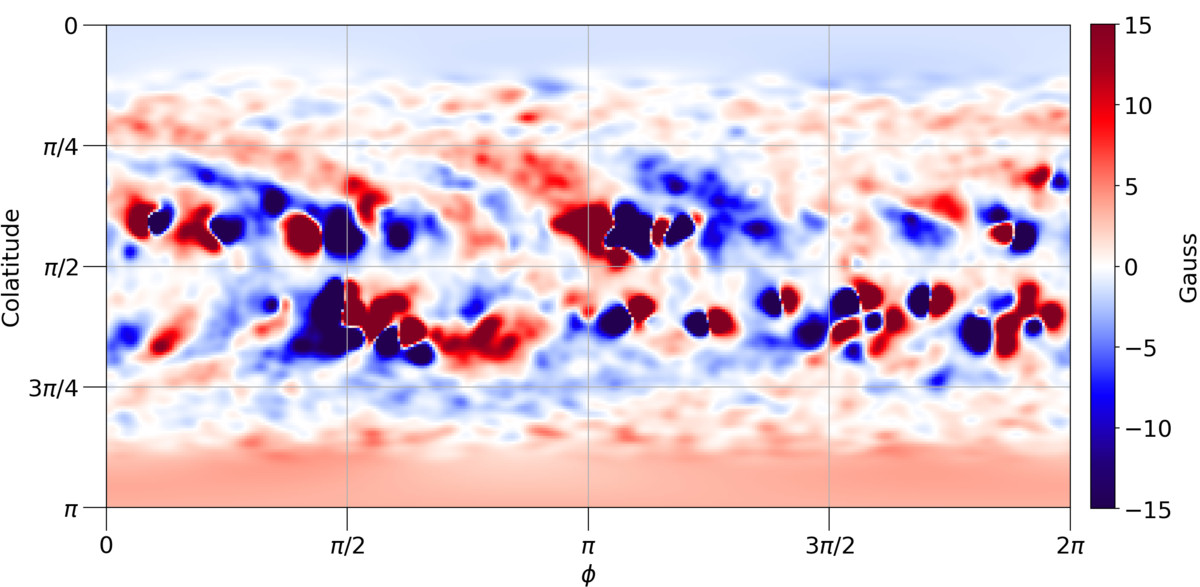}
\\
\rotatebox[origin=c]{90}{\mbox{{\bf PSI} (HMI data)}}
& 
\includegraphics[align=m,width=0.4\textwidth]{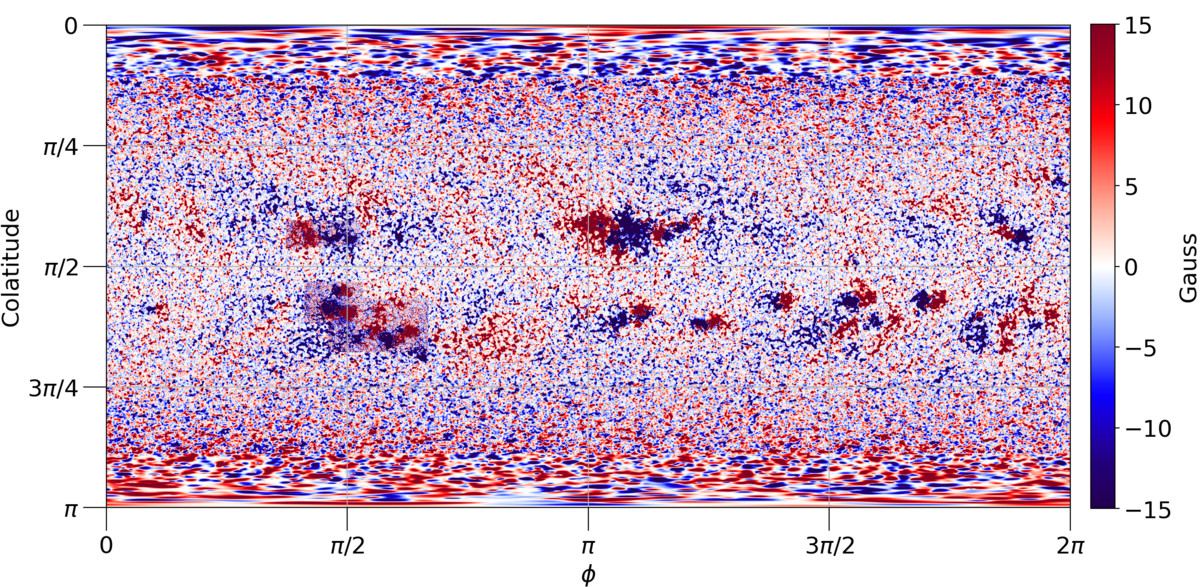}
&
\includegraphics[align=m,width=0.4\textwidth]{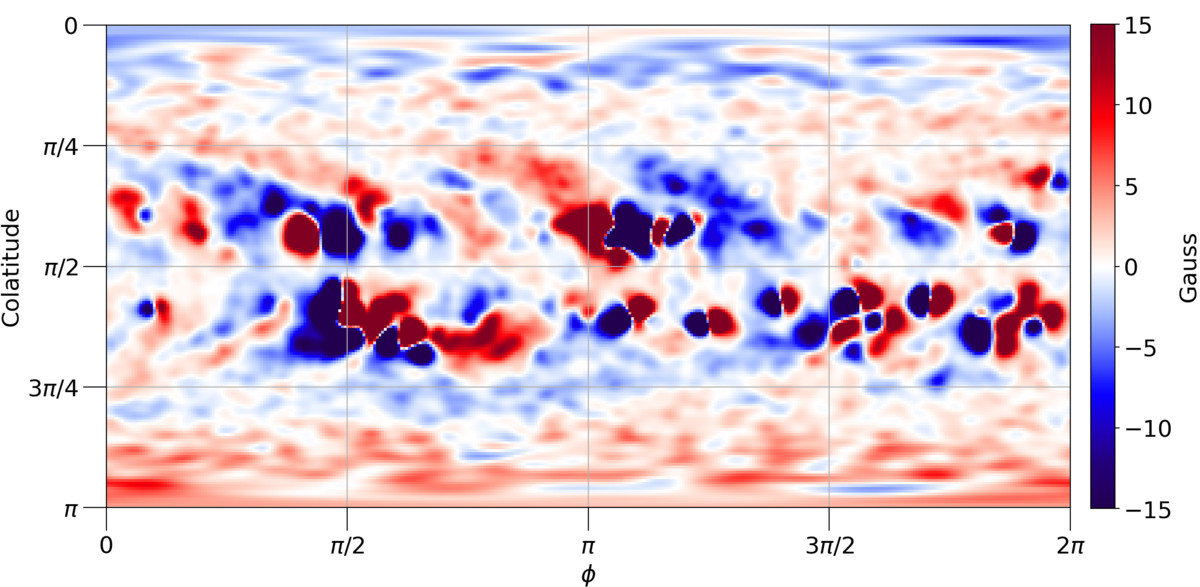}
\\
\; & \mbox{Original Map} & \mbox{Processed Map}
\end{array}$
\caption{Surface $B_r$ maps (original map on left, processed map at default $360\!\times\!180$ resolution on right) used for data source comparison.  Original maps {\bf (GH)}, {\bf (GJ)}, and {\bf (HD)} are shown after shifting their $\phi$ axis into Carrington coordinates.  Other details of the map sources are given in the text.\label{fig_maps}} 
\end{figure}

In Fig~\ref{fig_maps} we show both the original data product for each map, as well as the resulting default resolution map. 
We see that all maps appear similar after processing.  The {\bf (PSI)} map stands out as its poles have parasitic polarities and therefore exhibit more polar structure than the other maps.

\begin{figure}[htbp]
\centering
$\begin{array}{rc}
\rotatebox[origin=c]{90}{\mbox{{\bf G}ONG {\bf S}ynoptic}}
& 
\includegraphics[align=m,width=0.6\textwidth]{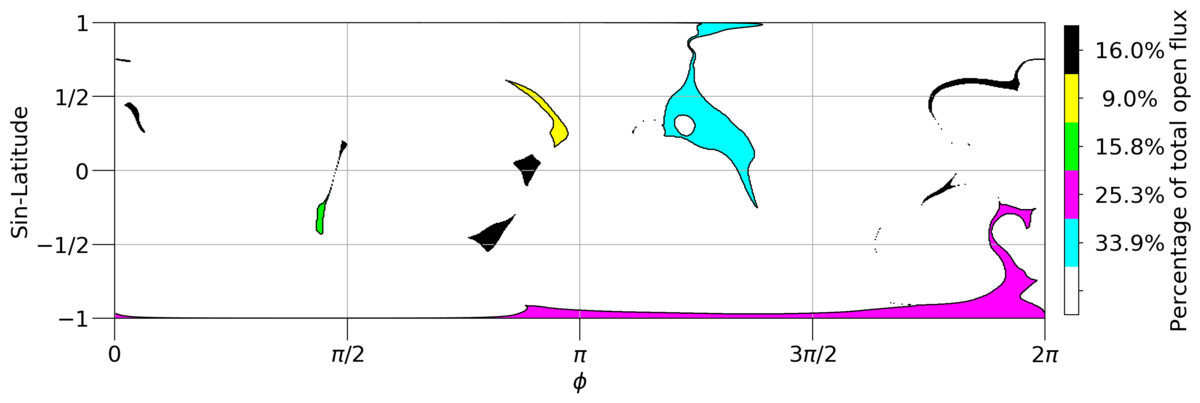}
\\
\rotatebox[origin=c]{90}{\mbox{{\bf G}ONG {\bf H}ourly}}
& 
\includegraphics[align=m,width=0.6\textwidth]{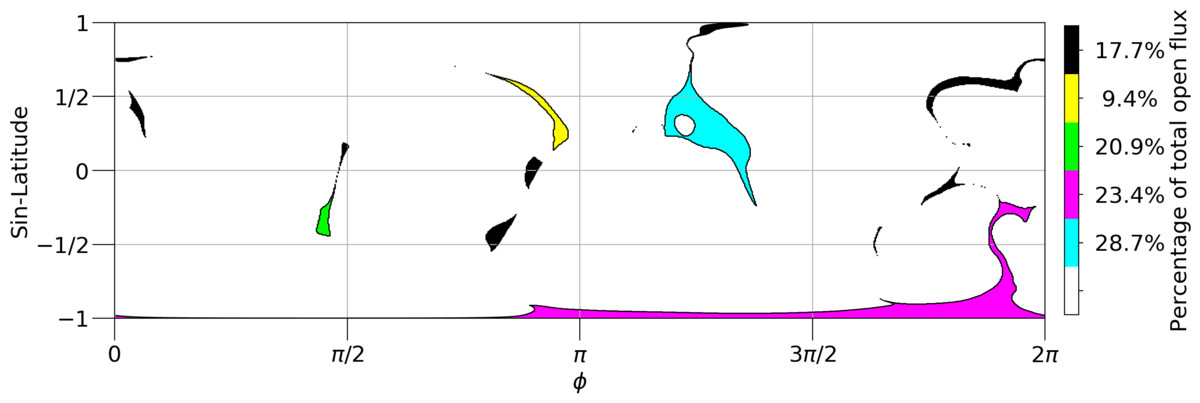}
\\
\rotatebox[origin=c]{90}{\mbox{{\bf G}ONG {\bf J}anis}}
&
\includegraphics[align=m,width=0.6\textwidth]{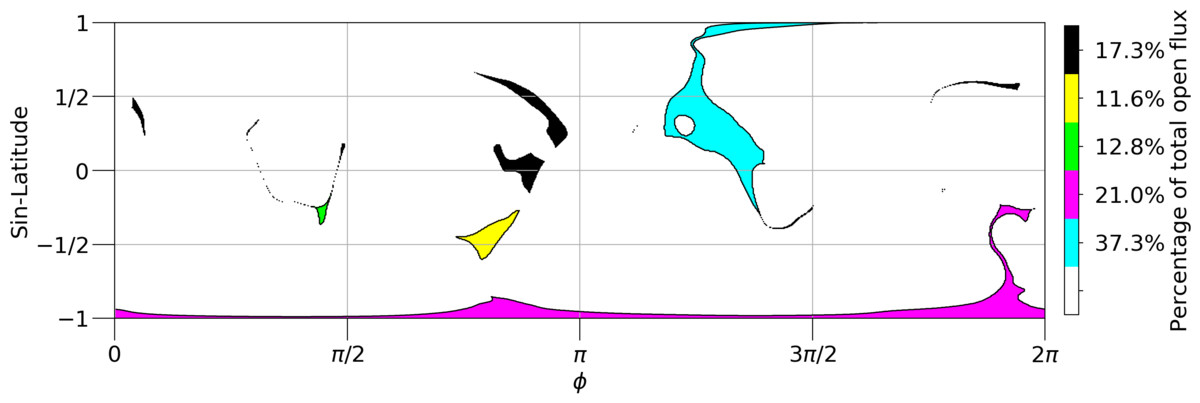}
\\
\rotatebox[origin=c]{90}{\mbox{{\bf H}MI {\bf S}ynoptic}}
&
\includegraphics[align=m,width=0.6\textwidth]{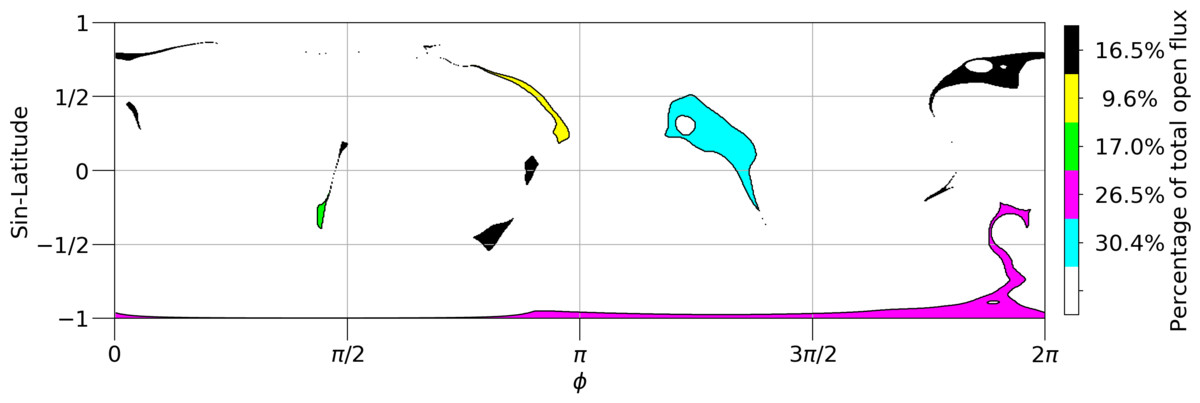}
\\
\rotatebox[origin=c]{90}{\mbox{{\bf H}MI {\bf D}aily}}
&
\includegraphics[align=m,width=0.6\textwidth]{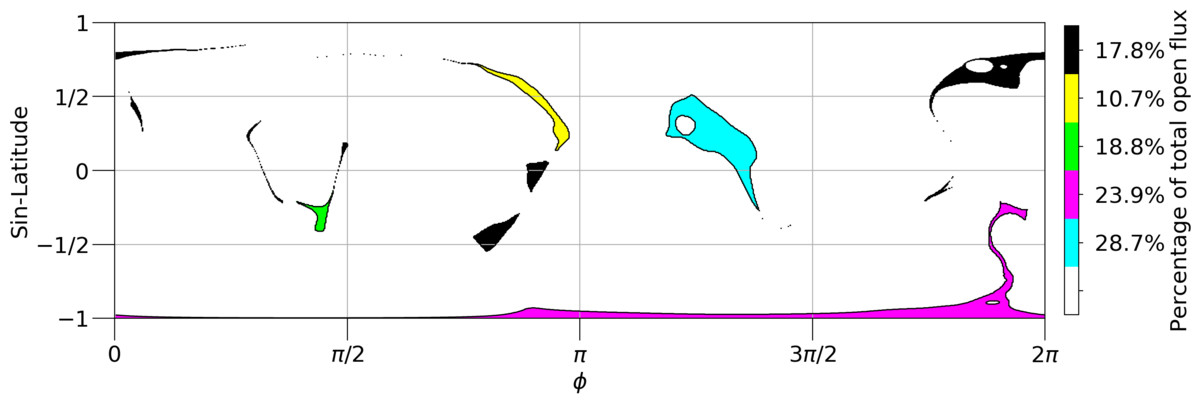}
\\
\rotatebox[origin=c]{90}{\mbox{{\bf PSI} (HMI data)}}
&
\includegraphics[align=m,width=0.6\textwidth]{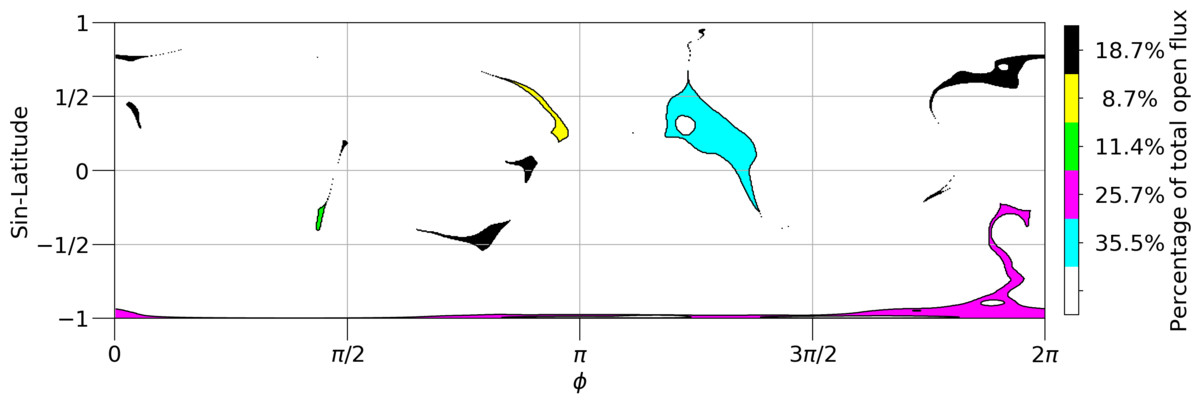}
\end{array}$
\caption{Open field maps from PF solutions for various input maps. Details of the map sources are given in the text.\label{fig_map_ch}} 
\end{figure}

In Fig~\ref{fig_map_ch} we show the maps of the open field at the surface or `open field maps', derived by tracing field lines through the PF solution for each $B_r$ map.  
The open regions are segmented and colored to indicate each region's percentage contribution to the total open flux. All maps yield somewhat similar open field maps, but with multiple small differences throughout.  A key observation is that a non-trivial amount of open flux is contained in scattered small regions of the open field. If these small open field regions are ubiquitous on the actual Sun, they may not be visible as coronal holes, and may contribute to the missing open flux identified by  \citet{linker2017open}.

\begin{figure}[htbp]
\centering
$\begin{array}{rcc}
\rotatebox[origin=c]{90}{\mbox{{\bf G}ONG {\bf S}ynoptic}}
& 
\includegraphics[align=m,width=0.45\textwidth]{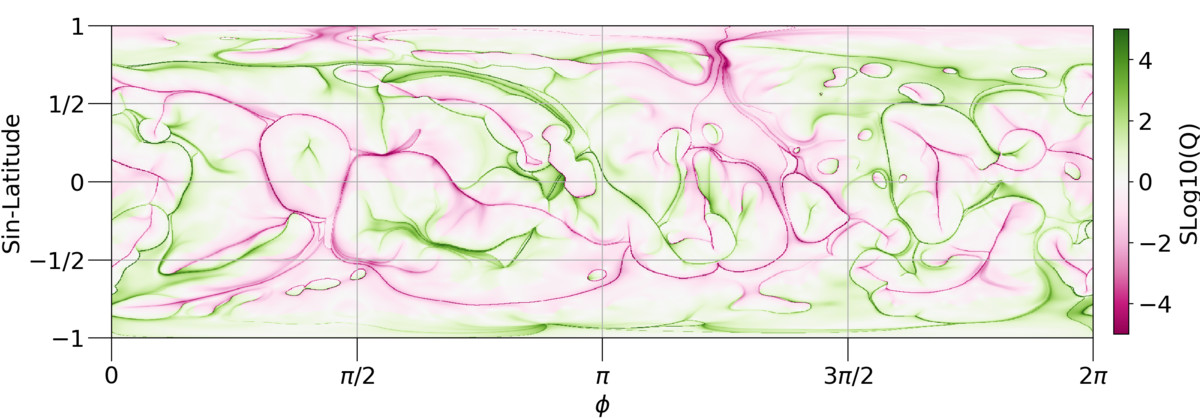}
&
\includegraphics[align=m,width=0.45\textwidth]{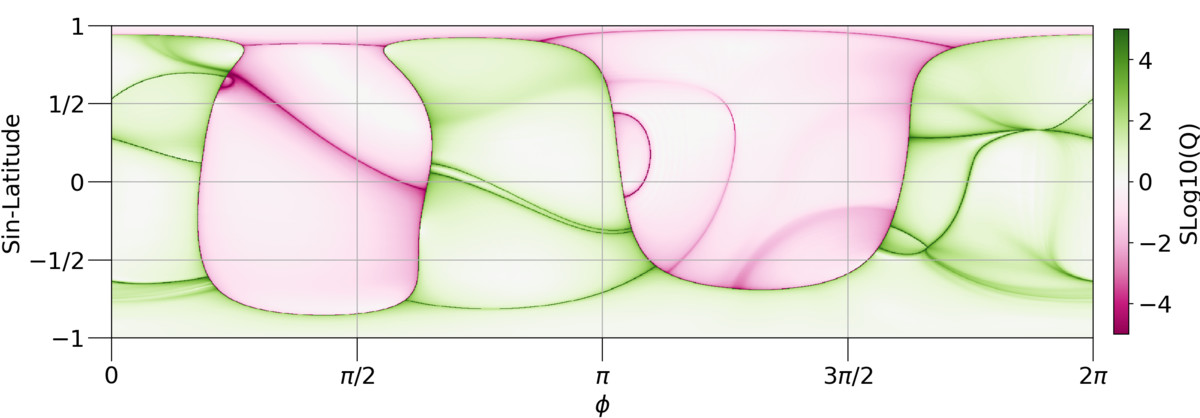}
\\
\rotatebox[origin=c]{90}{\mbox{{\bf G}ONG {\bf H}ourly}}
& 
\includegraphics[align=m,width=0.45\textwidth]{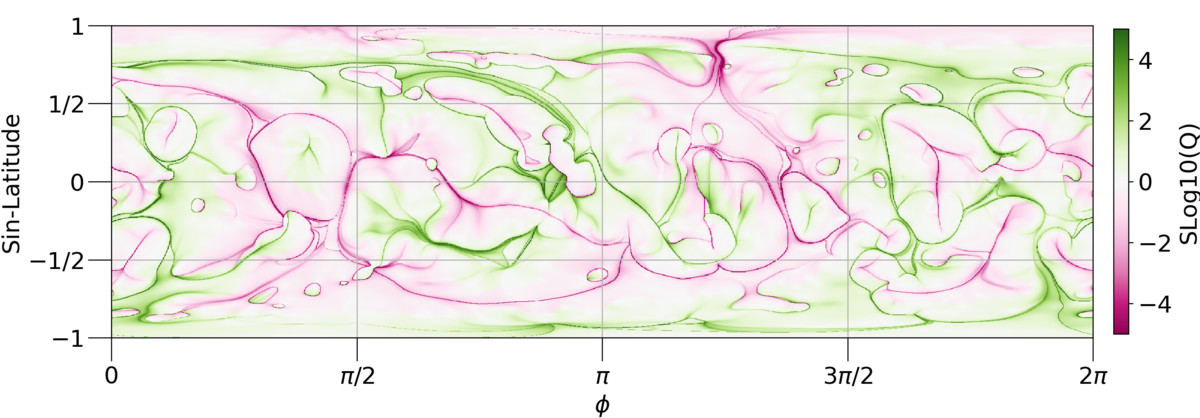}
&
\includegraphics[align=m,width=0.45\textwidth]{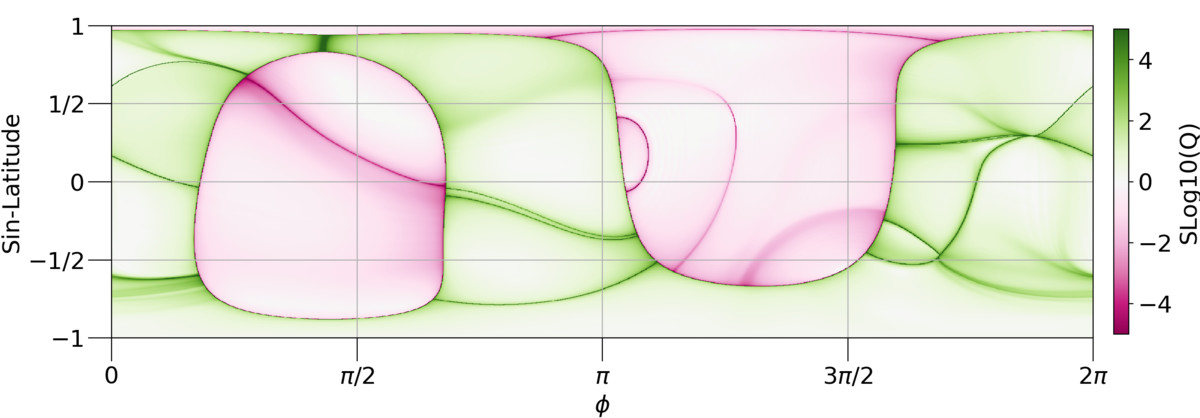}
\\
\rotatebox[origin=c]{90}{\mbox{{\bf G}ONG {\bf J}anis}}
& 
\includegraphics[align=m,width=0.45\textwidth]{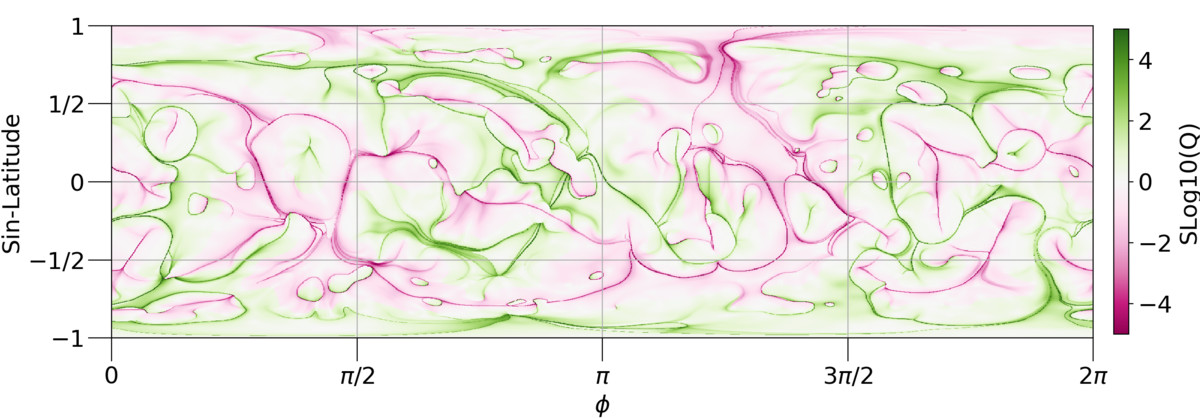}
&
\includegraphics[align=m,width=0.45\textwidth]{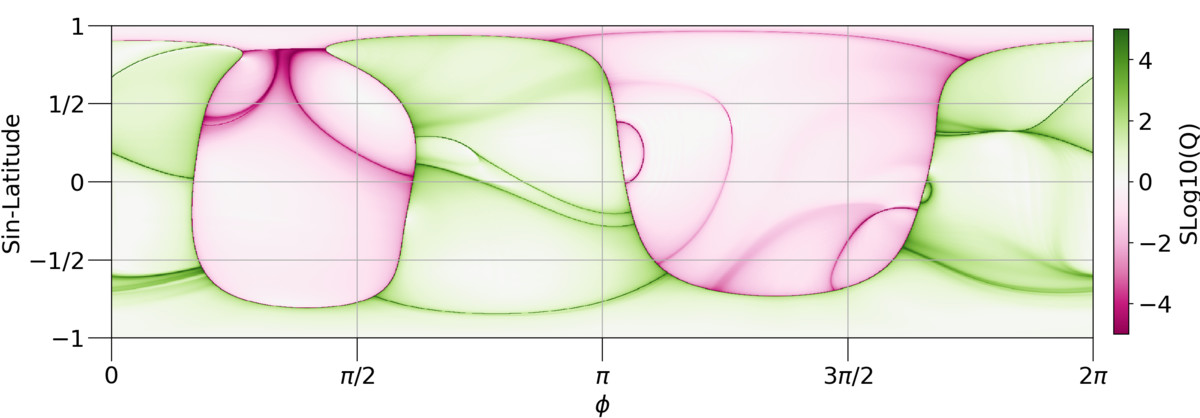}
\\
\rotatebox[origin=c]{90}{\mbox{{\bf H}MI {\bf S}ynoptic}}
& 
\includegraphics[align=m,width=0.45\textwidth]{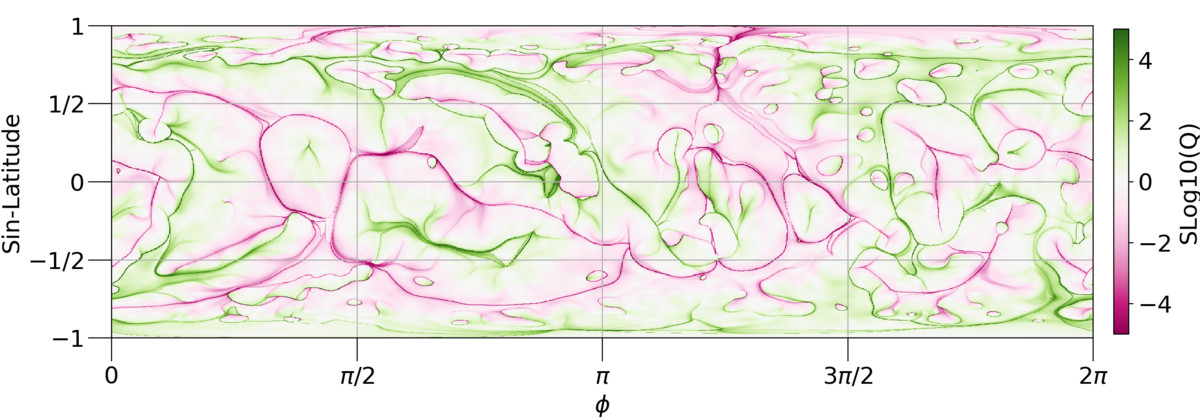}
&
\includegraphics[align=m,width=0.45\textwidth]{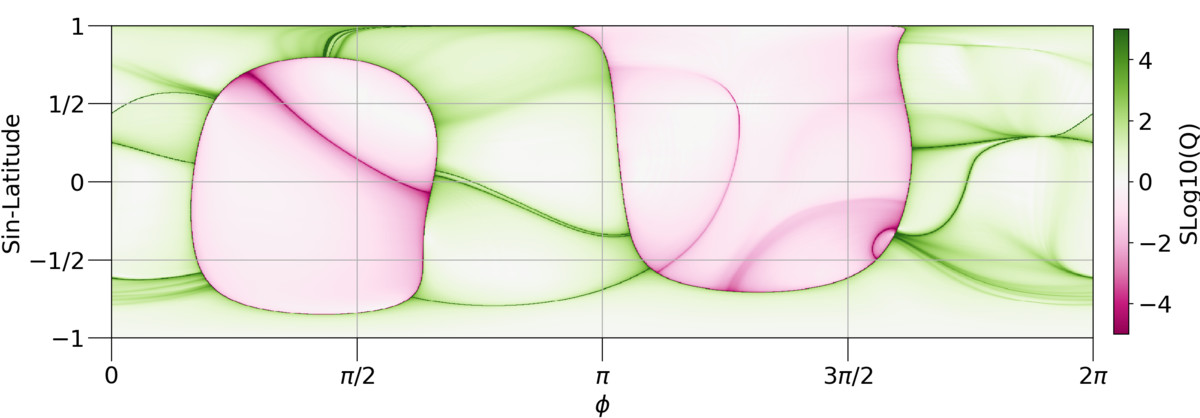}
\\
\rotatebox[origin=c]{90}{\mbox{{\bf H}MI {\bf D}aily}}
& 
\includegraphics[align=m,width=0.45\textwidth]{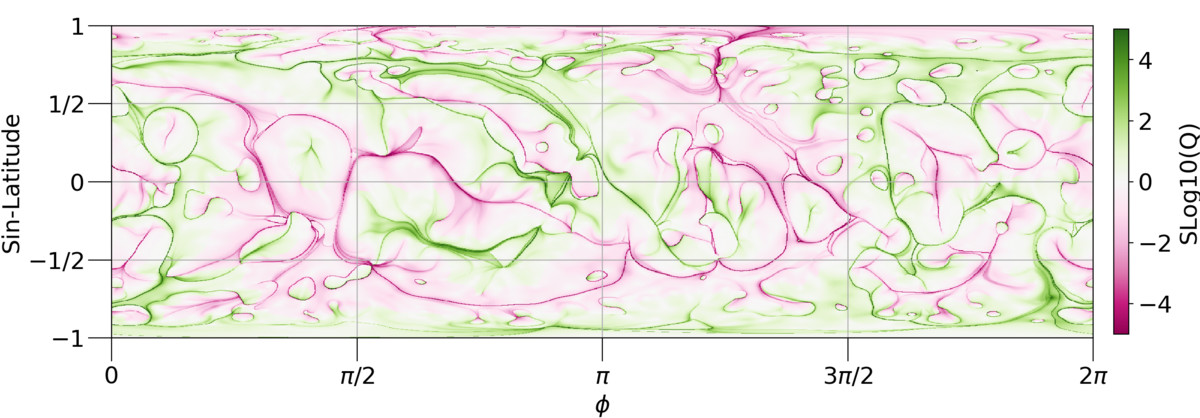}
&
\includegraphics[align=m,width=0.45\textwidth]{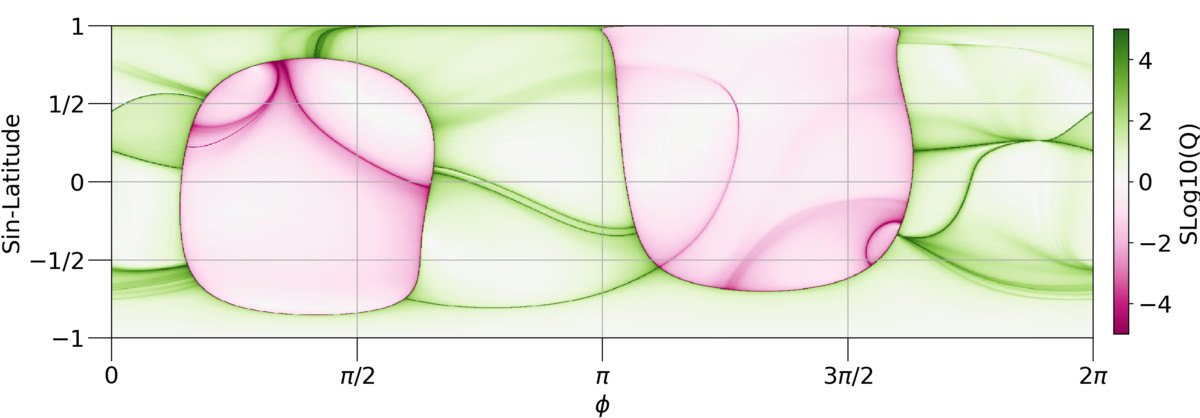}
\\
\rotatebox[origin=c]{90}{\mbox{{\bf PSI} (HMI data)}}
& 
\includegraphics[align=m,width=0.45\textwidth]{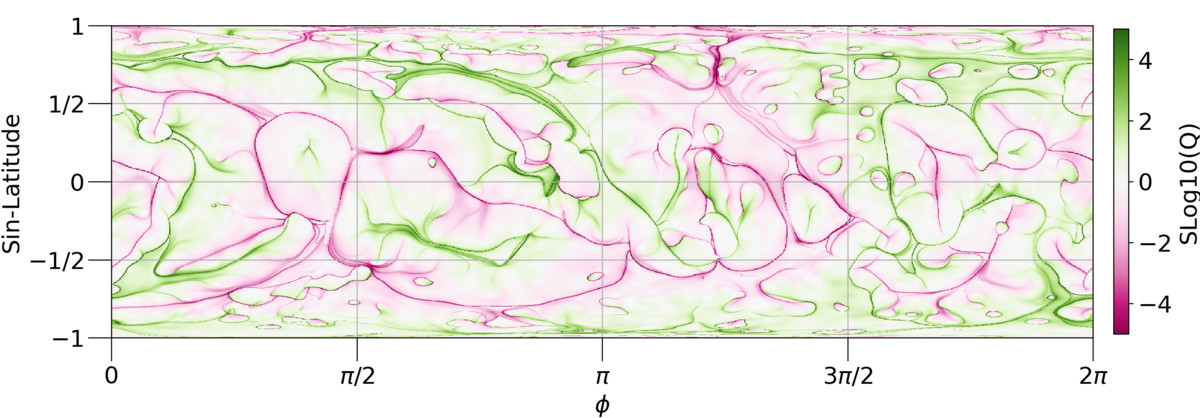}
&
\includegraphics[align=m,width=0.45\textwidth]{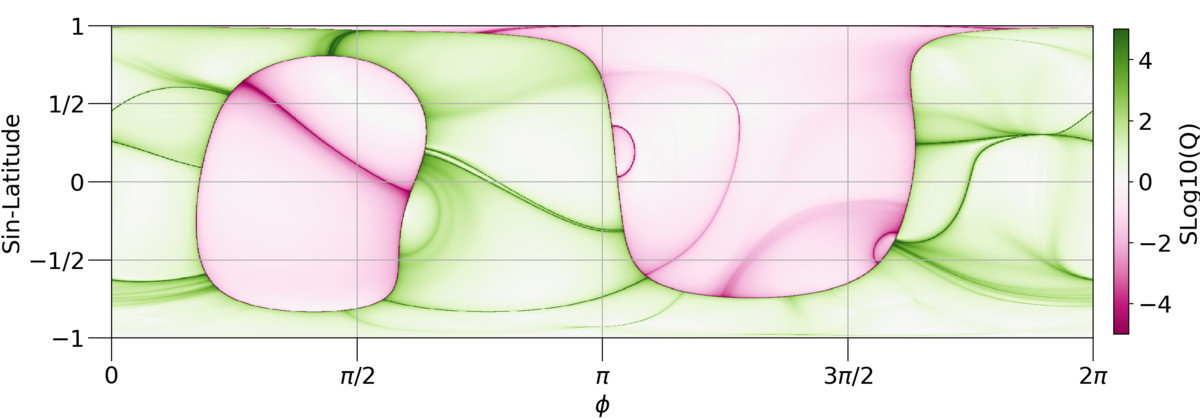}
\\
\, & Q\;\mbox{at}\;r=R_{\odot} & Q\;\mbox{at}\;r=r_{ss}
\end{array}$
\caption{Squashing factor from PF solutions for various input maps.  The squashing factor at the solar surface (left) and at the $r_{ss}$ outer radial boundary (right) are shown.  Details of the map sources are given in the text. \label{fig_map_q}} 
\end{figure}

In Fig~\ref{fig_map_q} we show maps of the squashing factor at the inner boundary and source surface. 
Although $Q$ appears similar in all maps, on close inspection, there exists small structural changes between the maps both at the surface and the outer radial boundary.  For example, the GONG Janis and HMI Daily maps at the outer radial boundary have a noticeably different $Q$ structure near $\phi=\pi/3$, $\cos\theta=1/2$  when compared to the rest of the maps. As these magnetograms are the data source for many coronal and solar wind models (empirical or MHD), these variations may be an important consideration.

\begin{figure}[htbp]
\centering
\includegraphics[align=m,width=0.45\textwidth]{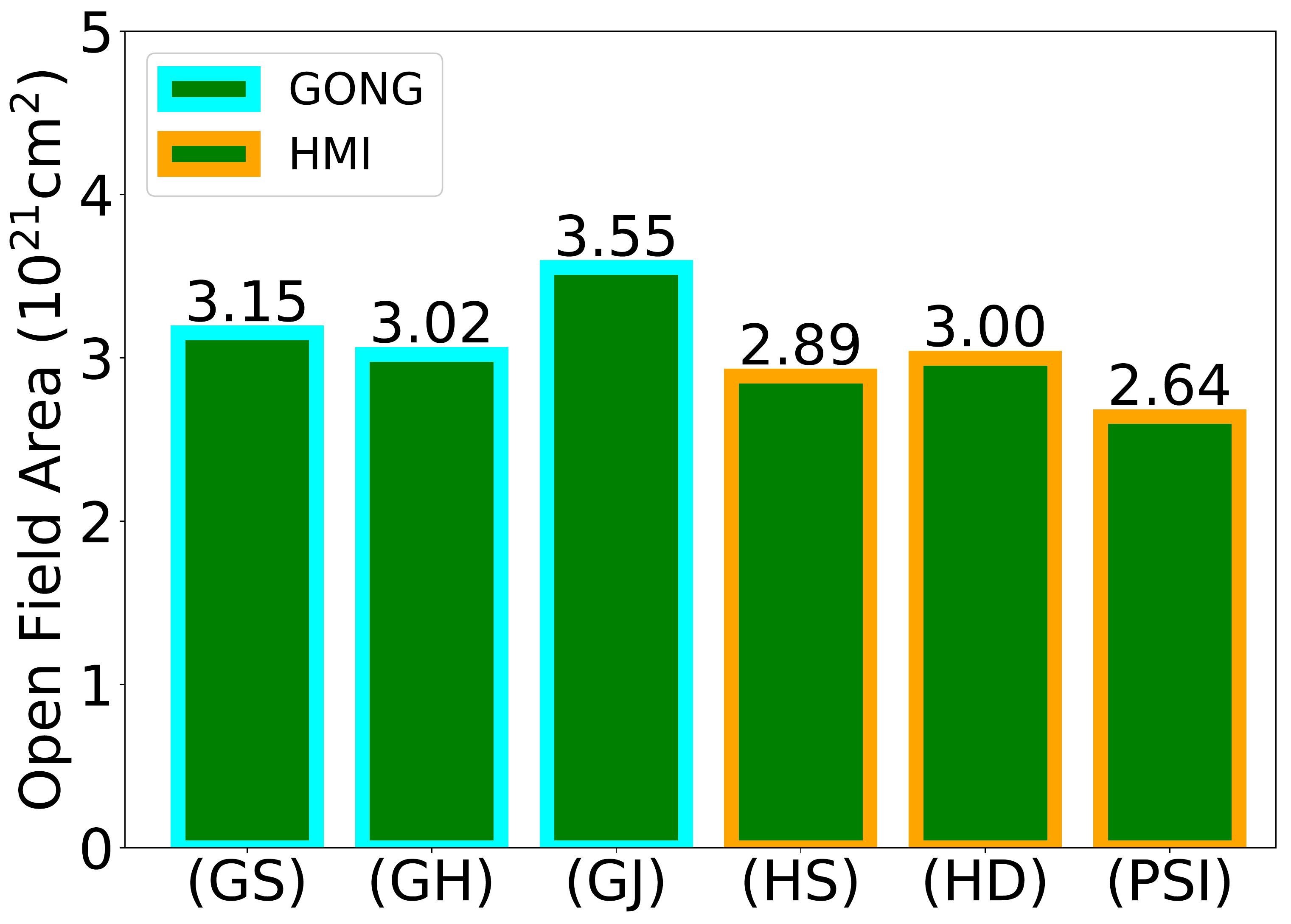}
\includegraphics[align=m,width=0.45\textwidth]{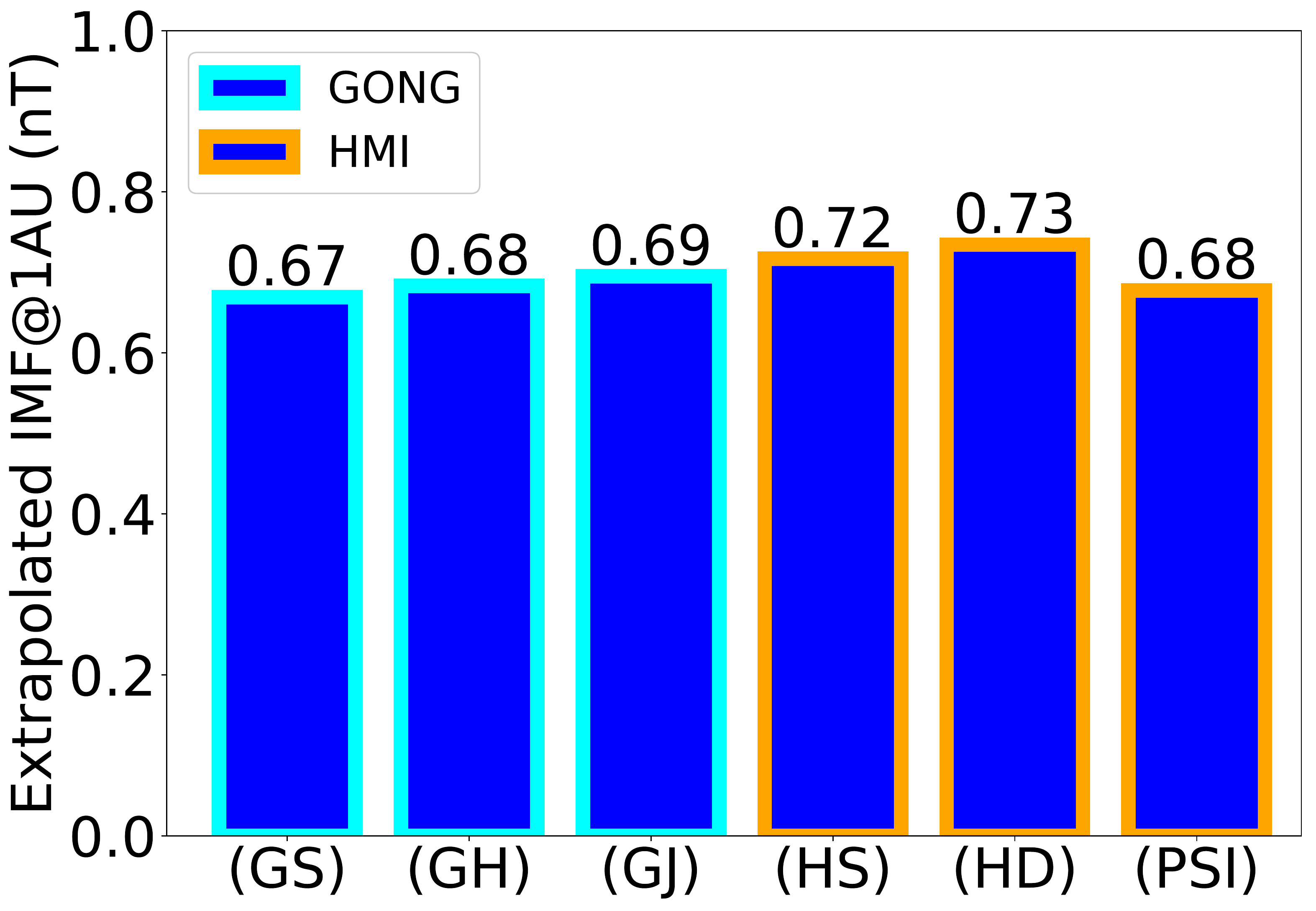}
\\
\includegraphics[align=m,width=0.45\textwidth]{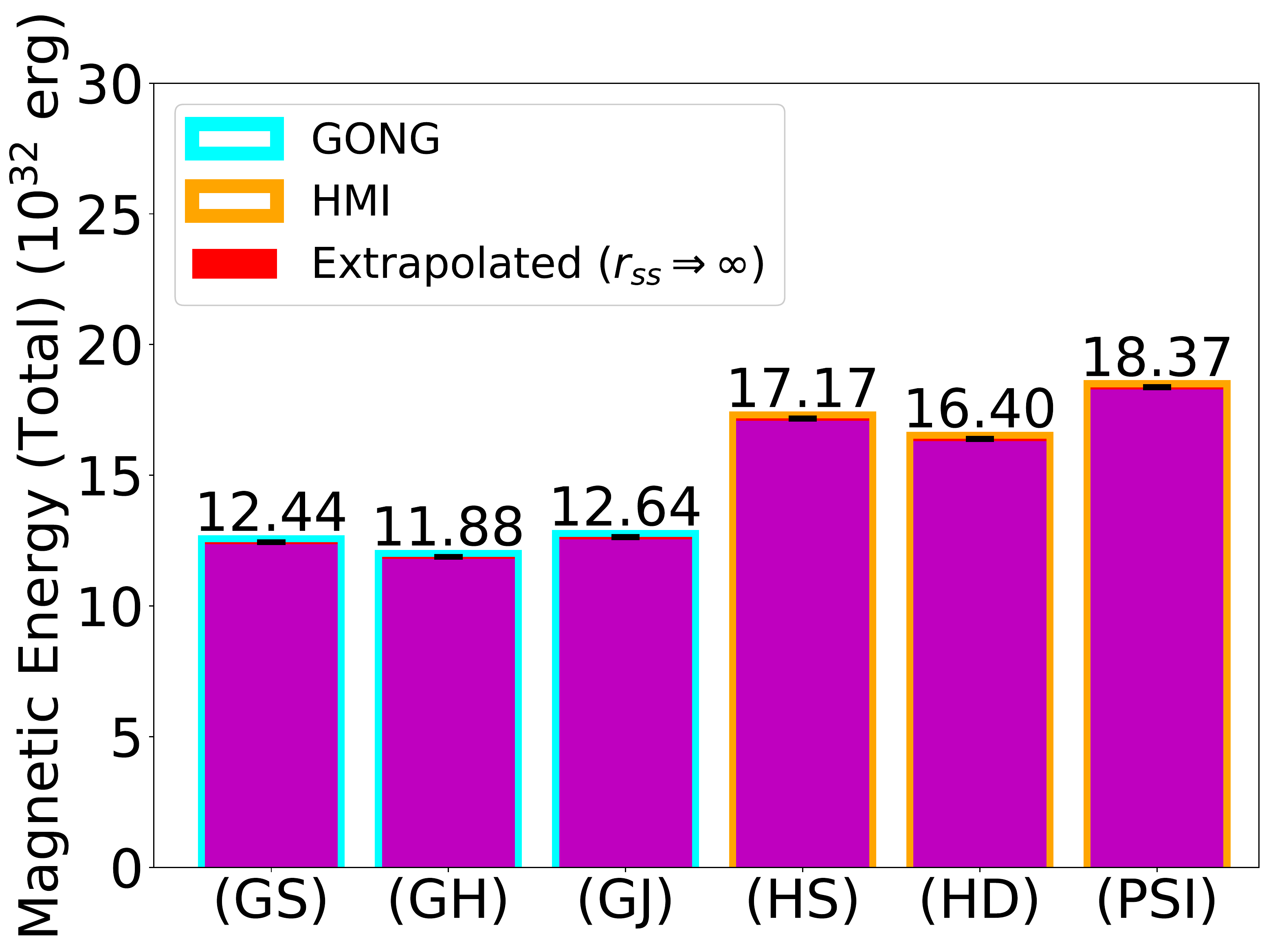}
\includegraphics[align=m,width=0.45\textwidth]{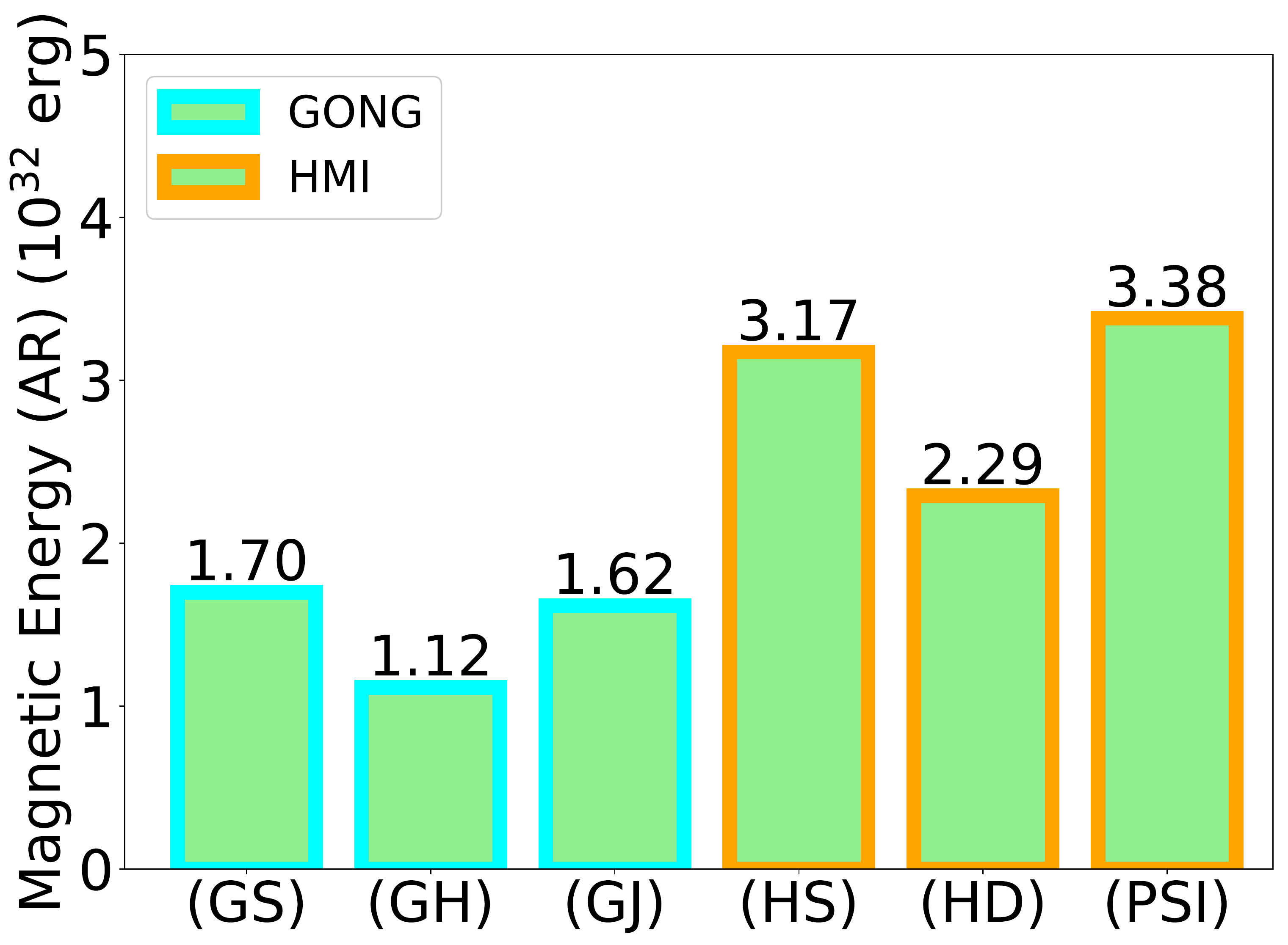}
\caption{Diagnostics (described in Sec.~\ref{sec:compare_method}) of potential fields varying the input surface magnetic map.  Descriptions of the map sources for each label is given in the text. \label{fig_map_diags}} 
\end{figure}

In Fig~\ref{fig_map_diags} we show the comparison quantities computed for each map as described in Sec.~\ref{sec:compare_method}. 
We see a clear grouping of the maps using GONG- or HMI data, with GONG-based maps exhibiting larger open field areas and smaller magnetic energies compared to HMI-derived maps.  However, the open flux for all maps are similar (within 10\%).   We also find that the magnetic energy (especially within the AR region) is highly variable even within the same instrument source.  This is likely due to the maps incorporating data in this region at different times (such as hourly vs. CR), and inherent differences between the observatories \citep{riley14}.   Therefore if one is interested in a region at a specific time, it is important to use data taken as close to that time as possible as the local properties can change significantly in a short amount of time due to surface evolution.

\section{Variations in Boundary Conditions}
\label{sec:var_bc}
Here we investigate how changing the outer boundary condition effects the results.  We compare a closed-wall outer boundary to a source surface boundary at various radii.  The selected options are
\begin{itemize}
\item[] {\bf (CW2.5)} Closed-wall with $r_1=2.5R_{\odot}$
\item[] {\bf (SS1.5)} Source-surface with $r_{ss}=1.5R_{\odot}$
\item[] {\bf (SS2.0)} Source-surface with $r_{ss}=2.0R_{\odot}$
\item[] {\bf (SS2.5)} Source-surface with $r_{ss}=2.5R_{\odot}$
\end{itemize}

\begin{figure}[htbp]
\centering
$\begin{array}{rcc}
\rotatebox[origin=c]{90}{$r_{ss}=1.5R_{\odot}$}
& 
\includegraphics[align=m,width=0.45\textwidth]{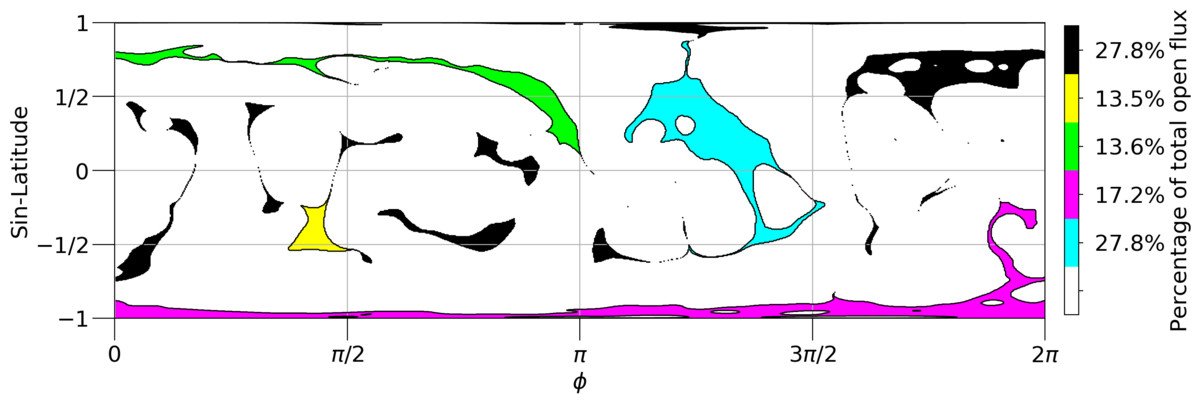}
&
\includegraphics[align=m,width=0.45\textwidth]{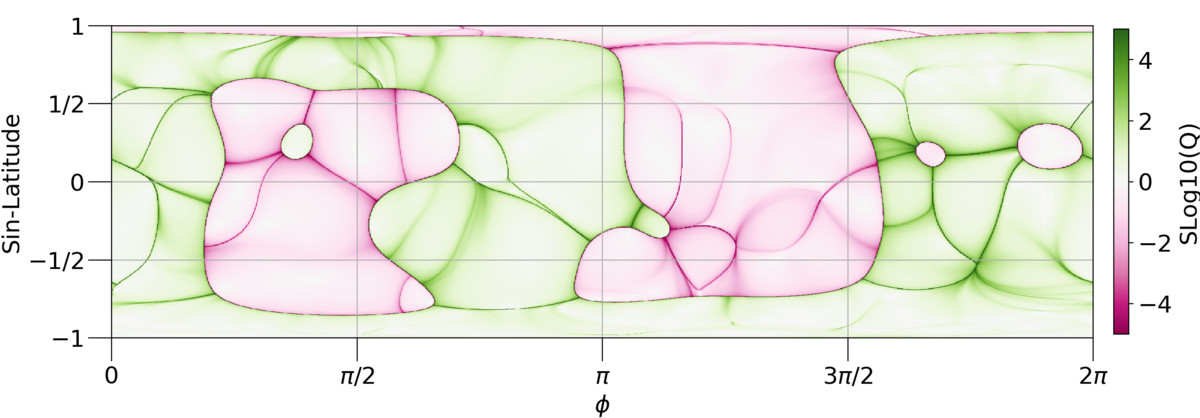}
\\
\rotatebox[origin=c]{90}{$r_{ss}=2.0R_{\odot}$}
& 
\includegraphics[align=m,width=0.45\textwidth]{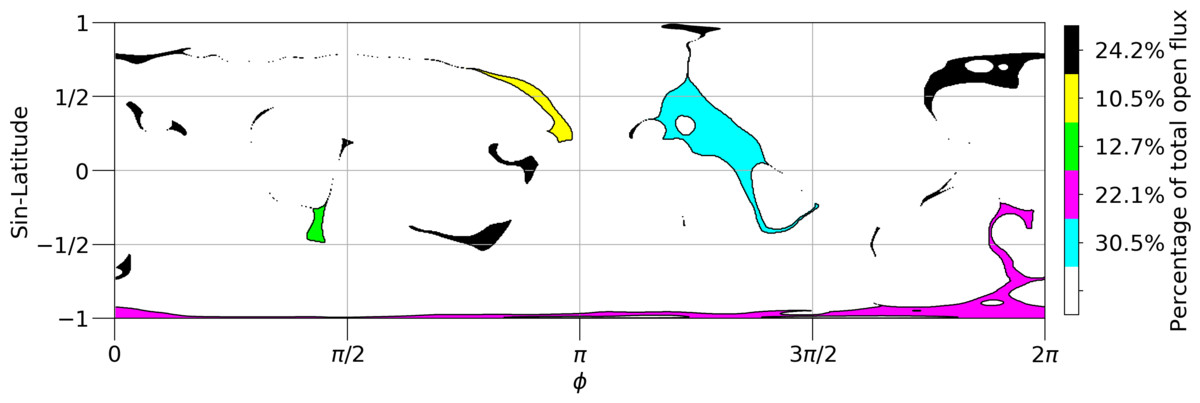}
&
\includegraphics[align=m,width=0.45\textwidth]{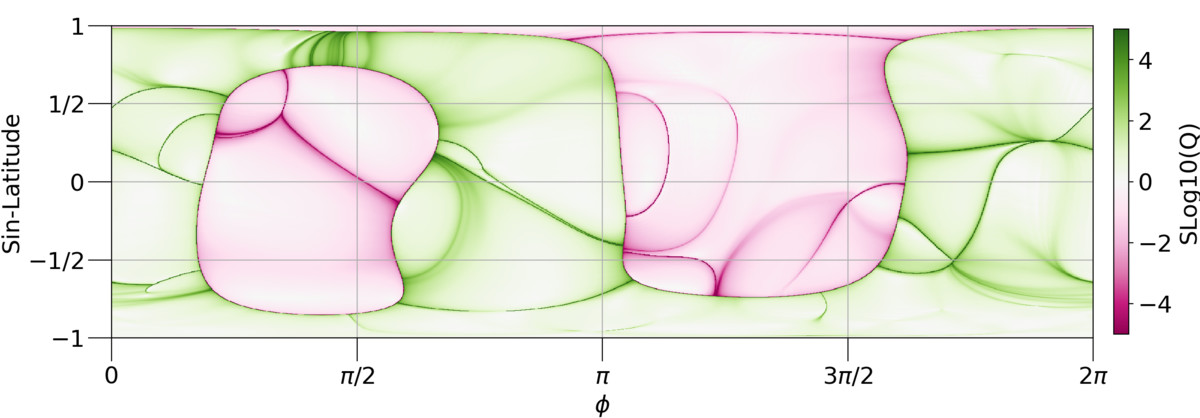}
\\
\rotatebox[origin=c]{90}{$r_{ss}=2.5R_{\odot}$}
& 
\includegraphics[align=m,width=0.45\textwidth]{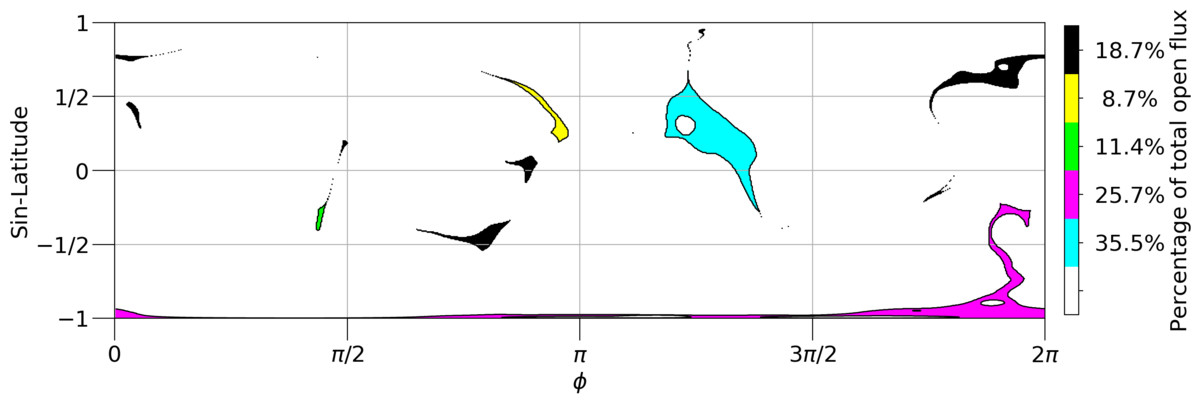}
&
\includegraphics[align=m,width=0.45\textwidth]{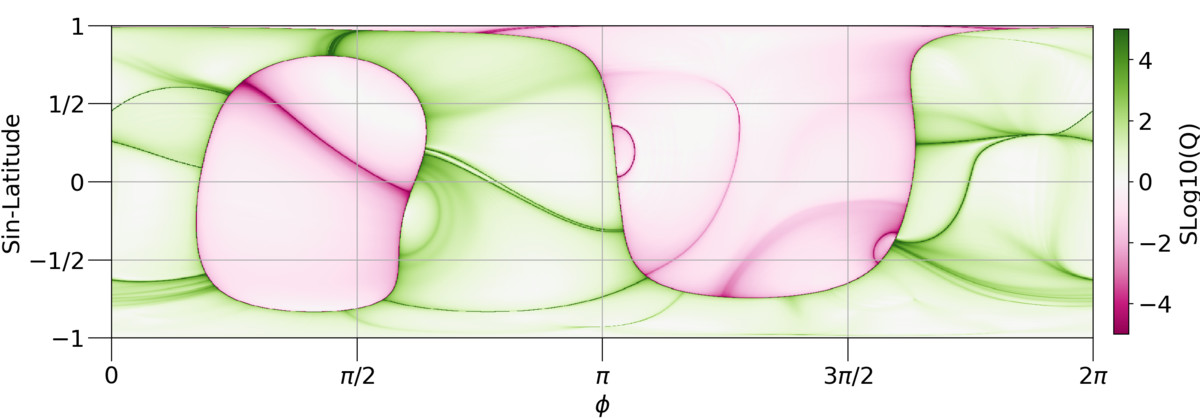}
\end{array}$
\caption{Open field maps (left) and outer radial boundary squashing factor (right) from PF solutions for various source-surface radii.\label{fig_bcq_ch}} 
\end{figure}

In Fig~\ref{fig_bcq_ch} we show the open field regions and the squashing factor at the upper radial boundary from the PF solution for each boundary condition option (excluding closed-wall, as it results in no open field).   
As expected, the open field maps vary significantly as the source-surface radius is changed, with an increase in open field area as the source-surface radius is lowered.  The $Q$ maps show a number morphological differences between them and even altered topology at some locations.  For example, at the lowest radius tested, opposite-polarity islands appear within uni-polar regions that are missing for the higher radii cases (i.e. new open field regions, new topological linkages).

\begin{figure}[htbp]
\centering
$\begin{array}{rc}
\rotatebox[origin=c]{90}{{\bf (CW)} $r_{1}=2.5R_{\odot}$}
& 
\includegraphics[align=m,width=0.6\textwidth]{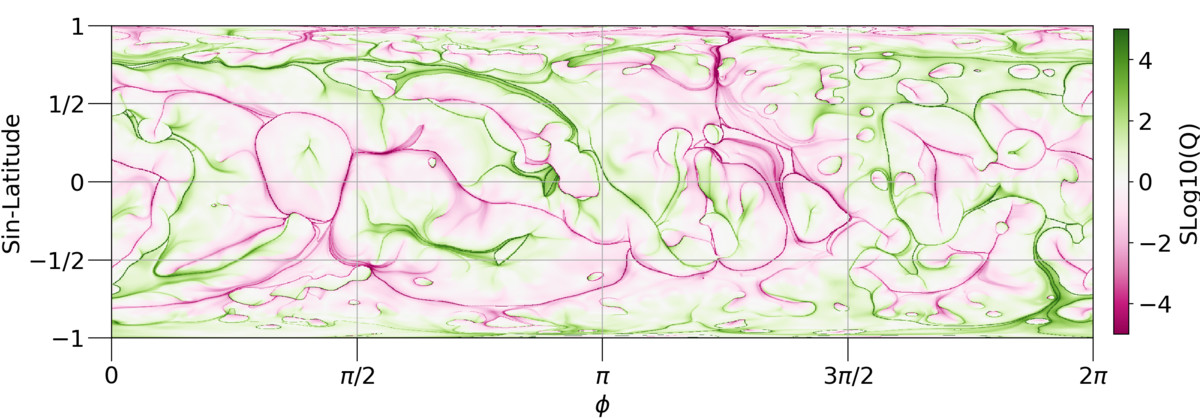}
\\
\rotatebox[origin=c]{90}{{\bf (SS)} $r_{ss}=1.5R_{\odot}$}
& 
\includegraphics[align=m,width=0.6\textwidth]{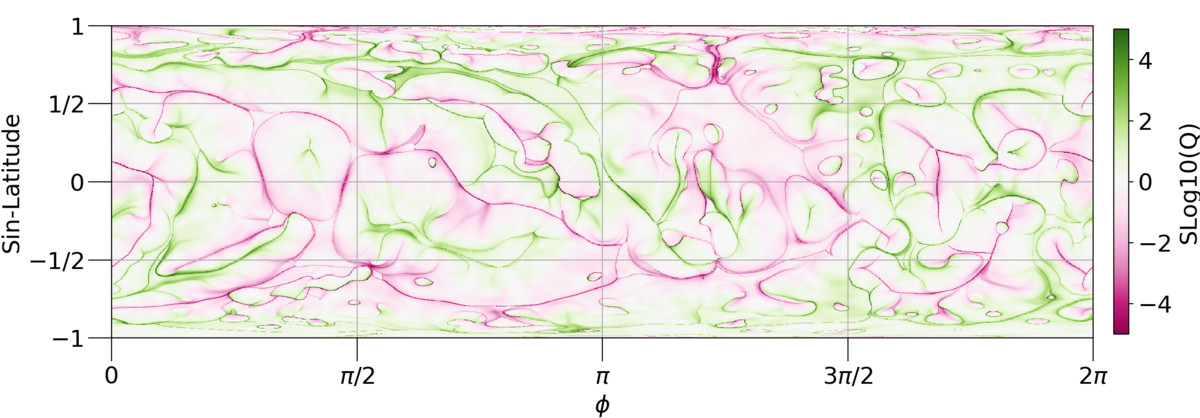}
\\
\rotatebox[origin=c]{90}{{\bf (SS)} $r_{ss}=2.0R_{\odot}$}
& 
\includegraphics[align=m,width=0.6\textwidth]{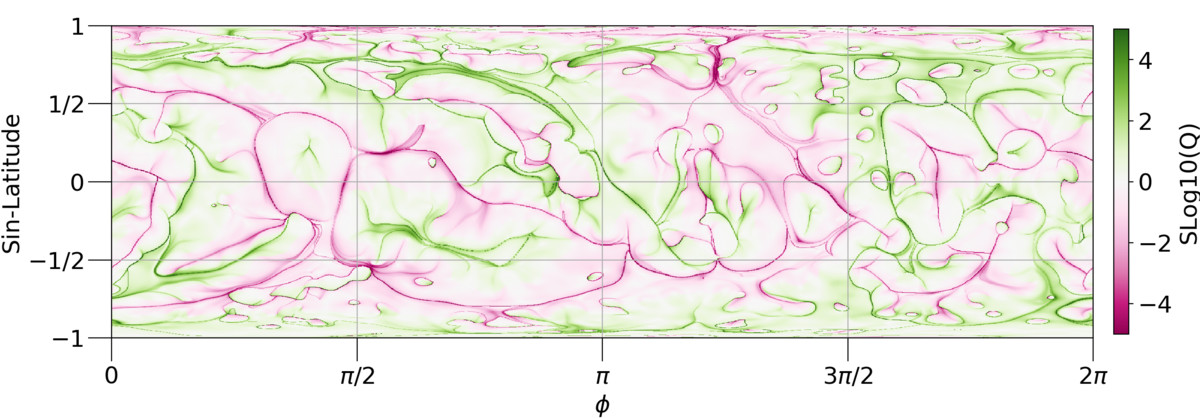}
\\
\rotatebox[origin=c]{90}{{\bf (SS)} $r_{ss}=2.5R_{\odot}$}
& 
\includegraphics[align=m,width=0.6\textwidth]{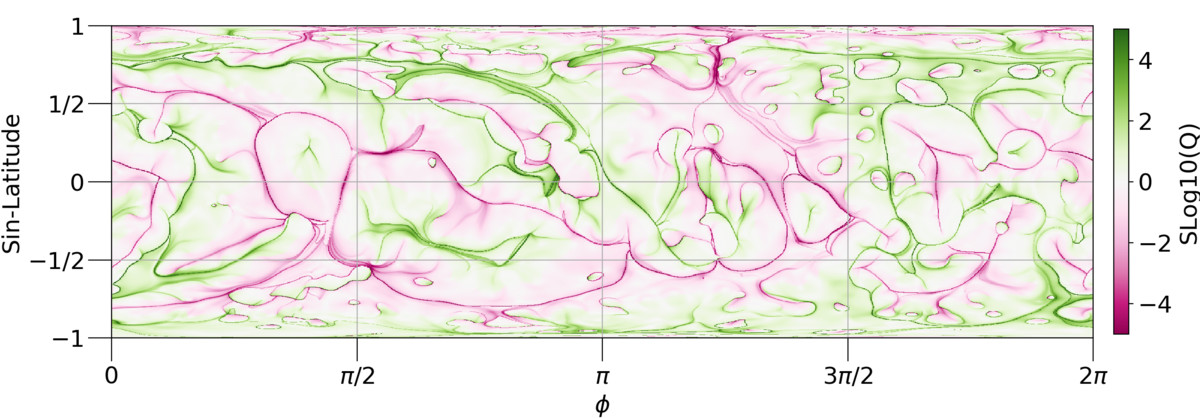}
\end{array}$
\caption{Squashing factor at the solar surface of PF solutions for various upper radial boundary conditions. \label{fig_bc_bq}} 
\end{figure}

In Fig~\ref{fig_bc_bq}, we show the squashing factor at the lower radial boundary for each upper-boundary condition option.  
Here we see that all maps that use a source-surface radial boundary exhibit similar features and morphology. This is true even for the closed-wall case, differing substantially only in places that would otherwise be open in a source-surface calculation.

\begin{figure}[htbp]
\centering
\includegraphics[align=m,width=0.45\textwidth]{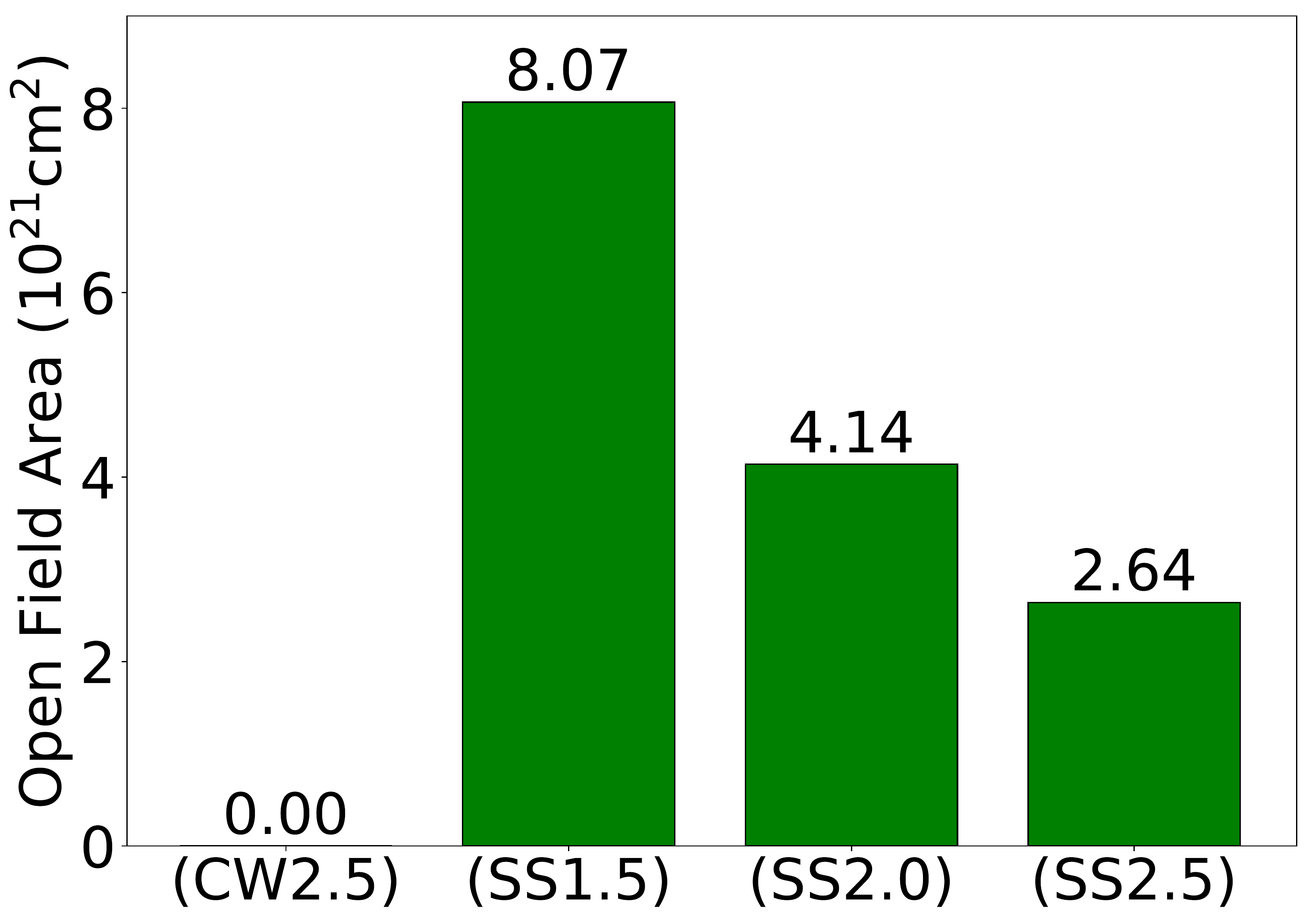}
\includegraphics[align=m,width=0.45\textwidth]{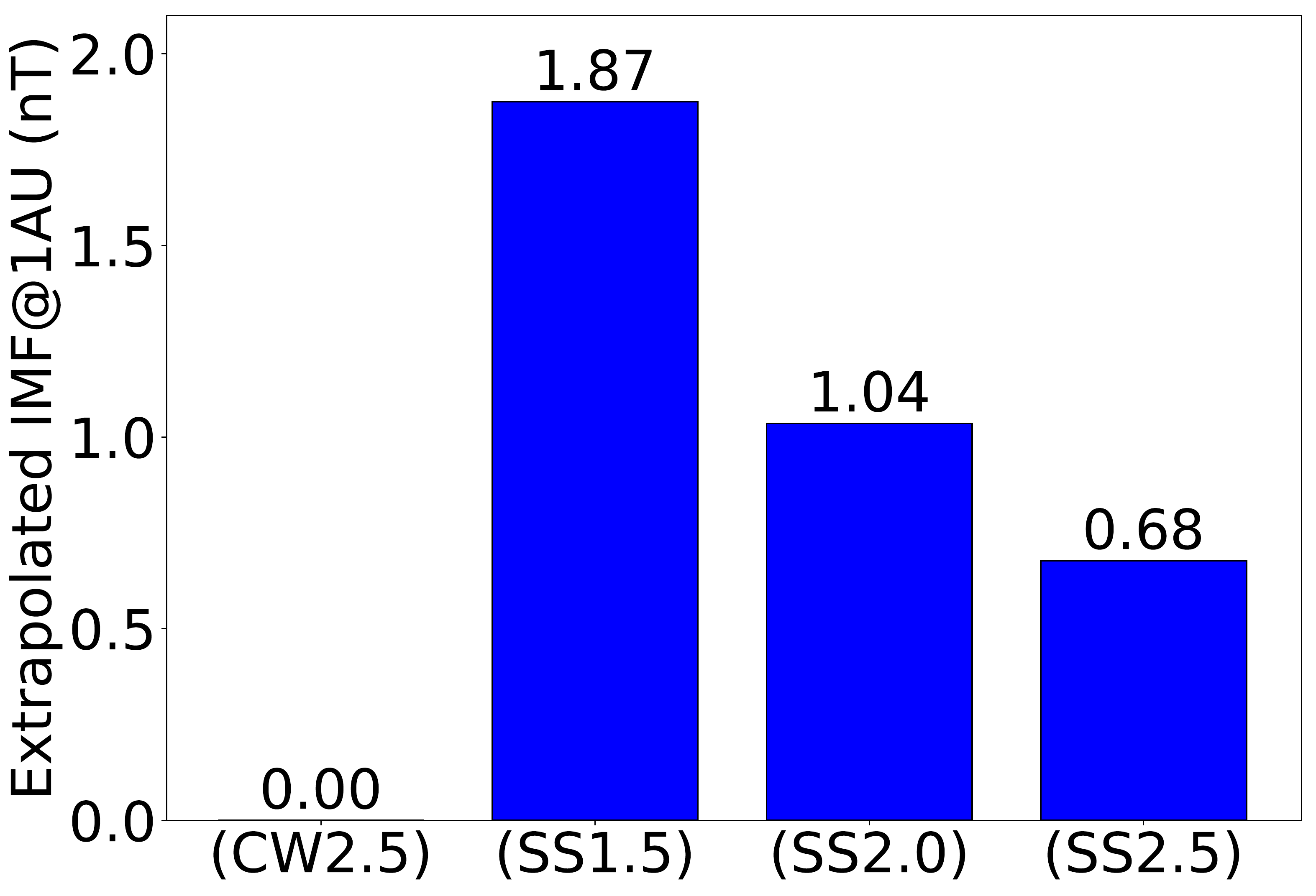}
\\
\includegraphics[align=m,width=0.45\textwidth]{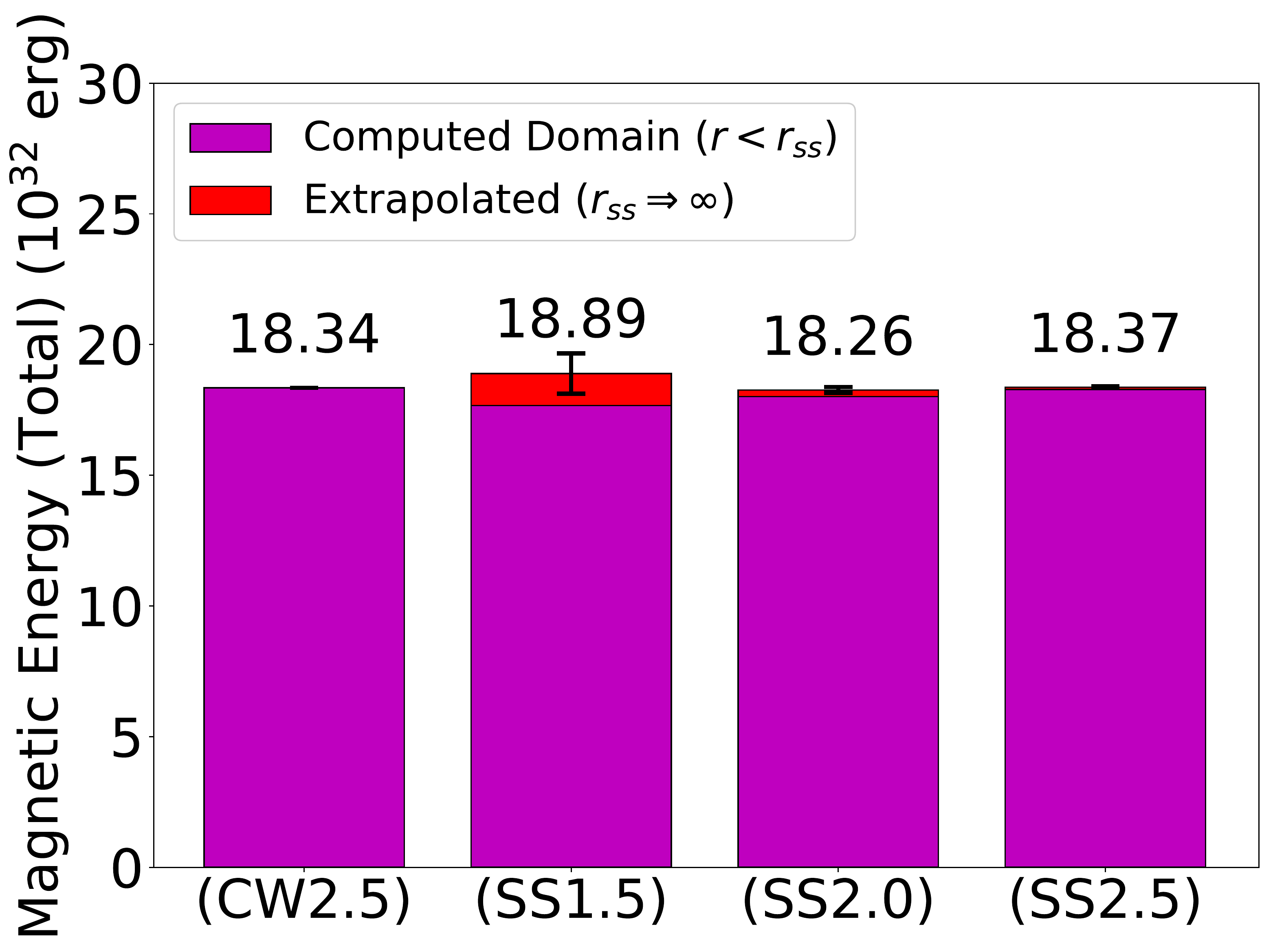}
\includegraphics[align=m,width=0.45\textwidth]{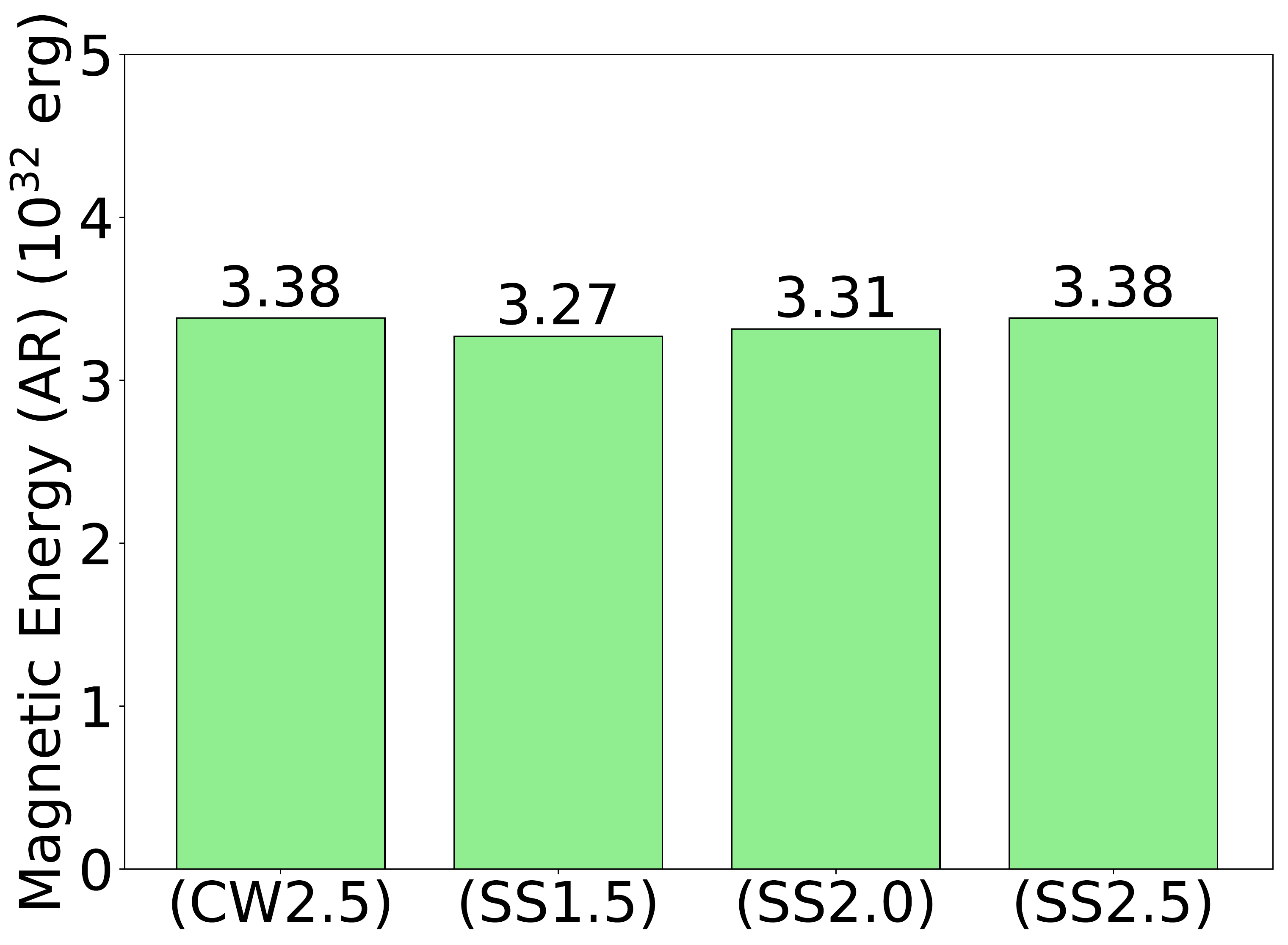}
\caption{Diagnostics of potential fields (described in Sec.~\ref{sec:compare_method}) varying the upper radial boundary condition.  Details of each boundary condition label is given in the text. \label{fig_bc_diags}} 
\end{figure}

In Fig~\ref{fig_bc_diags} we show the derived quantities described in Sec.~\ref{sec:compare_method}.  
We see an expected increase in open field area and open flux as the source-surface radius is lowered.  Both the global magnetic energy and AR energy is very similar across all cases. This is perhaps not surprising but worth noting, as these numbers are strongly dependent on the inner boundary (Secs.~\ref{sec:var_input}~\&~\ref{sec:var_res}) but not on the outer boundary condition, even for the closed wall case. Related to this, we also find that the extrapolated magnetic energy outside of the source-surface radius is very small compared to that within the domain (red bars, bottom left). Even with a low source-surface of $r_{ss}=1.5R_{\odot}$, it accounts for only $6.5\%$ of the total energy.  For higher $r_{ss}$, this rapidly decreases to less than $0.5\%$.

\section{Variations in Resolution}
\label{sec:var_res}
Here we investigate how the PF solutions change based on the resolution of the magnetogram and the corresponding 3D domain.  Using our HMI-derived custom magnetogram (map {\bf (PSI)} in Sec.~\ref{sec:var_input}), we compute PF solutions at increasing map resolutions up the native (full) resolution of a HMI synoptic map ($3600\!\times\!1440$ in longitude/sin-latitude).  As mentioned in Sec.~\ref{sec:compare_method}, for each map resolution, we set the number of radial grid points to be $N_r=N_{\phi}/6.67$.  The map resolutions we chose are listed below, showing the resolution of each dimension ($N_{\phi}\times N_{\theta}\times (N_r)$) and the resulting total number of grid cells.
\begin{itemize}
\item {\bf (TNY)} Tiny: $180\times 90$ $\times$ (27) $\approx 0.4$ million cells (uniform in $\theta,\phi$)
\item {\bf (SML)} Small: $360\times 180$ $\times$ (54) $\approx 3.5$ million cells (uniform in $\theta,\phi$) 
\item {\bf (MED)} Medium: $900\times 450$ $\times$ (135) $\approx 54.7$ million cells (uniform in $\theta,\phi$)
\item {\bf (LRG)} Large: $1800\times 900$ $\times$ (207) $\approx 335.3$ million cells (uniform in $\theta,\phi$)
\item {\bf (NAT)} Native: $3973\times 2012$ $\times$(827) $\approx 6.6$ \emph{billion} cells \\
(non-uniform in $\theta,\phi$ ranging from the native resolution of SHARP to that of HMI Synoptic)
\item {\bf (PSI)} $1095\times 742$ $\times$(216) $\approx 175.5$ million cells \\
(non-uniform in $\theta,\phi$ ranging from the native resolution of SHARP to that of {\bf (SML)})
\end{itemize}

\begin{figure}[htbp]
\centering
\includegraphics[align=m,width=0.325\textwidth]{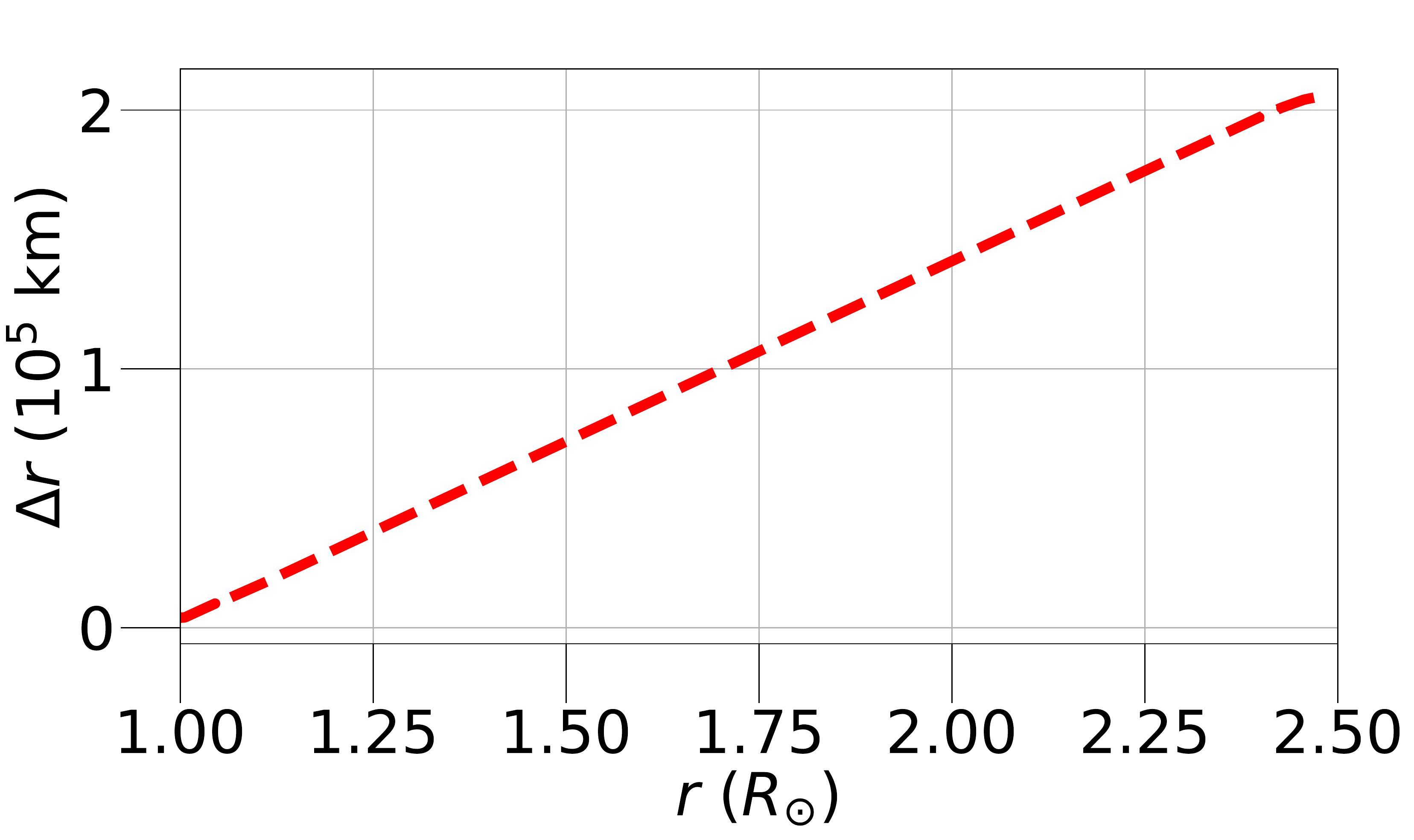}
\includegraphics[align=m,width=0.325\textwidth]{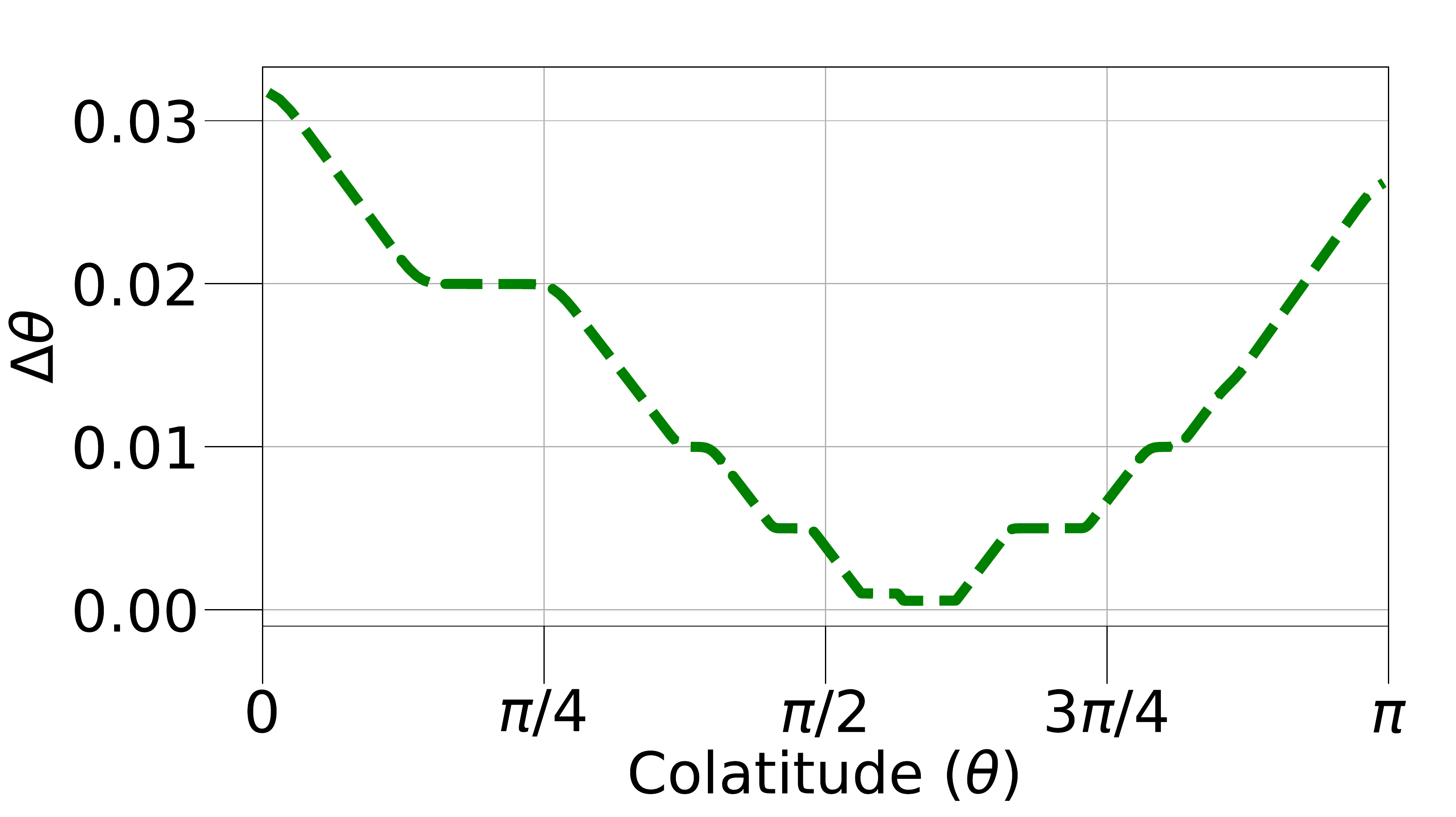}
\includegraphics[align=m,width=0.325\textwidth]{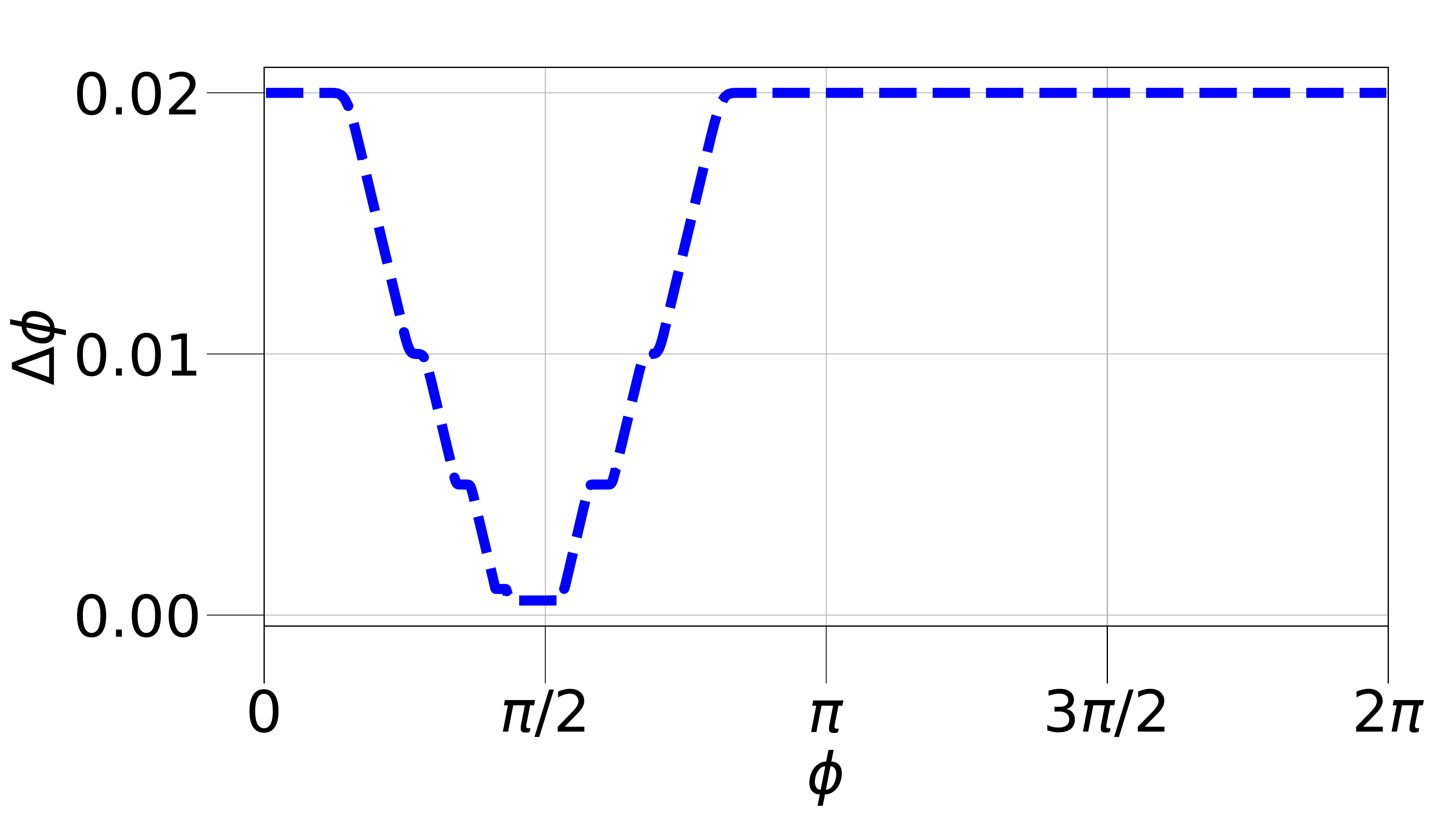}
\includegraphics[align=m,width=0.45\textwidth]{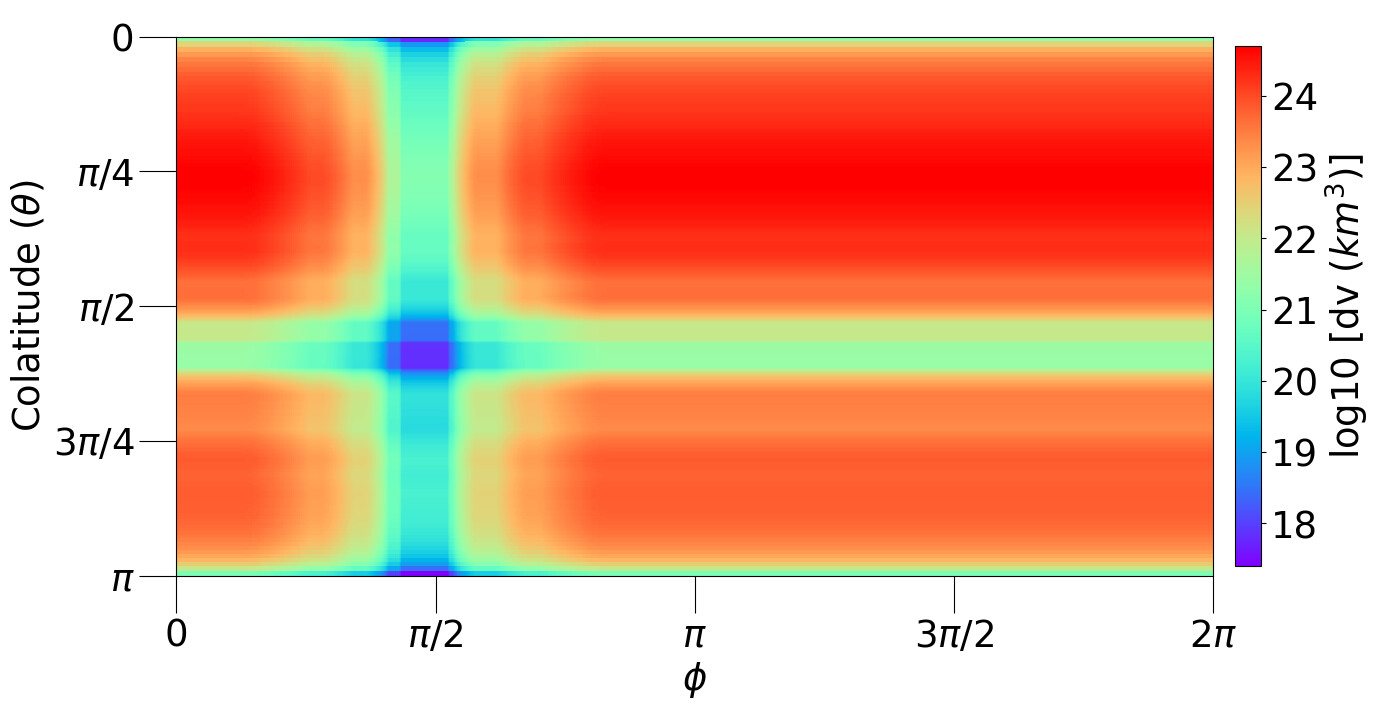}
\includegraphics[align=m,width=0.35\textwidth]{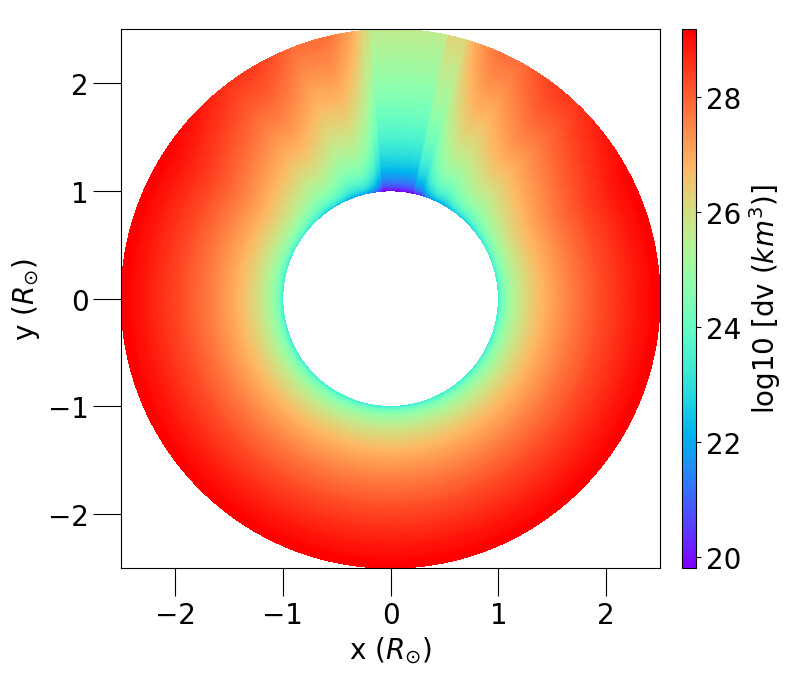}
\caption{Depiction of the non-uniformity of the grid for map resolution {\bf (PSI)}. Top row, left to right: $\Delta r$, $\Delta \theta$, $\Delta \phi$.  Bottom row: $dv=r^2\,\sin(\theta)\,\Delta r\,\Delta \theta\,\Delta \phi$ in the $\theta$--$\phi$ plane at $r=R_{\odot}$ (left) and the $r$--$\phi$ plane at $\theta=\pi/2$ (right). \label{fig_res_grid}} 
\end{figure}

As shown, maps {\bf (TNY)} through {\bf (LRG)} use a uniform resolution in co-latitude ($\theta$) and longitude ($\phi$).  Resolution {\bf (NAT)} is non-uniform in $\theta$ and $\phi$ in order to achieve high resolution equivalent to the native (full) resolution of the HMI SHARP data product near the AR ($\Delta \phi=5.6\!\times\!10^{-4}$). We then coarsen the grid to the native HMI synoptic map resolution outside the AR.  Our mesh in {\bf (PSI)} is very non-uniform in order to achieve the same high resolution near the AR, but coarsen dramatically outside the AR to the level of the {\bf (SML)} map.  This is useful for high-resolution  studies of a small region, while maintaining a global solution around it, eliminating the needs for ad-hoc or complicated boundary conditions in a localized `box' domain.  A depiction of the non-uniformity of the grid is shown in Fig.~\ref{fig_res_grid}.

\begin{figure}[htbp]
\centering
$\begin{array}{rcc}
\rotatebox[origin=c]{90}{\mbox{{\bf TNY}} ($180\!\!\times\!\!90$)}
& 
\includegraphics[align=m,width=4in]{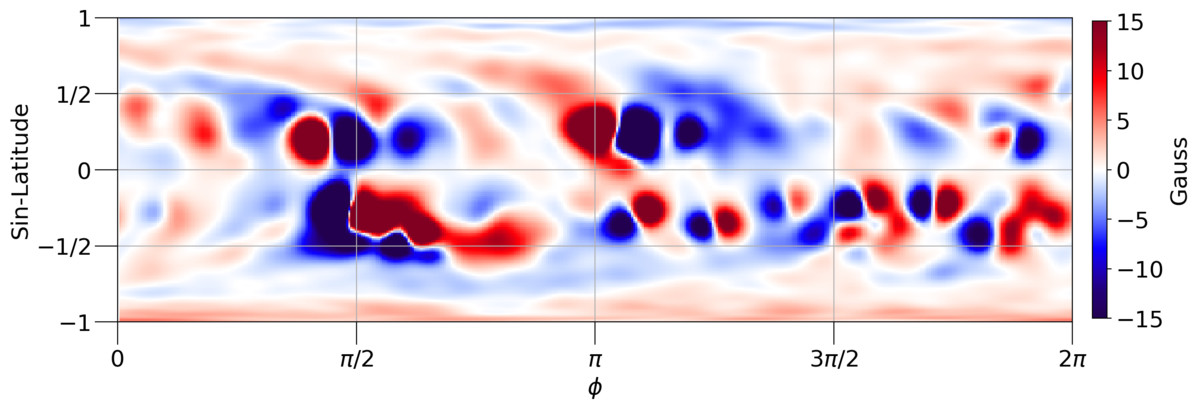}
&
\includegraphics[align=m,width=2.6in]{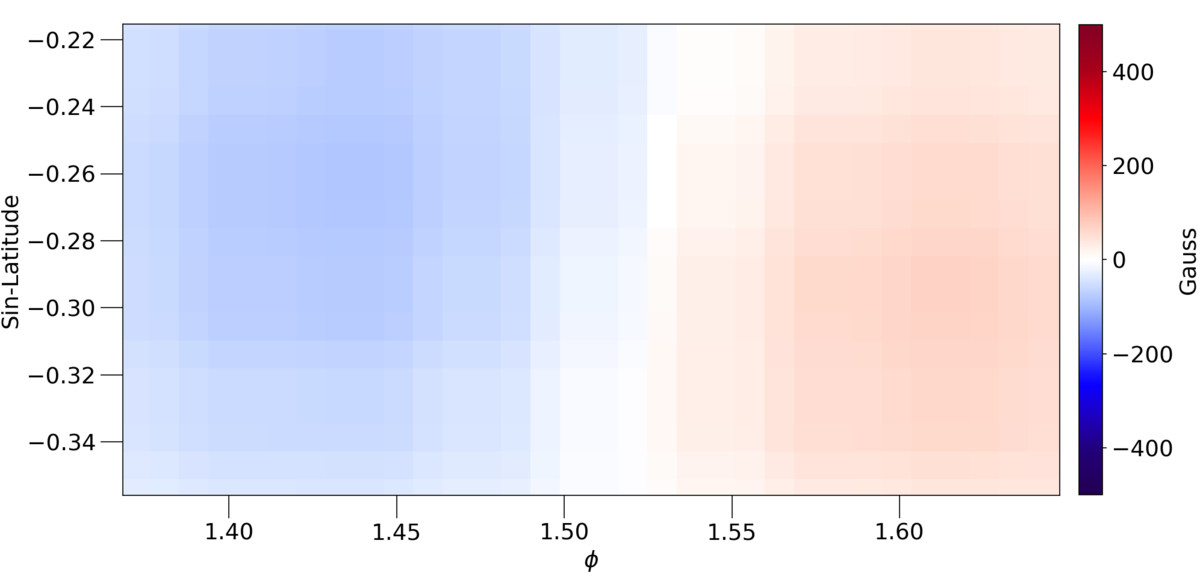}
\\
\rotatebox[origin=c]{90}{\mbox{{\bf SML}} ($360\!\!\times\!\!180$)}
& 
\includegraphics[align=m,width=4in]{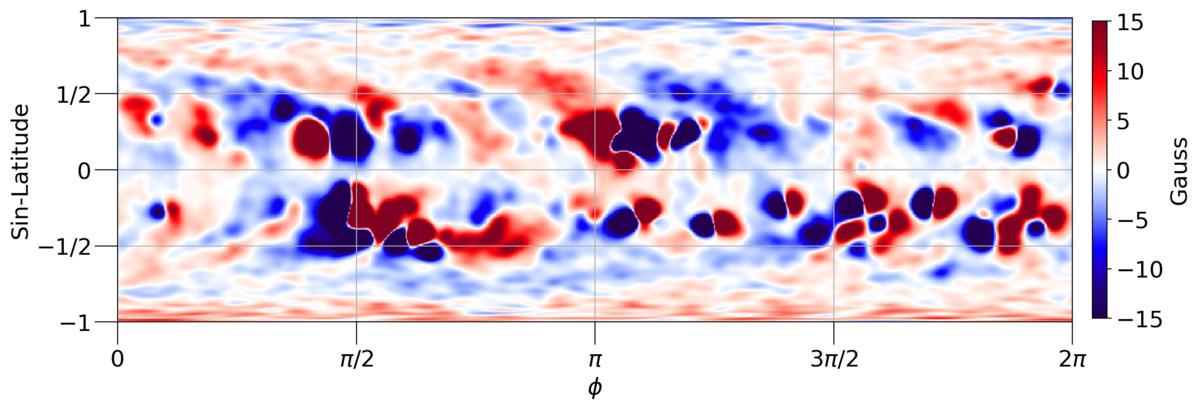}
&
\includegraphics[align=m,width=2.6in]{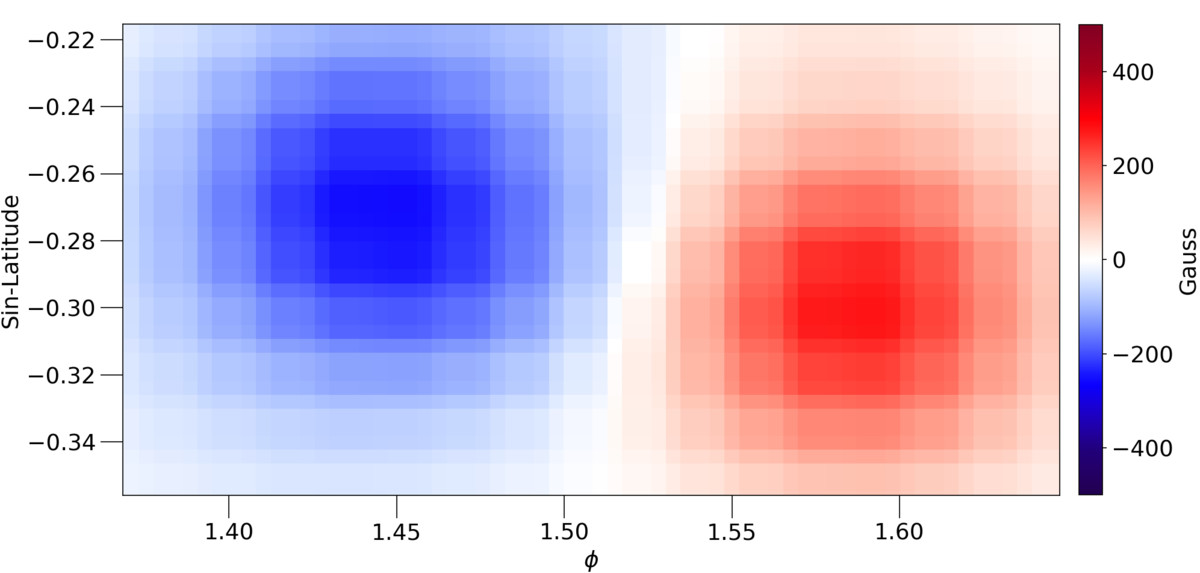}
\\
\rotatebox[origin=c]{90}{\mbox{{\bf MED}} ($900\!\!\times\!\!450$)}
& 
\includegraphics[align=m,width=4in]{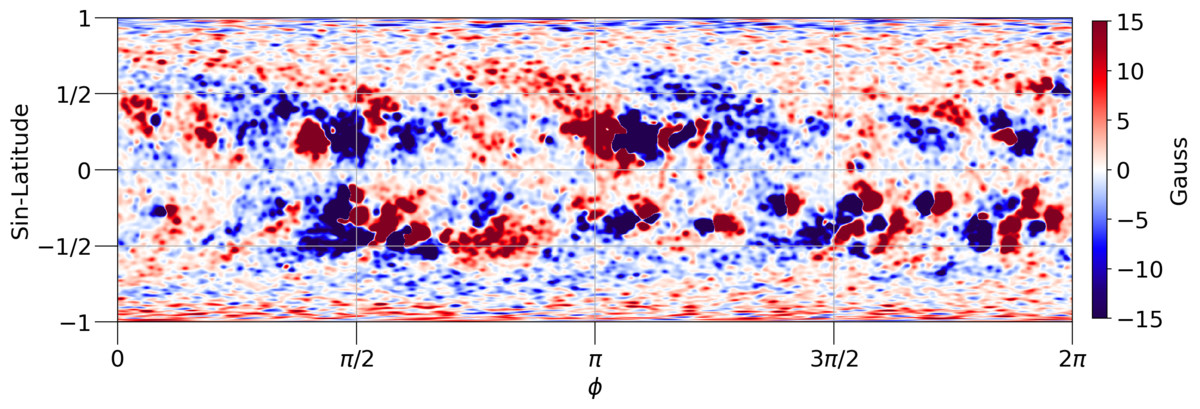}
&
\includegraphics[align=m,width=2.6in]{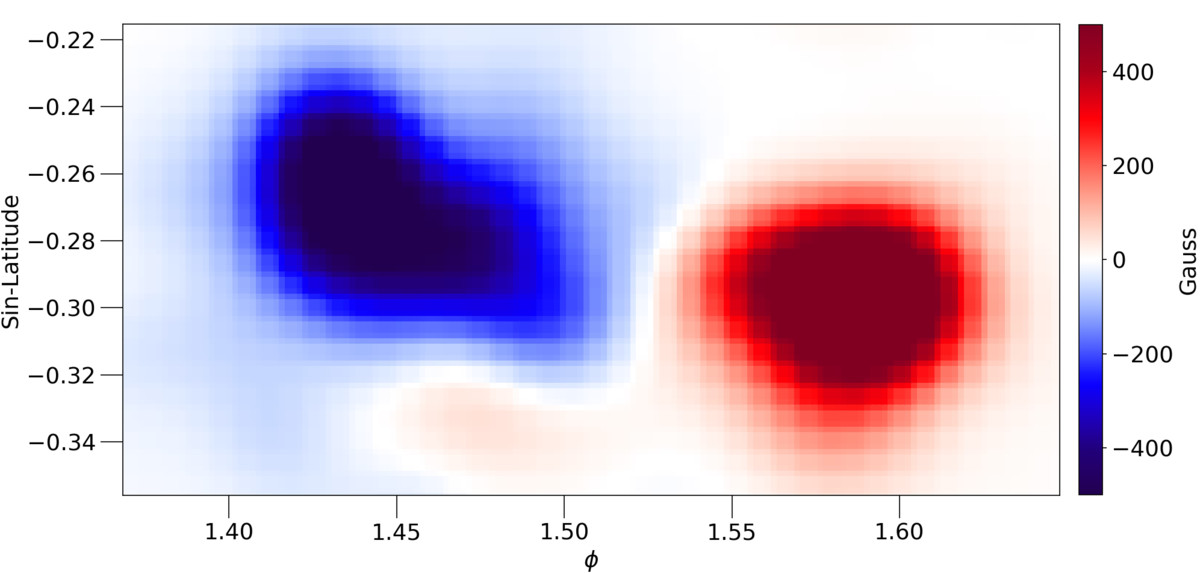}
\\
\rotatebox[origin=c]{90}{\mbox{{\bf LRG}} ($1800\!\!\times\!\!900$)}
& 
\includegraphics[align=m,width=4in]{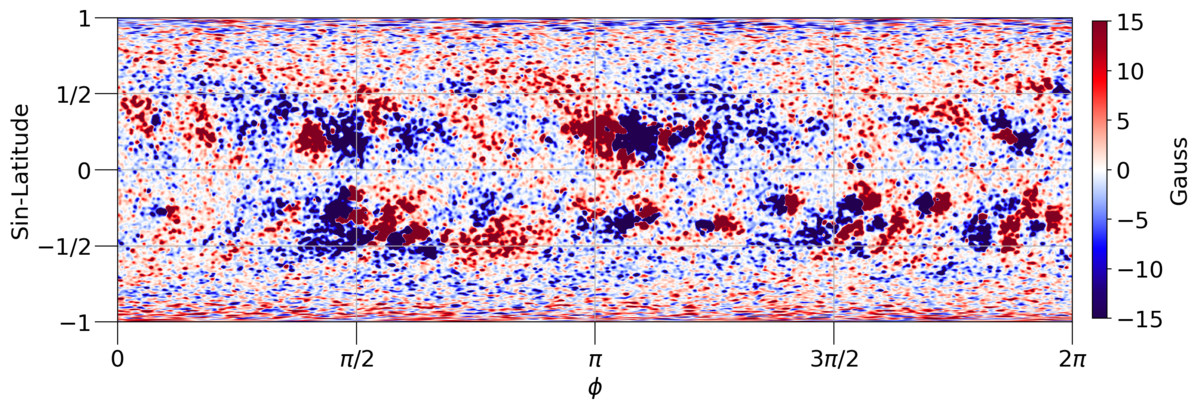}
&
\includegraphics[align=m,width=2.6in]{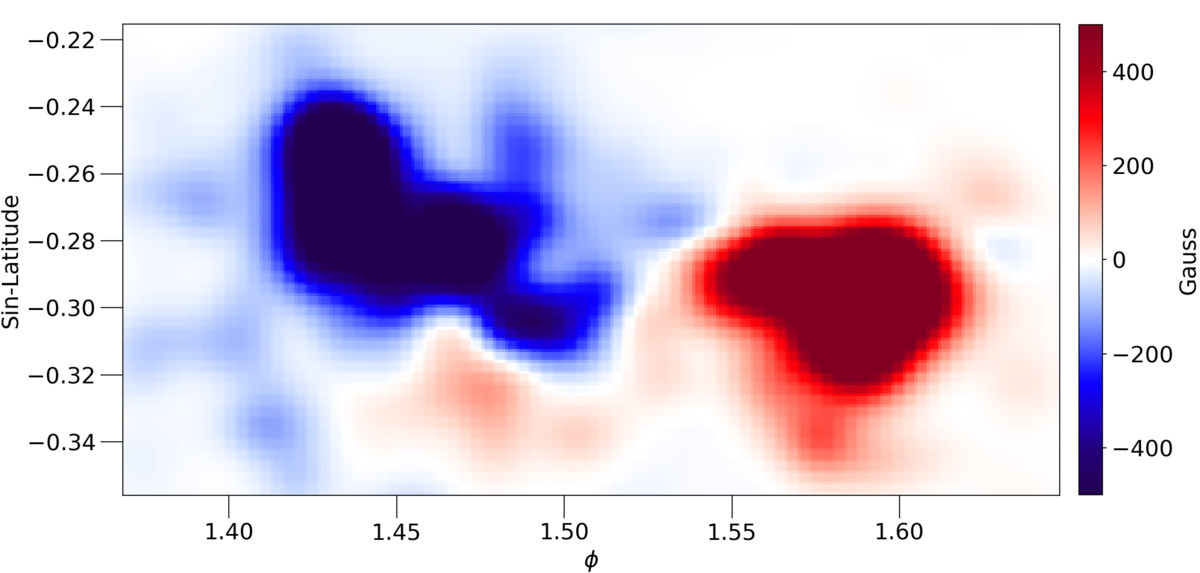}
\\
\rotatebox[origin=c]{90}{\mbox{{\bf NAT}} ($3973\!\!\times\!\!2012$)}
& 
\includegraphics[align=m,width=4in]{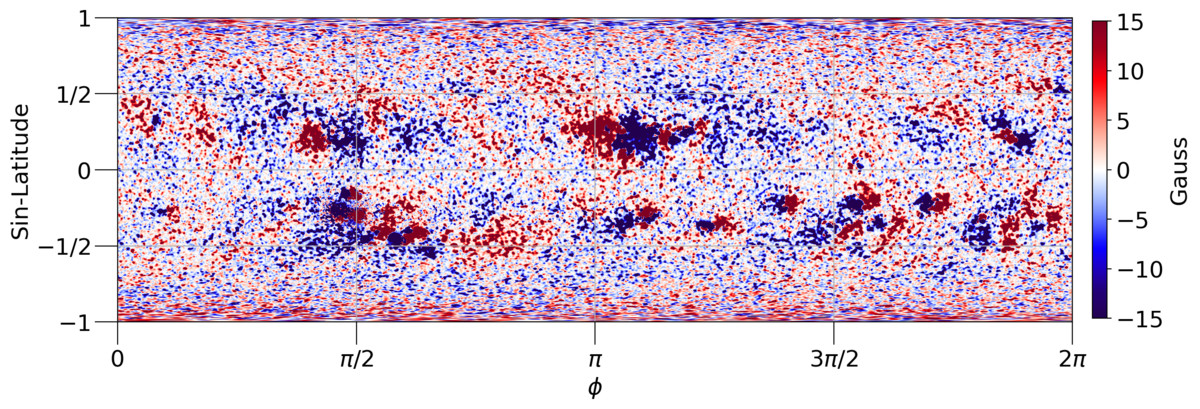}
&
\includegraphics[align=m,width=2.6in]{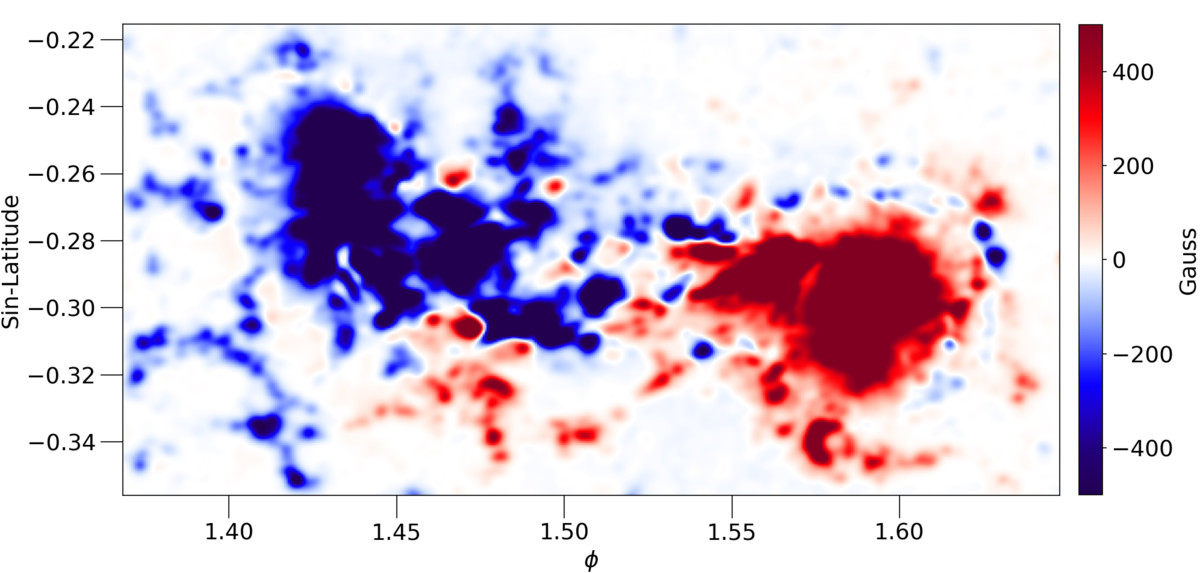}
\\
\rotatebox[origin=c]{90}{\mbox{{\bf PSI}} ($1095\!\!\times\!\!742$)}
& 
\includegraphics[align=m,width=4in]{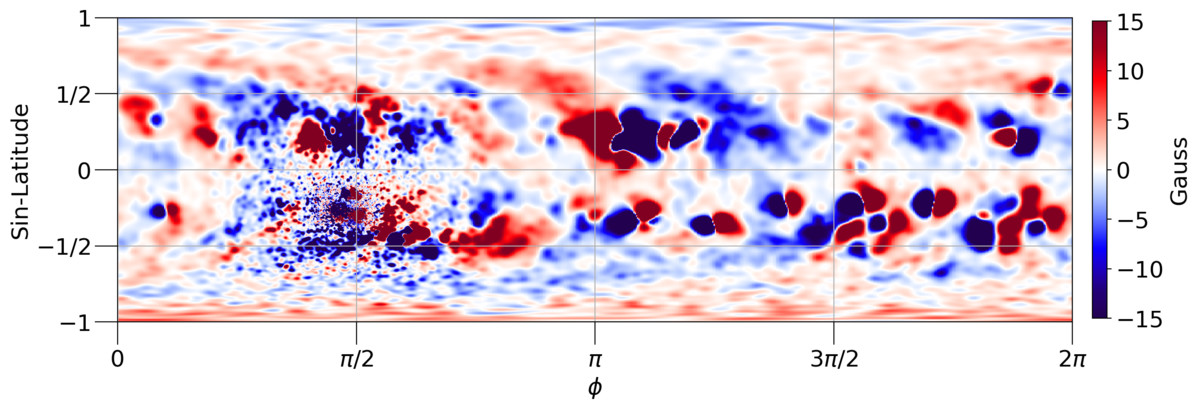}
&
\includegraphics[align=m,width=2.6in]{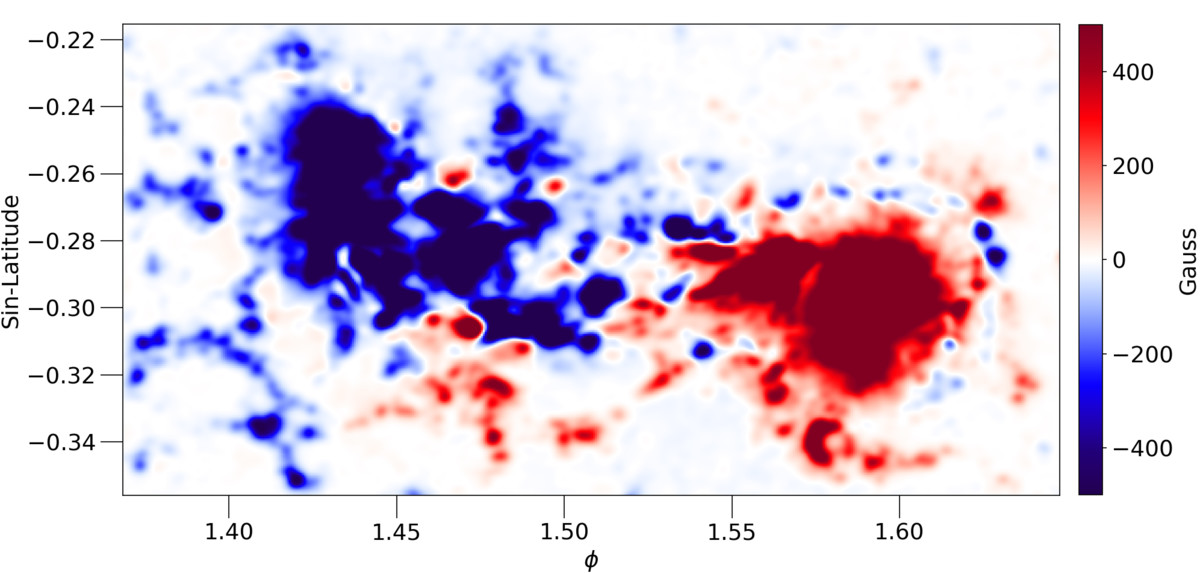}
\end{array}$
\caption{Processed surface $B_r$ maps used for each resolution test. Both the full map (left) and a zoomed view near the AR (right) are shown. The $\phi\times\theta$ resolution is indicated.  Additional details of each resolution are given in the text.\label{fig_res_br}} 
\end{figure}

In Fig~\ref{fig_res_br} we show the processed {\bf (PSI)} magnetic field map for each resolution along with a zoomed-in view of the AR.  
The AR goes from being overly smooth and featureless at {\bf (TNY)} resolution, to having a large amount of structure at higher resolutions.  The highly non-uniform resolution map {\bf (PSI)} retains the high-fidelity structure the AR even though it is quite coarse in the rest of the domain.

\begin{figure}[htbp]
\centering
$\begin{array}{rcc}
\rotatebox[origin=c]{90}{\mbox{{\bf TNY}} ($180\!\!\times\!\!90$)}
& 
\includegraphics[align=m,width=4in]{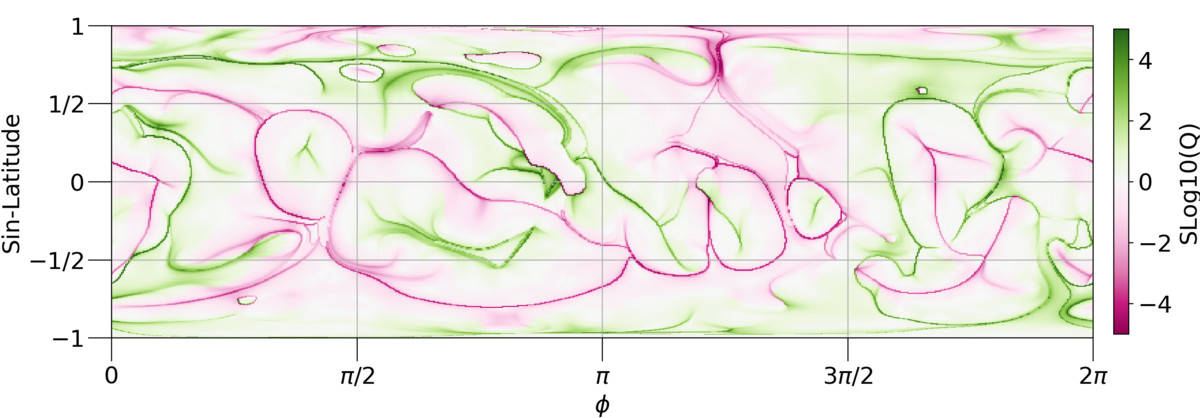}
&
\includegraphics[align=m,width=2.6in]{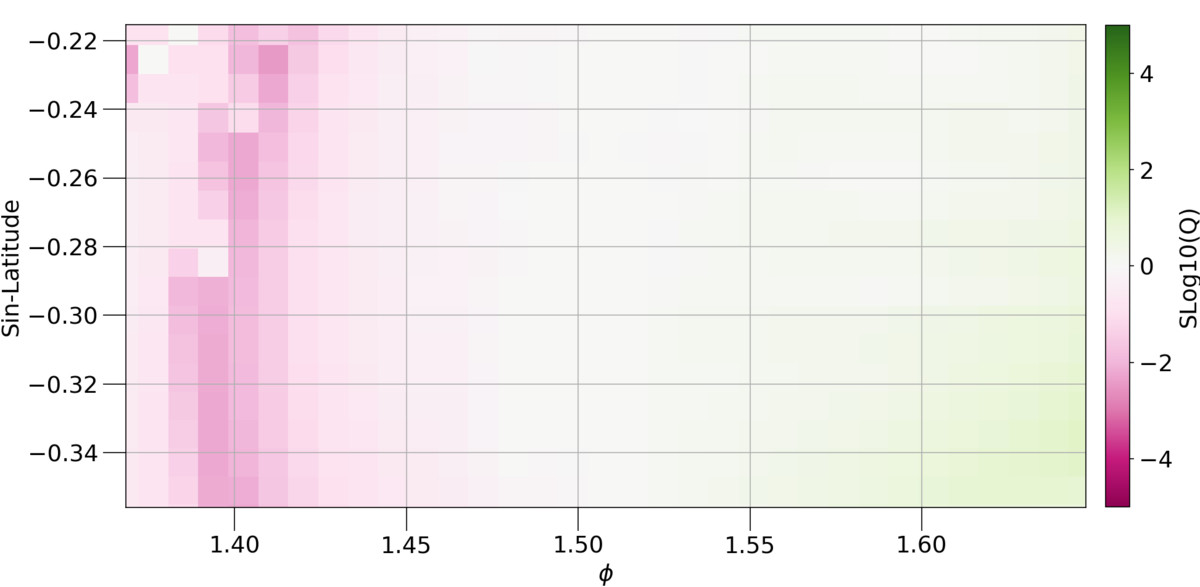}
\\
\rotatebox[origin=c]{90}{\mbox{{\bf SML}} ($360\!\!\times\!\!180$)}
& 
\includegraphics[align=m,width=4in]{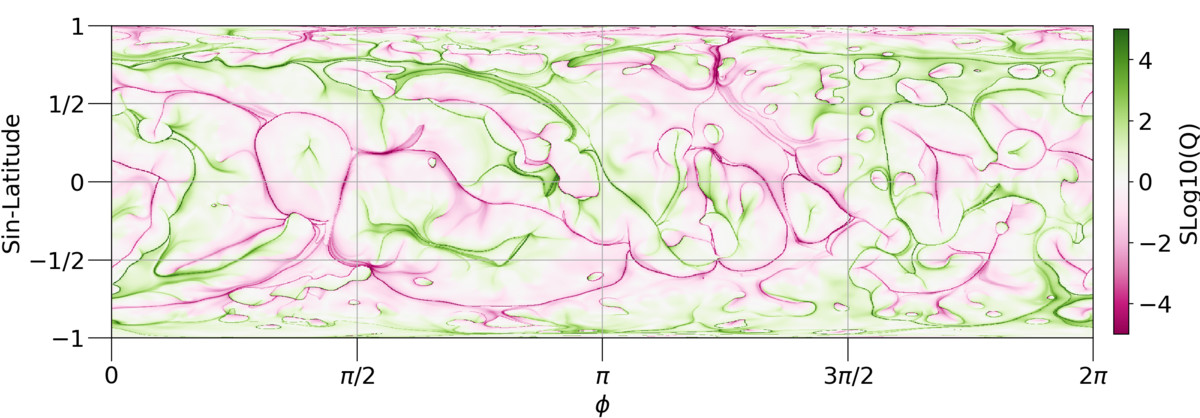}
&
\includegraphics[align=m,width=2.6in]{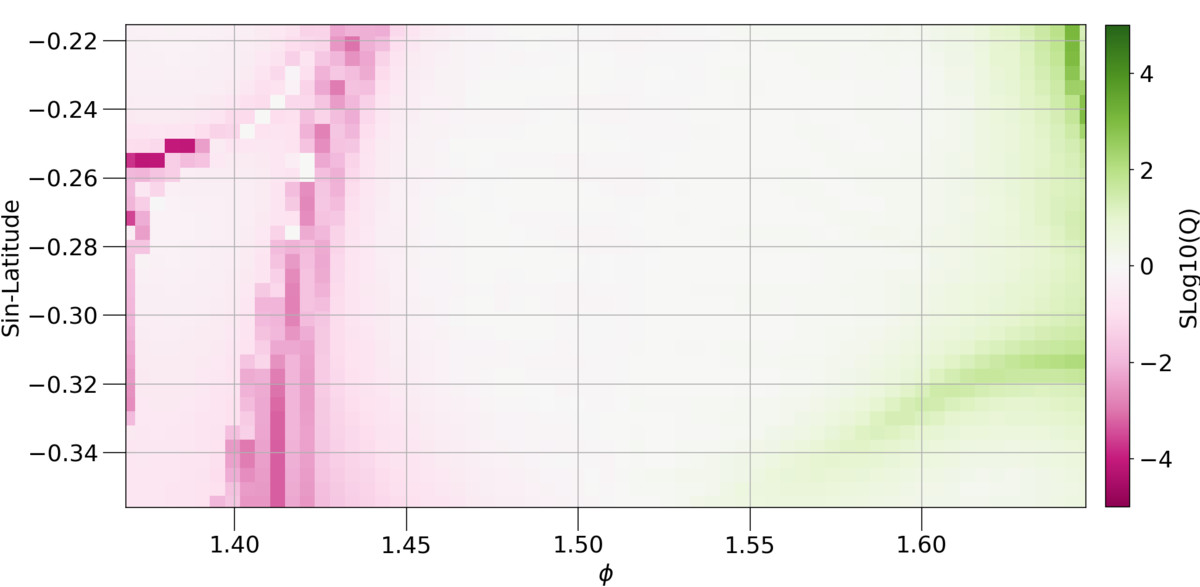}
\\
\rotatebox[origin=c]{90}{\mbox{{\bf MED}} ($900\!\!\times\!\!450$)}
& 
\includegraphics[align=m,width=4in]{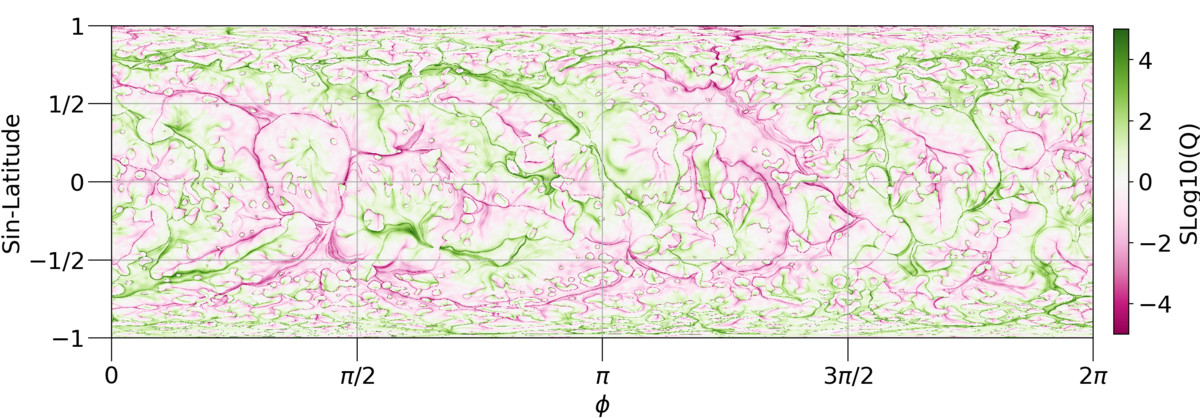}
&
\includegraphics[align=m,width=2.6in]{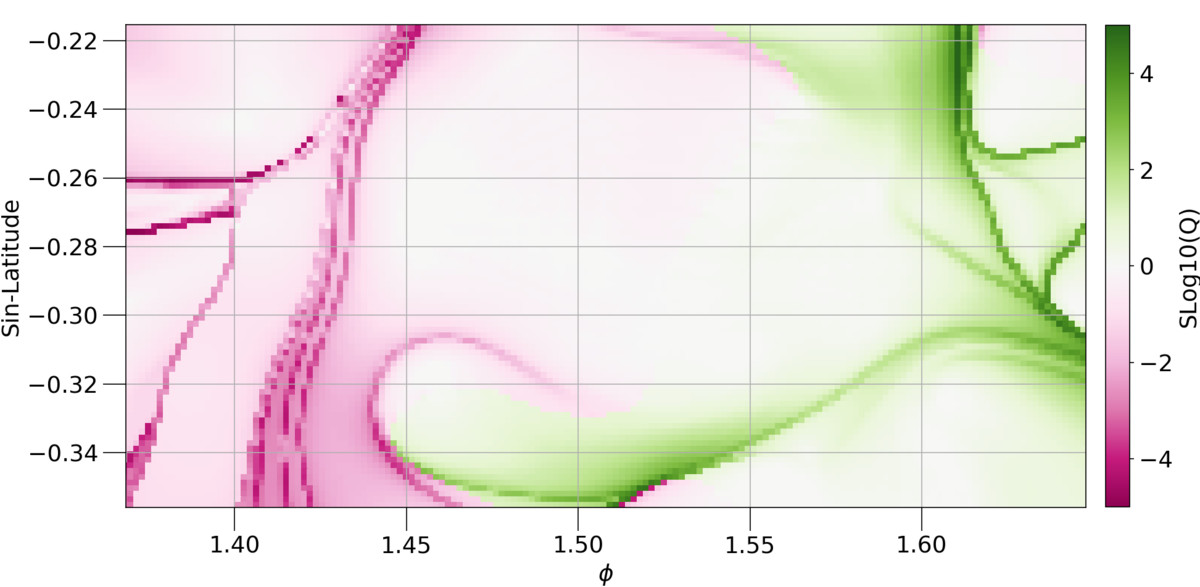}
\\
\rotatebox[origin=c]{90}{\mbox{{\bf LRG}} ($1800\!\!\times\!\!900$)}
& 
\includegraphics[align=m,width=4in]{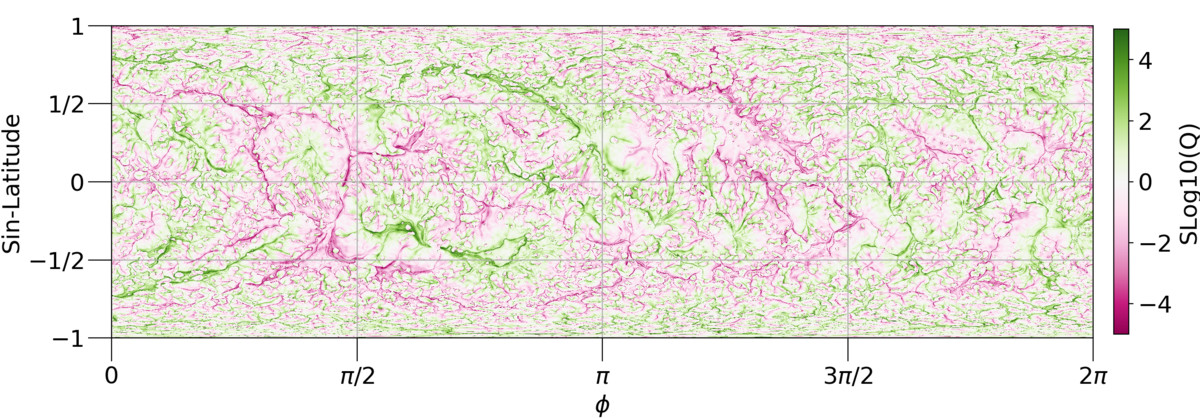}
&
\includegraphics[align=m,width=2.6in]{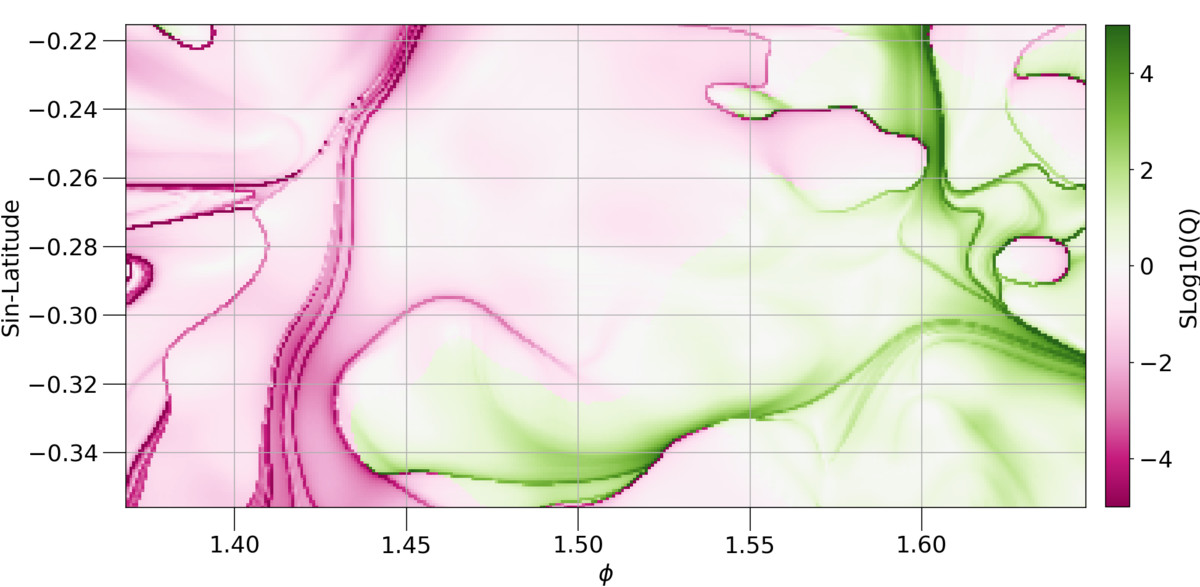}
\\
\rotatebox[origin=c]{90}{\mbox{{\bf NAT}} ($3973\!\!\times\!\!2012$)}
& 
\includegraphics[align=m,width=4in]{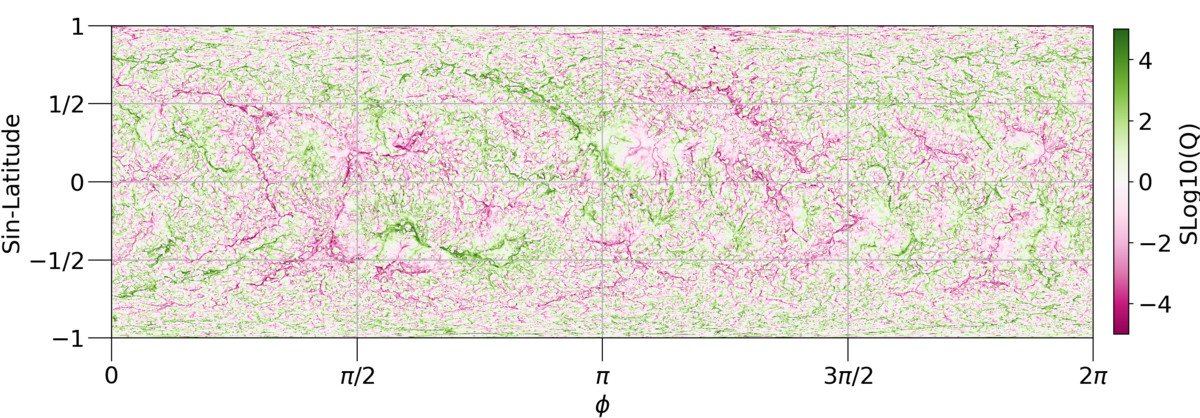}
&
\includegraphics[align=m,width=2.6in]{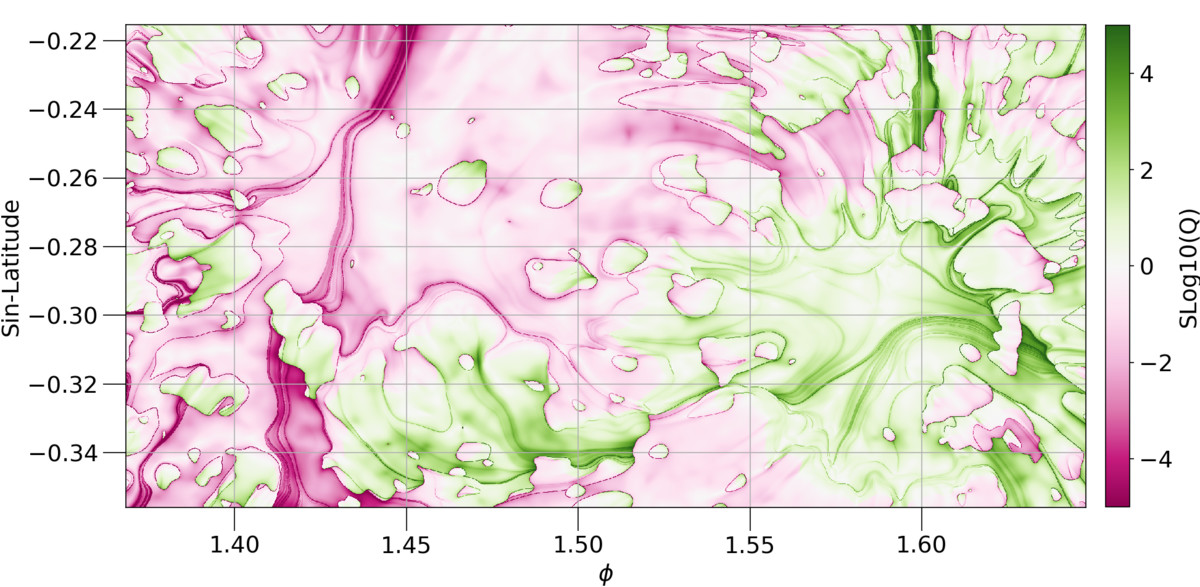}
\\
\rotatebox[origin=c]{90}{\mbox{{\bf PSI}} ($1095\!\!\times\!\!742$)}
& 
\includegraphics[align=m,width=4in]{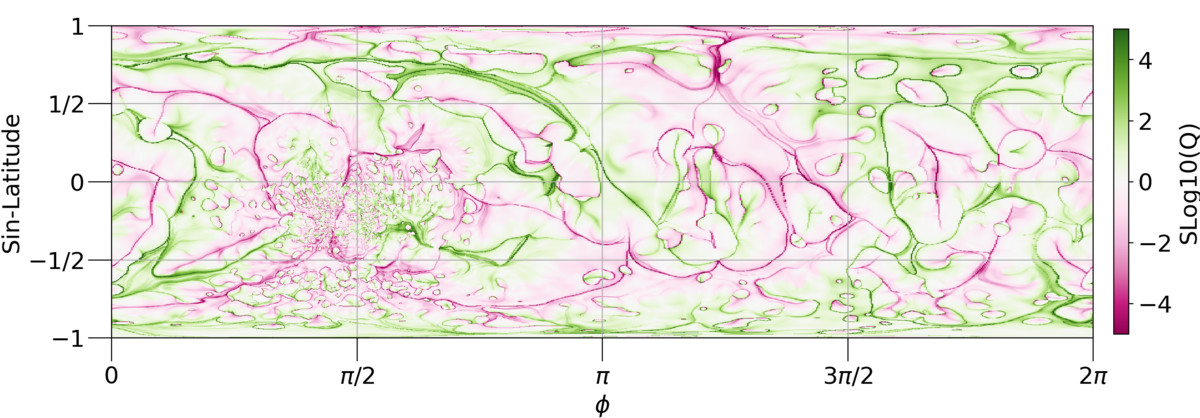}
&
\includegraphics[align=m,width=2.6in]{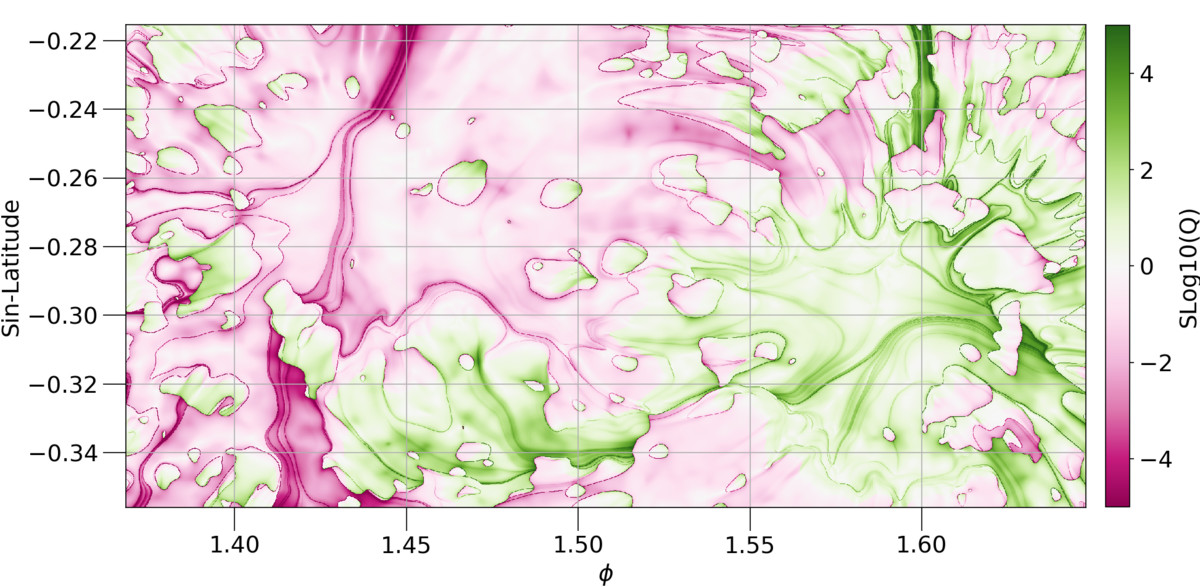}
\end{array}$
\caption{Squashing factor from PF solutions for various resolutions.  Left: Full Q-map at the solar surface, Right: Zoomed Q-map near the AR.  The $\phi\times\theta$ resolution is indicated.  Additional details of each resolution are given in the text.\label{fig_res_qr0}} 
\end{figure}

In Fig~\ref{fig_res_qr0} we show the squashing factor at the solar surface for each PF solution.  
Once again, in order to obtain high-resolution structure in the AR, one must use a high resolution map.  Also, even though $Q$ can be non-local (it is based on mappings), the $Q$ values in the AR for the non-uniform resolution {\bf (PSI)} are nearly identical to the native resolution {\bf (NAT)}. This is perhaps not surprising because the mapping still involves high-resolution field on one side and will largely be localized to the AR. On the other hand it confirms that multi-scale resolution calculations can reproduce such structure in a given region of interest.

\begin{figure}[htbp]
\centering
$\begin{array}{rcc}
\rotatebox[origin=c]{90}{\mbox{{\bf TNY}} ($180\!\!\times\!\!90$)}
& 
\includegraphics[align=m,width=0.46\textwidth]{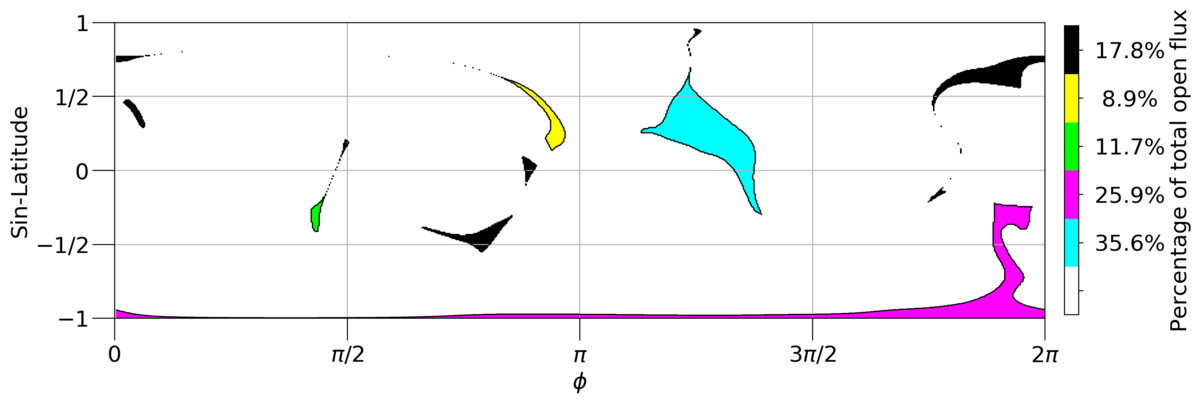}
&
\includegraphics[align=m,width=0.46\textwidth]{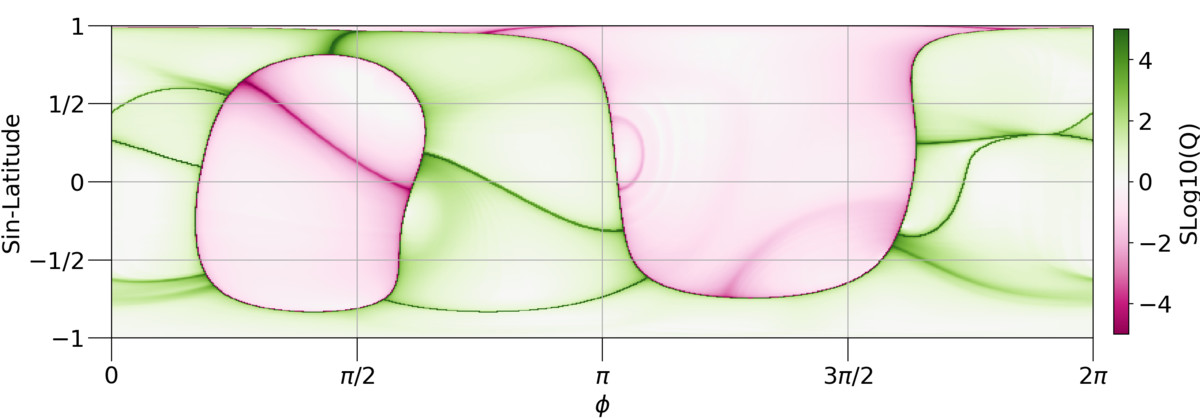}
\\
\rotatebox[origin=c]{90}{\mbox{{\bf SML}} ($360\!\!\times\!\!180$)}
& 
\includegraphics[align=m,width=0.46\textwidth]{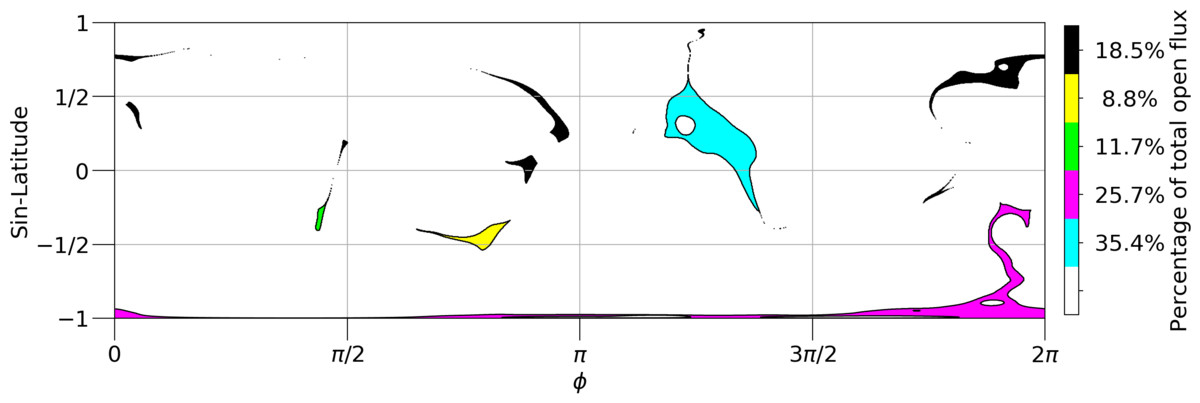}
&
\includegraphics[align=m,width=0.46\textwidth]{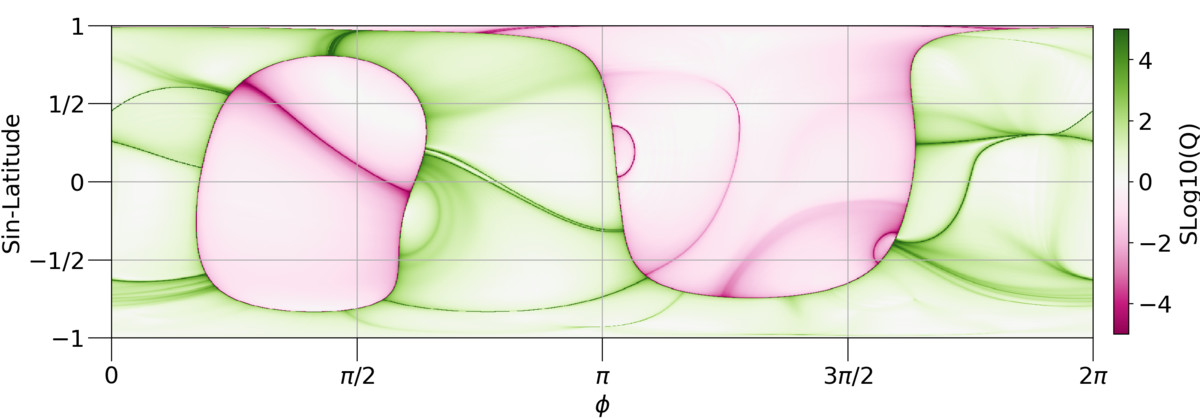}
\\
\rotatebox[origin=c]{90}{\mbox{{\bf MED}} ($900\!\!\times\!\!450$)}
& 
\includegraphics[align=m,width=0.46\textwidth]{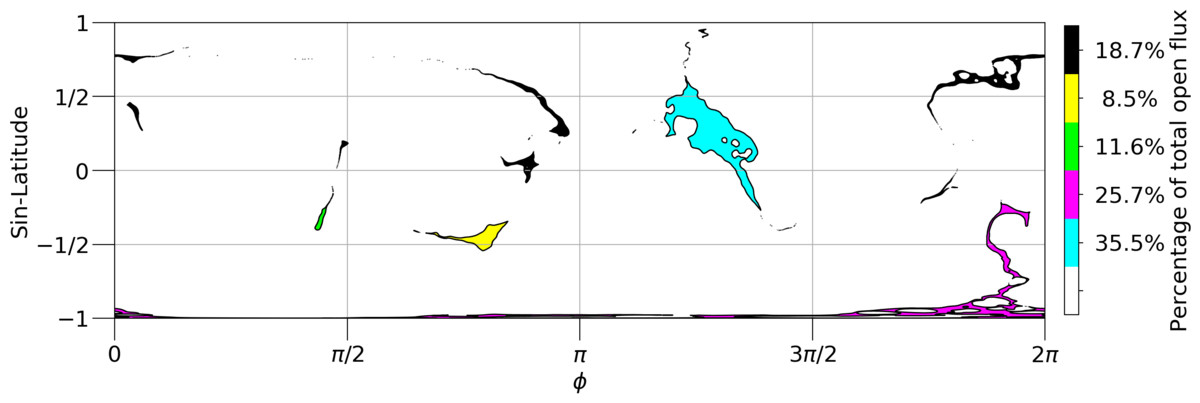}
&
\includegraphics[align=m,width=0.46\textwidth]{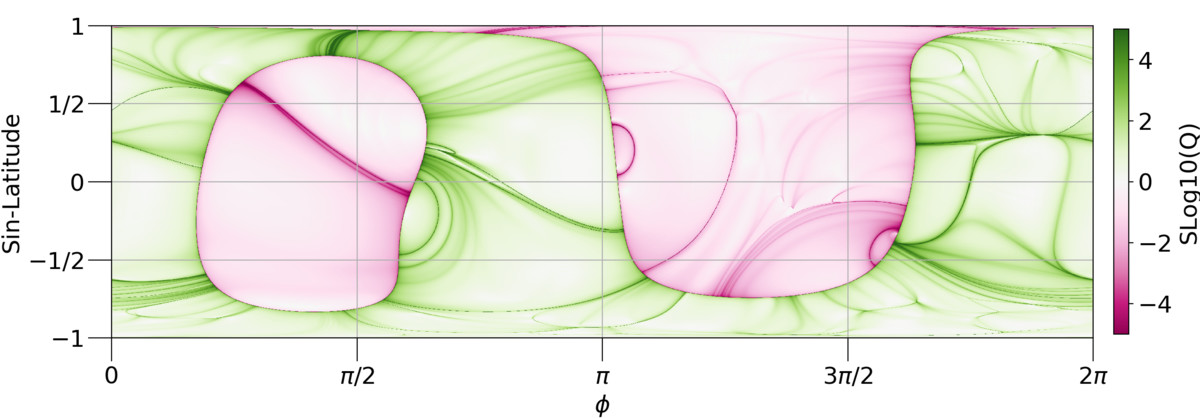}
\\
\rotatebox[origin=c]{90}{\mbox{{\bf LRG}} ($1800\!\!\times\!\!900$)}
& 
\includegraphics[align=m,width=0.46\textwidth]{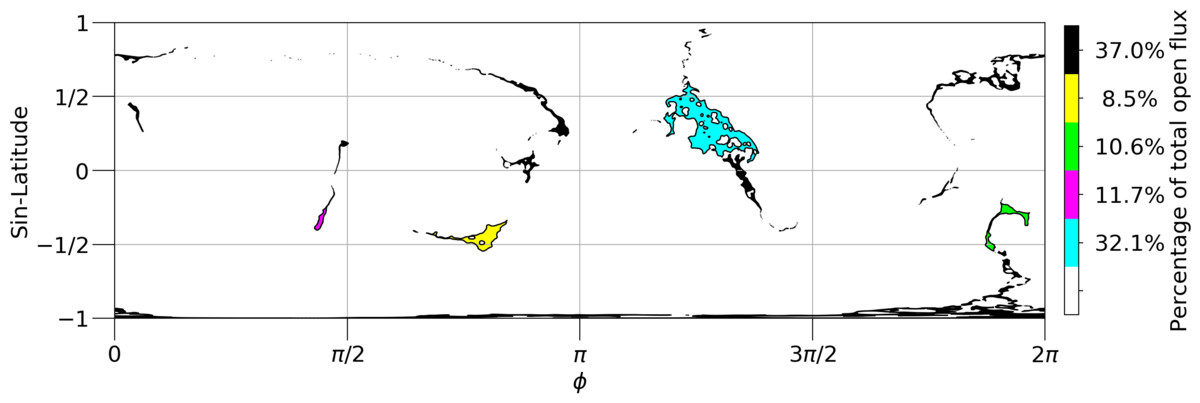}
&
\includegraphics[align=m,width=0.46\textwidth]{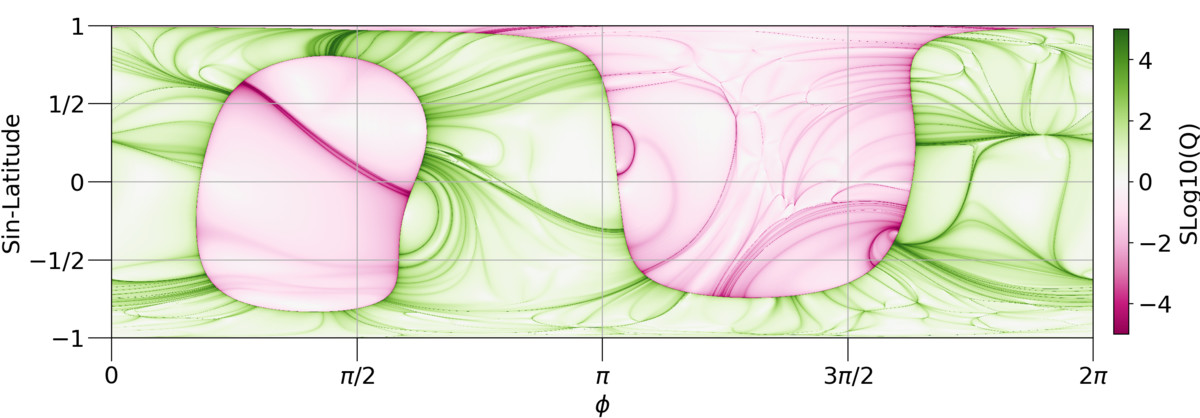}
\\
\rotatebox[origin=c]{90}{\mbox{{\bf NAT}} ($3973\!\!\times\!\!2012$)}
& 
\includegraphics[align=m,width=0.46\textwidth]{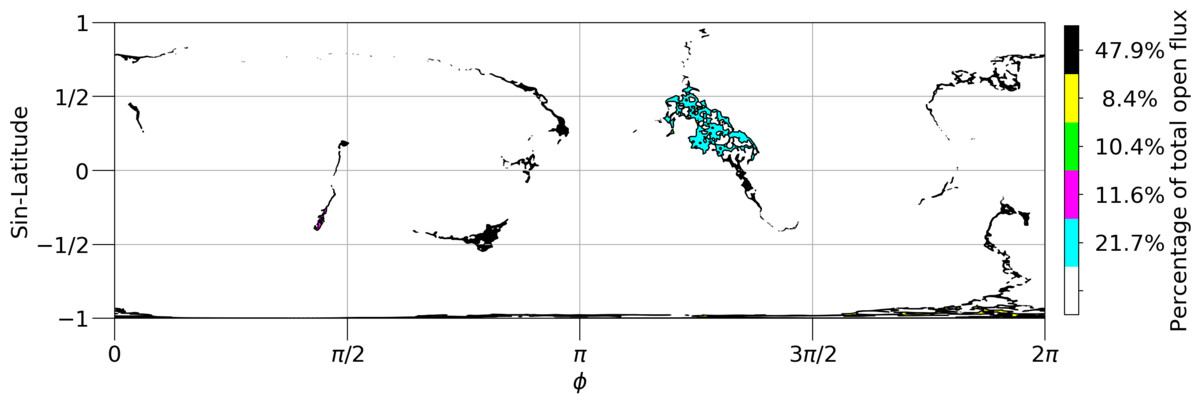}
&
\includegraphics[align=m,width=0.46\textwidth]{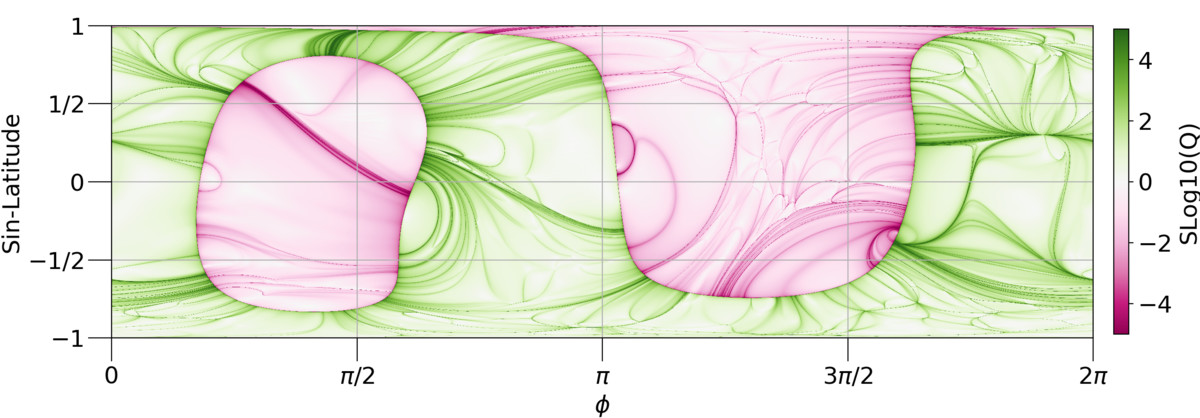}
\\
\rotatebox[origin=c]{90}{\mbox{{\bf PSI}} ($1095\!\!\times\!\!742$)}
& 
\includegraphics[align=m,width=0.45\textwidth]{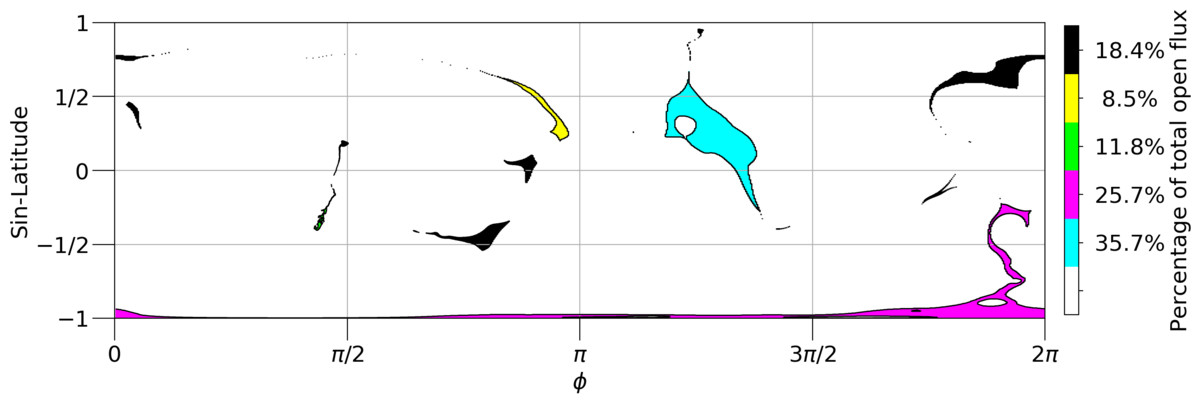}
&
\includegraphics[align=m,width=0.45\textwidth]{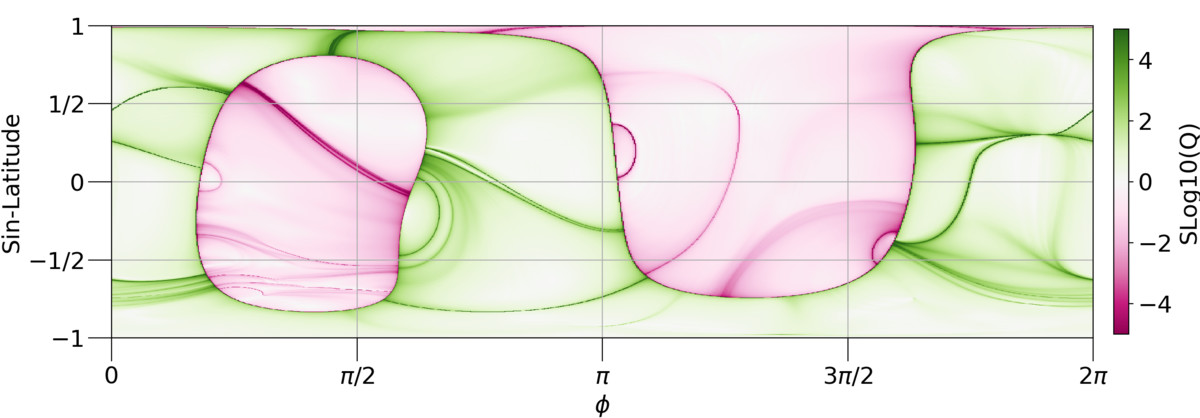}
\end{array}$
\caption{Left: Open field maps from PF solutions for various resolutions.  Right: Squashing factor at the $r_{ss}$ outer radial boundary. The $\phi\times\theta$ resolution is indicated.  Additional details of each resolution are given in the text.\label{fig_res_chqr1}} 
\end{figure}

In Fig~\ref{fig_res_chqr1} we show the open field areas along with the squashing factor at the outer radial boundary.  
We see that the higher resolution open field maps contain many small-scale parasitic polarities within the open field regions. This creates sort of a `Swiss cheese' effect where each low-lying closed-field dome leaves a hole in the open field map with the bounding footprint of the closed-open separatrix.  This implies that open flux-regions should not necessarily be considered monolithic entities with a smooth mapping in their interior.  Since empirical solar wind models rely on the distance to the open field boundary \citep{rileyetal2001,argeetal2003,rileyetal2015}, this might be an important consideration or potential pitfall as models naturally move towards higher resolution.

\begin{figure}[htbp]
\centering
\includegraphics[width=0.45\textwidth]{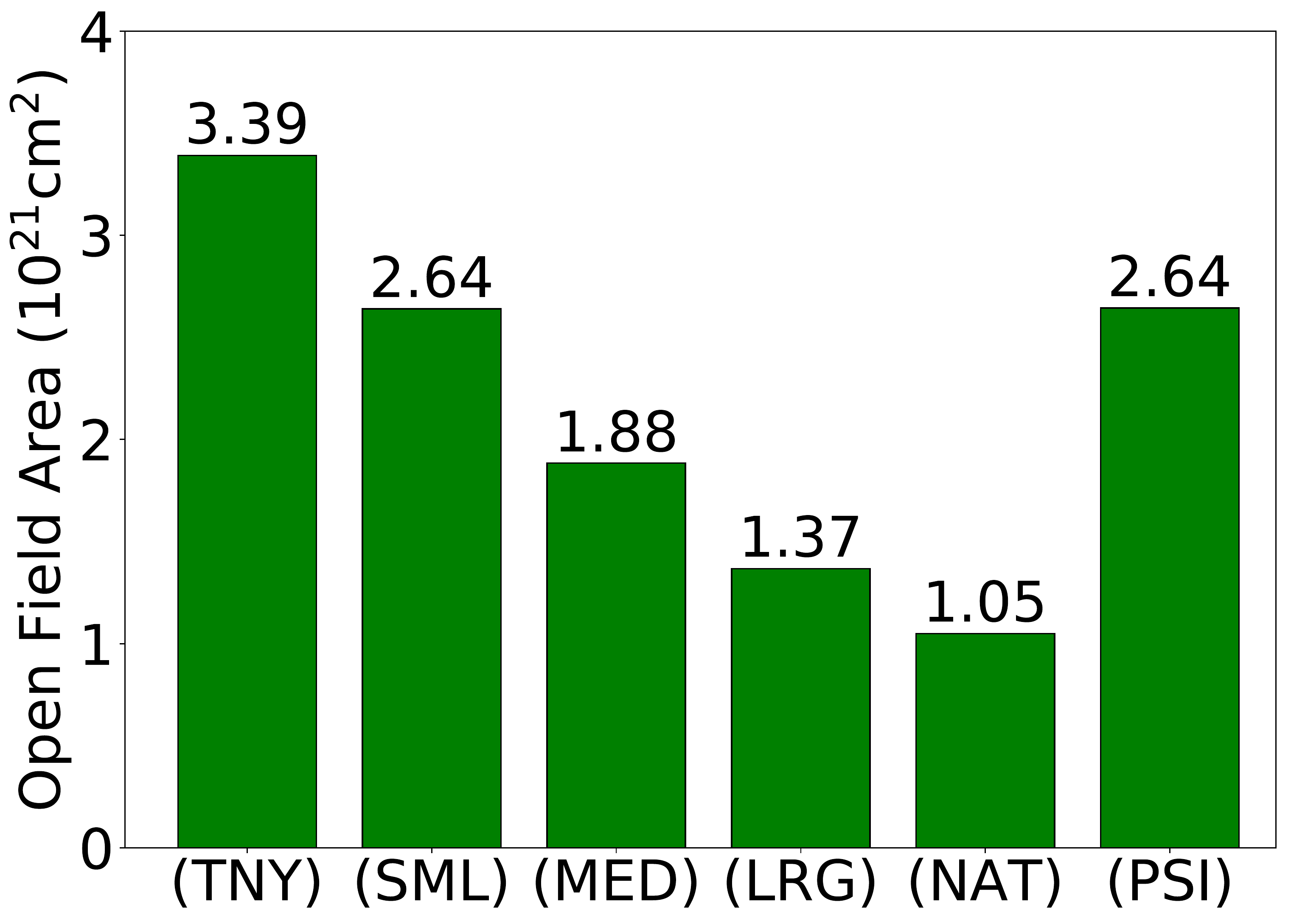}
\includegraphics[width=0.45\textwidth]{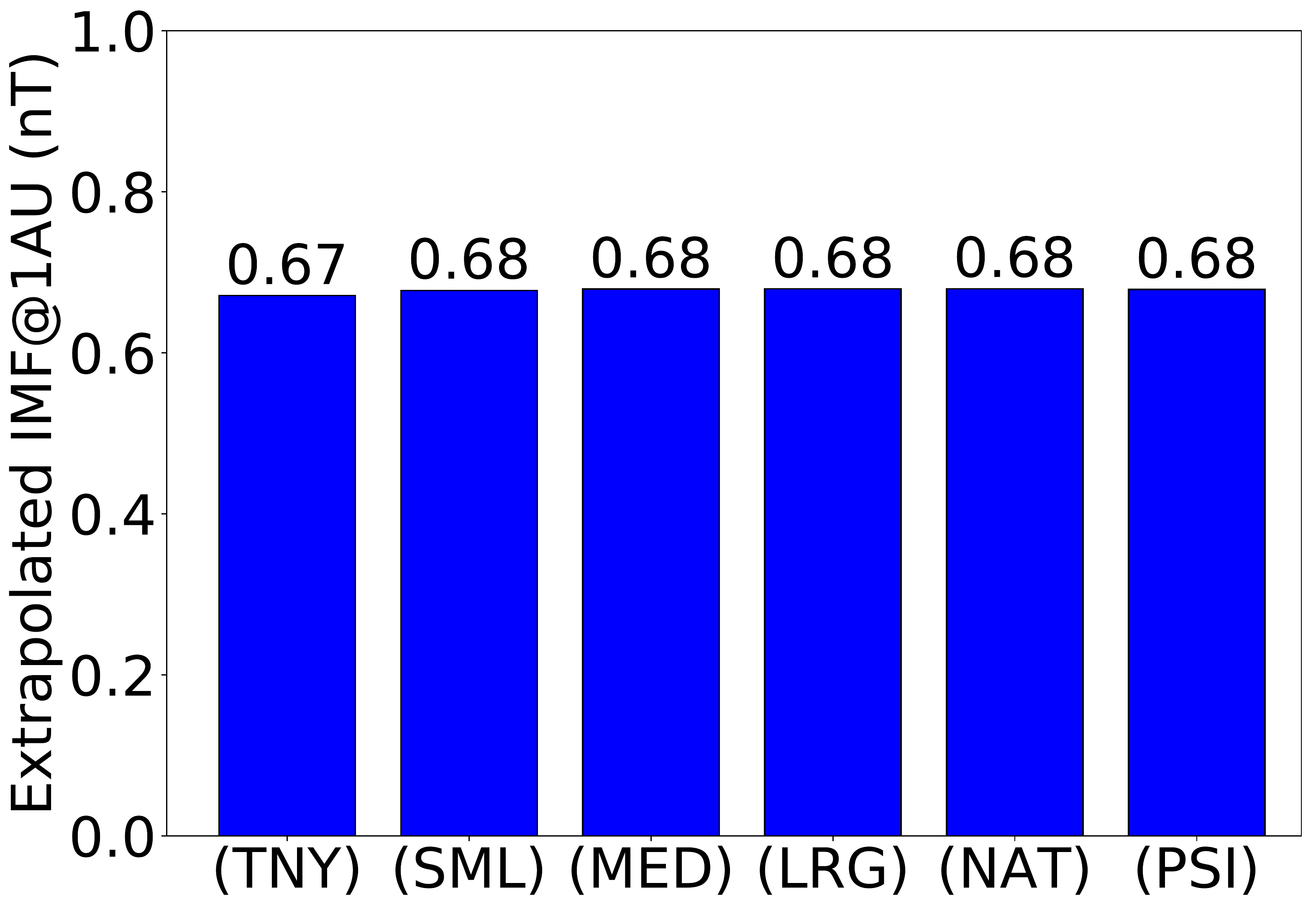}
\\
\includegraphics[width=0.45\textwidth]{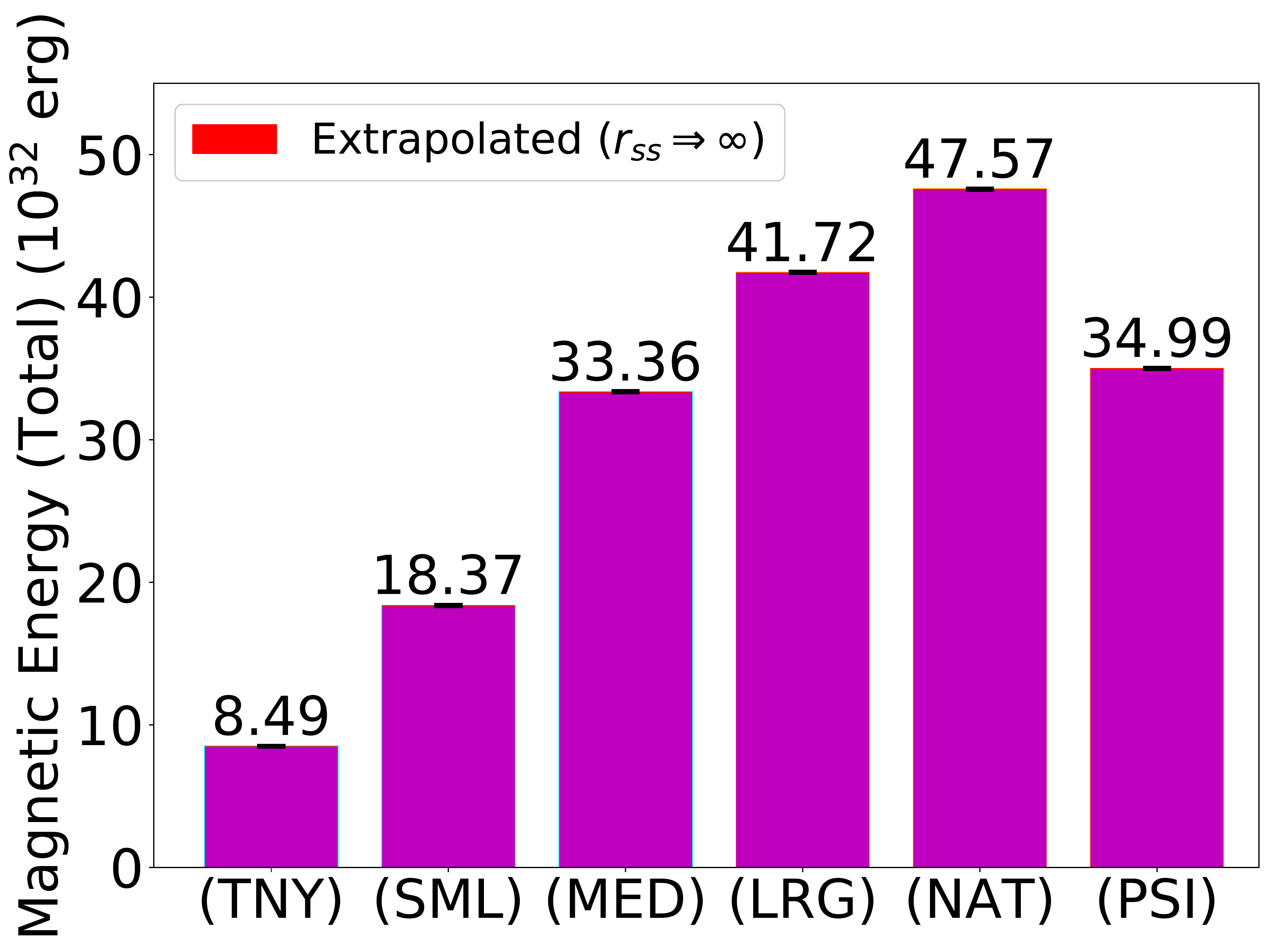}
\includegraphics[width=0.45\textwidth]{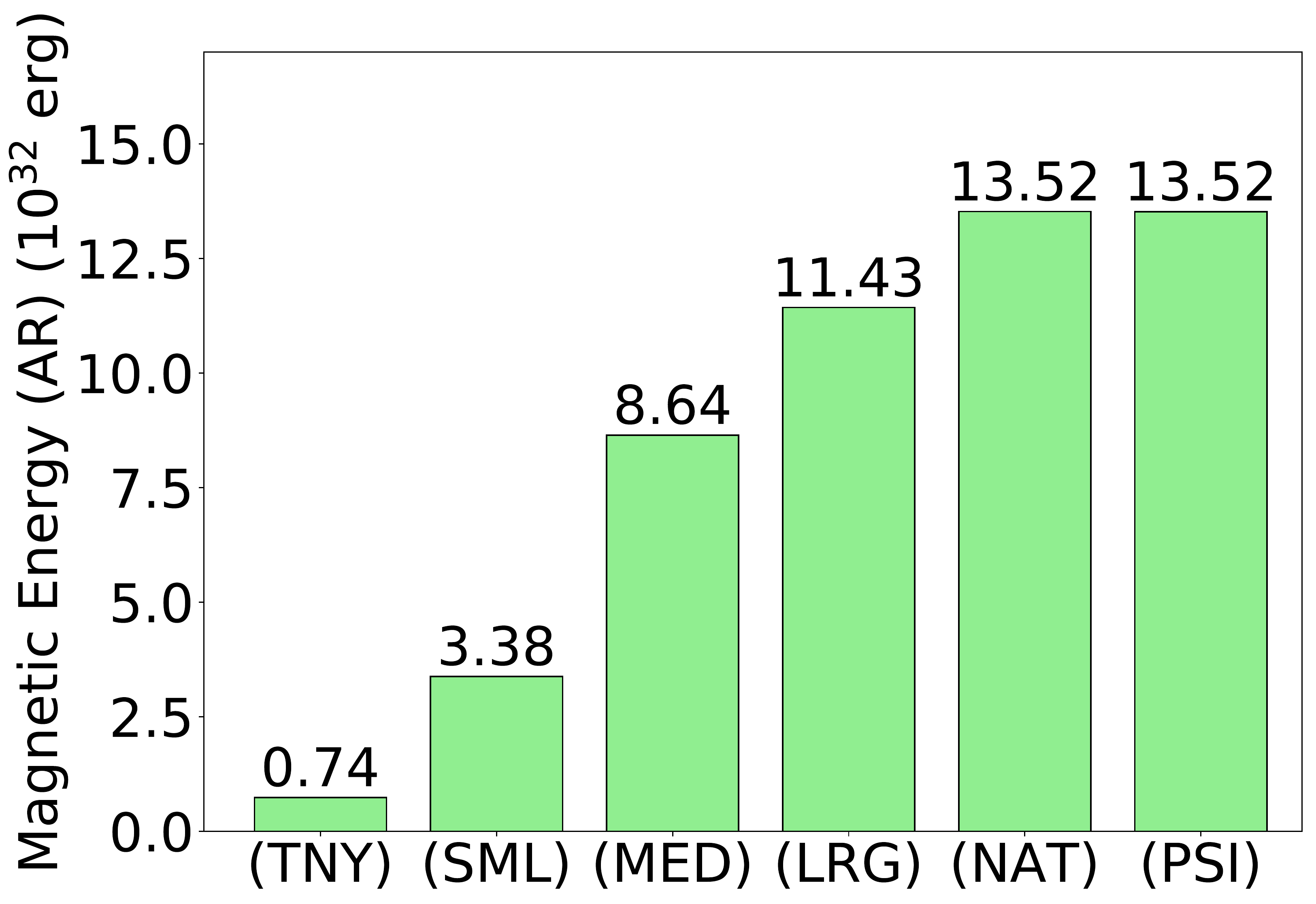}
\caption{Diagnostics of potential fields (described in Sec.~\ref{sec:compare_method}) varying the resolution of the input map and resulting potential field.  A description of each resolution label is given in the text. \label{fig_res_diags}} 
\end{figure}

In Fig~\ref{fig_res_diags} we show the derived quantities described in Sec.~\ref{sec:compare_method}.
Here we see that due to the increased detail in the open field areas, the total open field area reduces as the resolution increases.  However, the total open flux remains nearly constant at all resolutions (even with the very coarse {\bf (TNY)} resolution).  This is a key result as it demonstrates that one can rely on lower resolution magnetogram/PF solutions when calculating open flux.  We also see that the magnetic energy (both global and near the AR) increases as the resolution increases.  This is not surprising but emphasizes that absolute values of energy depend strongly on map/model resolution.

\begin{figure}[htbp]
\centering
\includegraphics[width=0.6\textwidth]{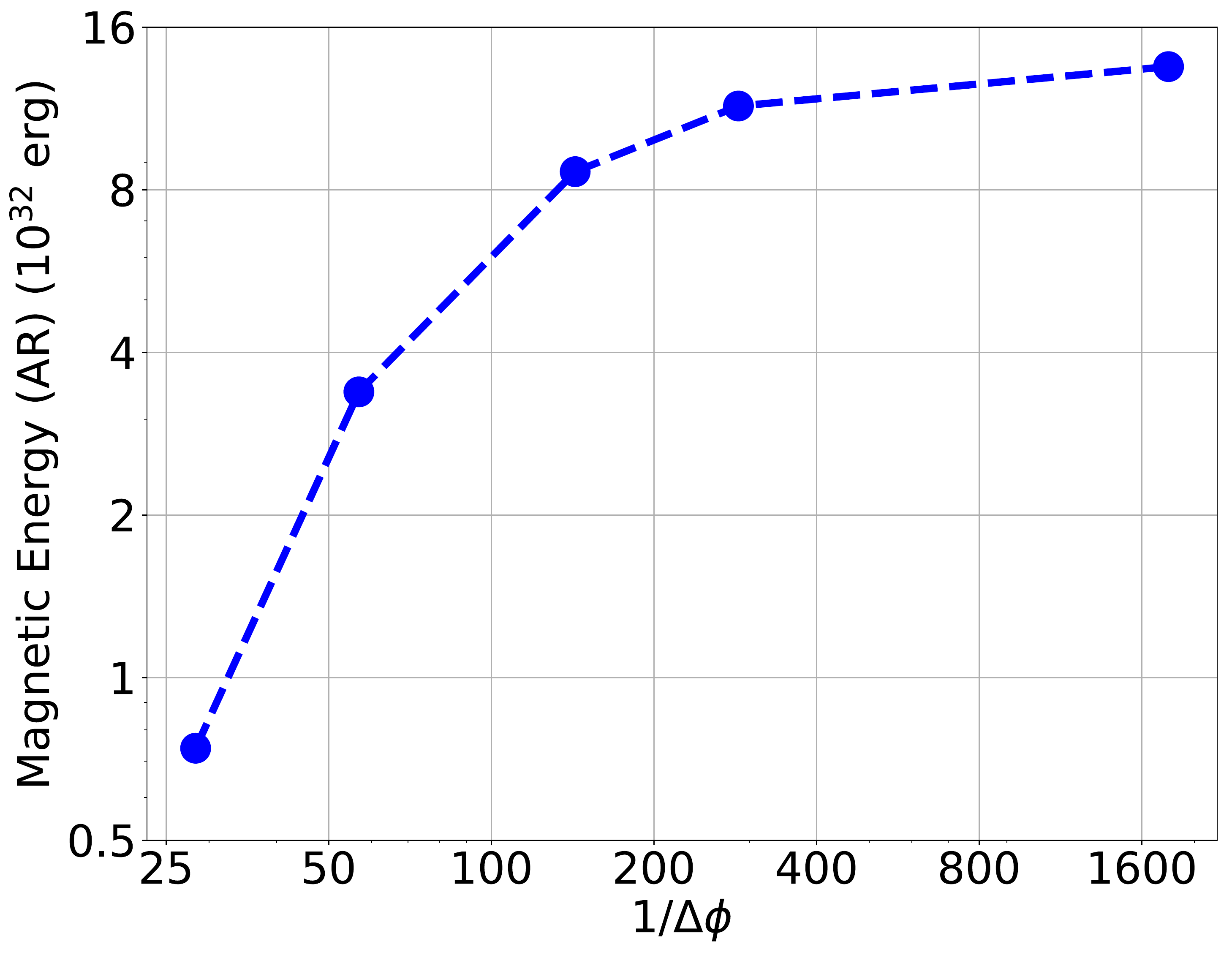}
\caption{Magnetic energy near the AR for multiple resolutions plotted as a function of $1/\Delta \phi$. The highest resolution is that of the HMI SHARP data product.\label{fig_res_magescale}} 
\end{figure}

We check the convergence of the magnetic energy near the AR as we approach HMI SHARP resolution by plotting the energy versus the $\Delta \phi^{-1}$ used in each resolution in Fig.~\ref{fig_res_magescale}.
We see that the magnetic energy of the AR is still increasing at the highest resolution available, but does seem to be saturating. This trend makes sense on an intuitive level considering the perspective of spherical harmonics: the higher-order scales not currently present in the highest-resolution HMI data will decay more quickly in height. In this way the integrated energy should eventually asymptote, even if higher resolution structure is present. 

We also find that the non-uniform resolution case {\bf (PSI)} yields the same magnetic energy for the AR as the native resolution case {\bf (NAT)}.  Given that case {\bf (PSI)} is nearly forty times smaller, the computation is considerably faster than {\bf (NAT)} (see Appendix \ref{sec_appx_comp}). This demonstrates how useful non-uniform grid capabilities are when studying ARs at high-resolution within global coronal domains.

\begin{figure}[htbp]
\centering
\includegraphics[angle=90,width=0.55\textwidth]{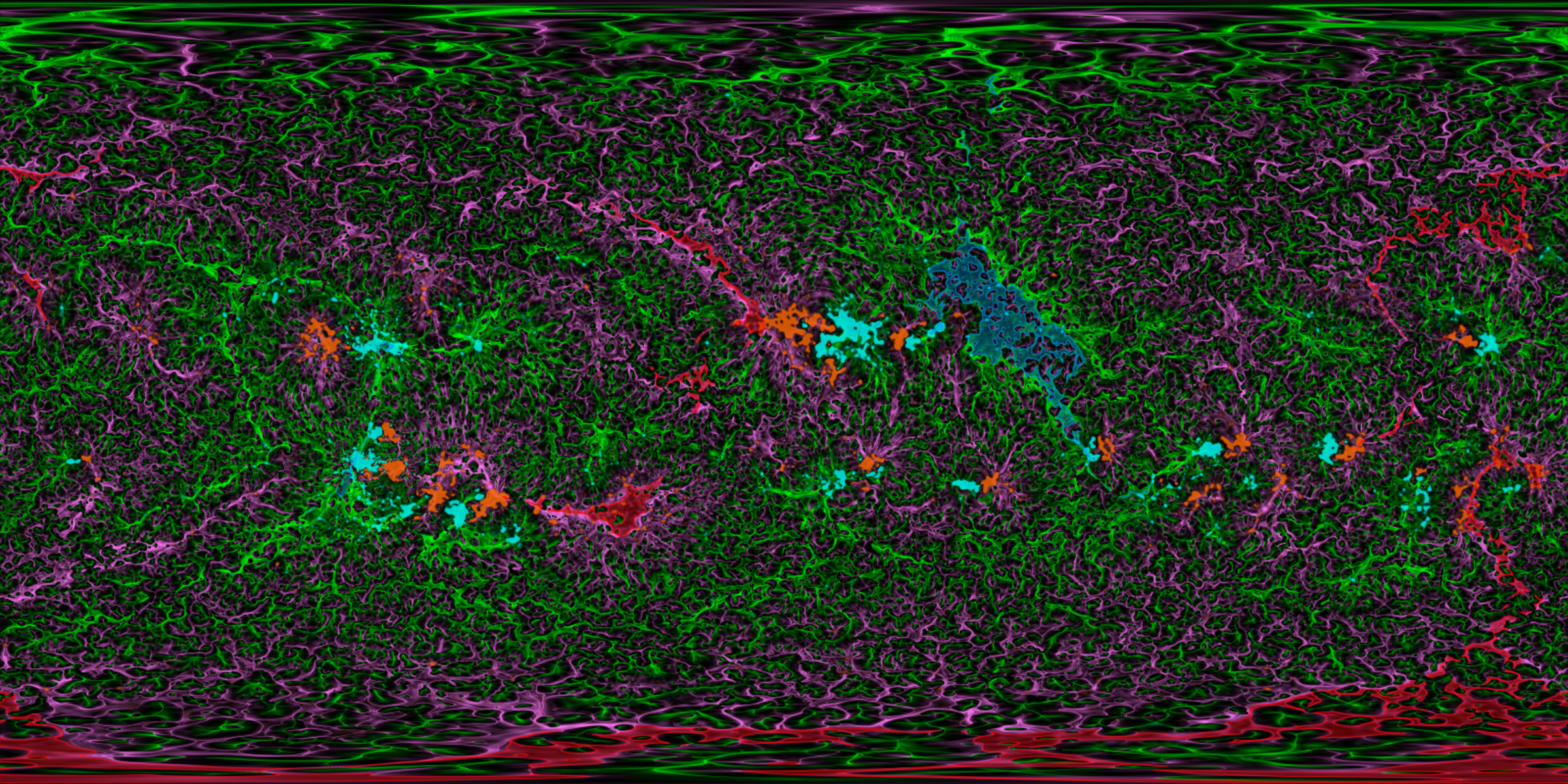}
\caption{Composite image of $B_r$, the squashing factor, and the open field regions from the PF solution using the very high resolution {\bf (NAT)}.  The squashing factor is shown in a signed-log color scale (green and purple), while the magnetic field is colored by orange and cyan.  The signed open field is also indicated using a blue-red color table. 
\label{fig_res_highq}} 
\end{figure}

The PF solution for resolution {\bf (NAT)} is still quite interesting though. This calculation is likely the largest resolution PF ever computed for the solar corona, and it is worth examining the inherent complexity present in such a high-resolution, high-fidelity calculation. In Fig~\ref{fig_res_highq} we show a composite map of several quantities. 
The squashing factor is shown in a signed-log color scale ($+$ in green, $-$ in purple) while the magnetic field is overlaid as proportionally scaled, semi-transparent orange ($+$) and cyan ($-$) colors.  The signed open field is also indicated in semi-transparent red ($+$) and blue ($-$) overlays.  The combined image illustrates how \emph{all} areas of the surface map exhibit extreme amounts of complexity. This is particularly true where the magnetic field is weak, and the random network flux dominates. Here myriad topological boundaries are formed by each fragmented, small-scale flux system. This inherent complexity of the magnetic field mapping may also have important consequences for coronal heating and plasma structuring, as is explored in a related paper \citep{downs21_complexity_in_prep}. To convey the extent of structural detail present, we provide the full resolution image ($16000\times 8000$) online\footnote{\url{http://www.predsci.com/papers/pot3d}}.

\section{Discussion}
\label{sec:discussion}
We have computed potential field solutions for 2012 June 13, varying the source and type of input magnetic data, the outer radial boundary condition, and the resolution.  In doing so, we took advantage of the high-performance, flexible grid, multi-platform (GPUs and CPUs) aspects of POT3D and our magnetogram preparation pipeline to explore PFs in ways that were not previously possible, including at extreme resolution (6.6 billion points). For each solution, we looked at maps of the open field, squashing factor, magnetic energies, open field area, and open interplanetary flux.  

We found that variations in the source magnetogram had only a minor effect on the total open flux, with a more noticeable, but small effect on the magnetic energy and field topology, especially near the AR of interest.  On the other hand, changing the outer radial boundary condition and source surface height had nearly no effect on the magnetic energy (both global and near the AR), but a very large effect on the open flux and open field structure.  The lower the source-surface radius, the larger the open field area and open flux. Since the source-surface radius is a free parameter in PF models,  its large effect on both the open flux and open field suggests the need to find independent sources of validation (such as correlating in-situ magnetic field comparisons with comparisons between open field and coronal hole structure).

Varying the resolution of the PF solutions from very coarse to very fine was quite illuminating. Our experiments show how model resolution has essentially no effect on the open flux. This is because the unsigned flux might change considerably with resolution, but the net flux in a given region does not. However the open field area decreased significantly as resolution increased. Here the same net-flux is captured over ever-smaller flux concentrations (stronger fields) at high resolution, emphasizing that flux, not area, is an essential quantity when studying coronal holes. Most importantly our results confirm that modest resolution potential fields can do a fine job of capturing the overall open flux and morphology, provided that the data is smoothed and averaged properly.  

The resolution experiments and $Q$ maps also illustrate how magnetic complexity changes as function of resolution. By capturing more detail at the surface, the complexity of the surface mapping shows new information at commensurate scales. This may be crucially important when studying active regions or coronal heating within a global model, where the solution properties (energies, mapping, unsigned flux) are inherently connected to both the smoothing of the input data and model resolution. This complexity is also increasing and present in all regions of the coronal base, particularly in the weak-field quiet-sun and open flux regions.

These experiments have also served to show the value the feature set of the POT3D code is for high-resolution, high-performance PF calculations.   For example, the use of a non-uniform grid allows us to capture an AR of interest at the highest resolution of the data source within a global model, while keeping the total problem size reasonable.  In this case, the grid size was nearly forty times smaller than a comparable calculation with uniform resolution, enabling a fast-computation with nearly identical results within the AR.  Additionally, the scalability of the code allows extreme resolution PFs to be computed quickly using large HPC systems. An earlier version of POT3D is presently available for runs on demand as part of the WSA model in CORHEL hosted at the Community Coordinated Modeling Center\footnote{\url{https://ccmc.gsfc.nasa.gov}}.  The version used in this paper is currently released as part the Standard Performance Evaluation Corporation's beta version of the SPEChpc\textsuperscript{TM} 2021 benchmark suites\footnote{\url{https://www.spec.org/hpc2021}}, and, with the publication of this paper, as an open-source release on GitHub\footnote{\url{https://github.com/predsci/POT3D}}.

Overall, the experiments in this work emphasize how much can still be learned from PF calculations, particularly now that high resolution, high-quality data is routinely available. They also show that caution should be taken when deriving conclusions using PF solutions; that the resolution is appropriate, the boundary conditions are validated, and solution variations between magnetic field input sources is considered.

All of the processed maps and POT3D input files used to generate the results in this paper are available at \url{http://www.predsci.com/papers/pot3d}.


\acknowledgements
This work was supported by the NASA Heliophysics Guest Investigator program (grants NNX17AB78G and 80NSSC19K0273), AFOSR (contract \# FA9550-15-C-0001), the NSF PREEVENTS program (grant ICER1854790)), the NSF/NASA SWQU program (grants AGS 2028154, NNX12AB30G), and the NASA HSR program (grant 80NSSC18K0101).  Computational resources were provided by NASA's NAS (Pleiades) and NSF's XSEDE (TACC \& SDSC).
We also wish to thank the organizers (KD Leka, Graham Barnes, and James McAteer), and participants of a 2019 SHINE session dedicated to PF modeling for stimulating discussion and prompts that helped precipitate some of this work. 

\appendix

\section{Additional Model Capabilities of POT3D}
\label{sec_appx_pot3d_models}
This paper has focused on source surface potential fields (PFSS).  Here we briefly detail additional uses of the POT3D code, namely the PFCS and Open Field (OF) models.

The PFCS model is used to extend a PFSS model farther out into the corona (typically to 20-30~$\text{R}_{\odot}$) while reproducing the latitudinally independent behavior of the radial magnetic field magnitude observed by Ulysses.  To do this, the magnitude of the outer (source-surface) boundary magnetic field of a PFSS solution ($|B_r(r=r_{\mbox\scriptsize ss})|$) is used as the lower boundary condition of the potential field solver, with a source-surface outer boundary condition (see Sec.~\ref{sec:pot3d_model}).  The resulting field is unsigned.  To obtain a signed field, we can post-process the field through tracing field lines from each point in the domain to the lower boundary, and setting the sign at that point according to the sign of $B_r$ where the field line mapped to.

It is well known that the combination of the PFSS field and PFCS field is discontinuous at the source-surface boundary \citep[e.g.,][]{mcgregoretal2008}.  This occurs because the PFSS field is forced to be radial at the source-surface (upper boundary of the PFSS), while in general the resulting solution for the PFCS will have small non-radial components at its lower boundary.   \citet{mcgregoretal2008} describe a procedure for reducing the magnitude of this discontinuity by using the radial field from the PFSS at a slightly lower height than the source surface for the PFCS boundary condition.  While this is not an explicit option in POT3D, it is easily implementable.

The Open Field (OF) for a given boundary magnetic flux distribution is the field where all the field lines are open, connected to the solar surface, and stretch to infinity \citep{barnes_sturrock1972}.  It has important theoretical significance, as the Aly-Sturrock ``theorem'' \citep{aly1984,aly1991,sturrock1991} identifies the OF as the upper energy bound for simply connected force-free magnetic fields.  While a rigorous proof of the theorem has not been shown, the results are consistent with a number of simulation studies \citep[e.g.,][]{mikiclinker1994,amarietal1996,antiochosetal1999,amarietal2000,linkeretal2003,hu2004}.
The OF model is computed using the same procedure as the PFCS model, but with the lower boundary at the photosphere, using the observational radial magnetic field boundary conditions ($B_r(r=R_{\odot})\leftarrow |B_r(r=R_{\odot})|$). The resulting solution yields a fully open unsigned field which can be signed through field tracing as in the PFCS case.

\section{Computational Performance of POT3D}
\label{sec_appx_comp}

A key advantage of using PF models is that they are computationally efficient compared to MHD models.  However, in this work, we have computed very large PF solutions up to 6.6 billion points. At this size, such calculations can still be quite computationally expensive, especially when using script-based or non-optimal solvers.  POT3D is designed to be a high-performance PF solver.  Here we show its performance on the high-resolution PF solutions from Sec.~\ref{sec:var_res} and highlight the computational advantage of its ability to use non-uniform grids.  For solutions computed on CPUs, we use the dual-socket Xeon Platinum 8160 CPU nodes (48 cores per node) on the Stampede 2 system at TACC. For solutions computed on GPUs, we use the GPU nodes of the NASA NAS HECC Pleiades machine which contain 4 NVIDIA V100 GPUs per node.

\begin{figure}[htbp]
\centering
\includegraphics[width=0.7\textwidth]{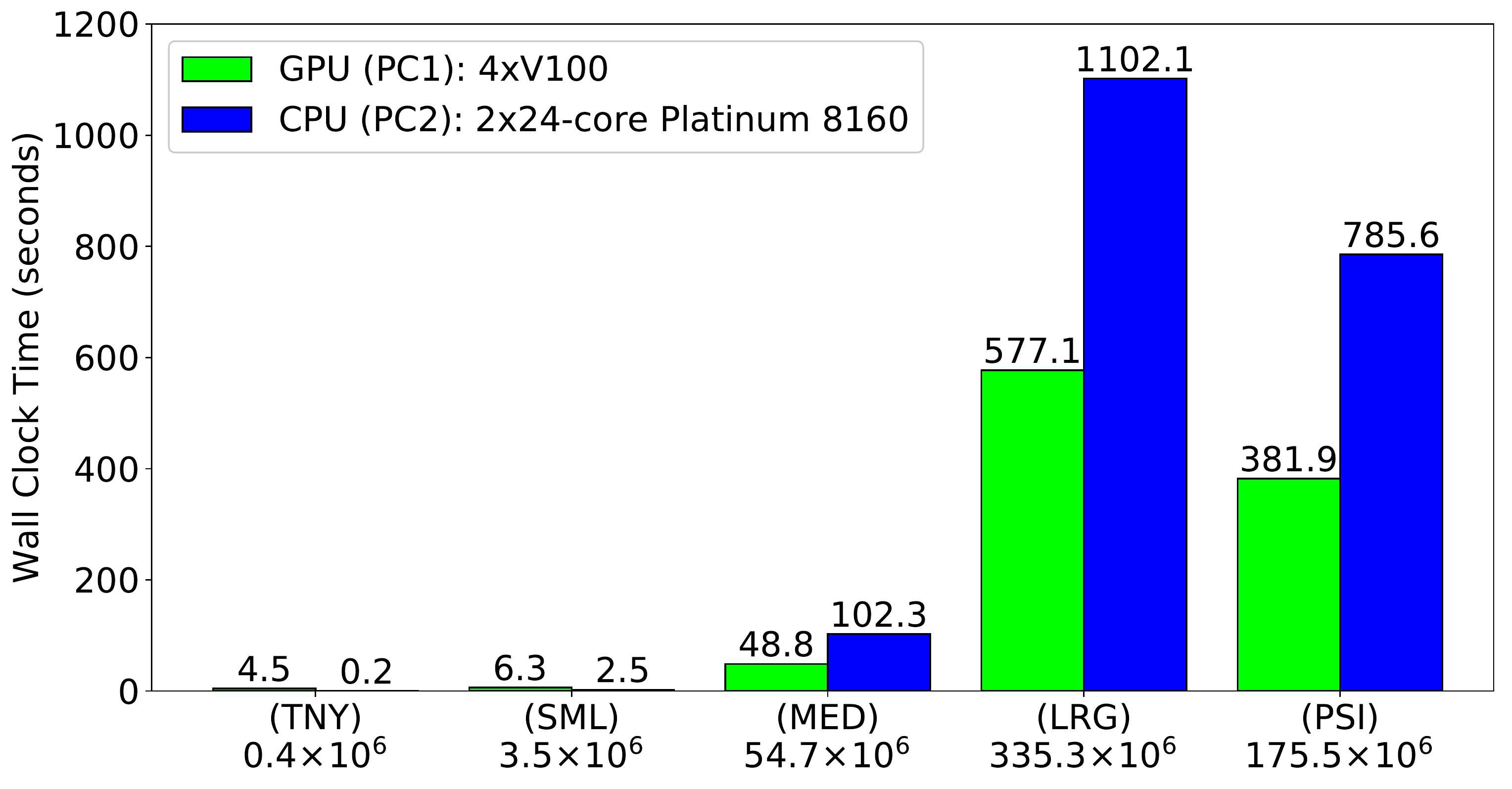}
\caption{Single-server timing results for PF solutions using POT3D for the resolutions described in Sec.~\ref{sec:var_res}.  The total number of grid cells for each resolution is indicated.\label{fig_timings}} 
\end{figure}

In Fig.~\ref{fig_timings}, we show wall clock times for computing PF solutions for the resolutions described in Sec.~\ref{sec:var_res} on a single CPU or GPU compute node/server (except for {\bf (NAT)} as it does not fit into the memory of a single node).
We see that resolutions {\bf (TNY)} and {\bf (SML)} compute very quickly and are too small for the GPU to exhibit speedup.  For resolution {\bf (MED)} through {\bf (PSI)}, we see that the GPU computes the PF about twice as fast as the CPU.  This speedup (while very beneficial) may appear lower than expected considering the hardware capabilities of the GPUs versus the CPU.  However, this is because when running on the CPU, POT3D can use the much more efficient PC2 preconditioner described in Sec.~\ref{sec:pot3d_model}, while when running on GPUs, it can only currently utilize the less efficient PC1 algorithm (see \cite{POT3DACC} for details).  Running POT3D on GPUs still has a great advantage in this case since one can configure a relatively inexpensive workstation with multiple GPUs, allowing for fast `in-house' PF computations.

For the largest resolution PF (resolution {\bf (NAT)}, yielding 6.6 billion points), we used 120 CPU nodes (5760 total CPU cores), yielding a solution time of 14.5 minutes (including all setup and I/O).  The  non-uniform resolution {\bf (PSI)} has the same resolution as {\bf (NAT)} near the active region of interest, but coarsens greatly outside the region, resulting in a resolution $\sim40$ times smaller (175 million cells).  Running this on the same number of CPUs finds the solution in only 18 seconds (50x faster). These results demonstrate both the large scalability of POT3D, as well as the advantage of its ability to use non-uniform grids in cases where a localized region needs to be highly resolved.

\section{Estimation of Magnetic Energy Outside of the Source-Surface Radius}
\label{sec:mage}
To estimate the magnetic energy outside of a source-surface potential field model, we assume that the radial magnetic field falls off as $1/r^2$. This reflects the fact that from the source surface to the inner heliosphere ($\sim30~\text{R}_{\odot}$) the variations in $B_r$ relax away from the HCS, with $B_r$ becoming essentially independent of latitude the farther out one goes. We can bound the excess energy then by integrating the surface field at the outer boundary to infinity, first by keeping the spatial dependence of $B_r(r_{ss},\theta,\phi)$ and second by integrating the mean flux only ($\overline{|B_r(r_{ss})|}$). The former is an upper bound on the open field energy (from $r=r_{ss}$ to infinity) while the latter is a lower bound.

The magnetic energy (in CGS units) from $r=r_{ss}$ to $r=r_1$ is defined as
\begin{equation}
\label{eq_me1}
W_{r_{ss}}^{r_1} = \frac{1}{8\,\pi}\,\int_0^{2\,\pi}\int_0^{\pi}\int_{r_{ss}}^{r_1} |{\bf B}|^2\,r^2\,\sin\theta\,dr\,d\theta\,d\phi.
\end{equation}
Assuming a purely radial field at the source-surface boundary ($B_r(r_{ss})$) with a $1/r^2$ fall-off of the field strength, we have
\[
B_r(r,\theta,\phi)=B_r(r_{ss},\theta,\phi)\,\frac{r_{ss}^2}{r^2}.
\]
Inserting this into Eq.~\ref{eq_me1} and letting $r_1\rightarrow \infty$ yields
\begin{equation}
\label{eq_me2}
W_{r_{ss}}^{\infty} \approx \frac{r_{ss}^3}{8\,\pi}\,\int_0^{2\,\pi}\int_0^{\pi} \left|B_r(r_{ss},\theta,\phi)\right|^2\,\sin\theta\,d\theta\,d\phi.
\end{equation}
For the upper bound estimate, we use $B_r(r_{ss},\theta,\phi)$ directly from the model.  For the lower bound estimate, we use $\overline{|B_r(r_{ss})|}$ in place of $B_r(r_{ss},\theta,\phi)$, where
\[
\overline{|B_r(r_{ss})|}=\frac{1}{4\,\pi\,r_{ss}^2}\,\int_0^{2\,\pi}\int_0^{\pi}|B_r(r_{ss},\theta,\phi)|\,r_{ss}^2\,\sin\theta\,d\theta\,d\phi,
\]
which simplifies Eq.~\ref{eq_me2} to
\begin{equation}
\label{eq_me3}
W_{r_{ss}}^{\infty} \approx \frac{r_{ss}^3}{2}\, \overline{|B_r(r_{ss})|}^2.
\end{equation}

\def\myitemsep{5pt}
\bibliographystyle{aasjournal}
\bibliography{POT3D}

\begin{thebibliography}{}
\expandafter\ifx\csname natexlab\endcsname\relax\def\natexlab#1{#1}\fi
\providecommand{\url}[1]{\href{#1}{#1}}
\providecommand{\dodoi}[1]{doi:~\href{http://doi.org/#1}{\nolinkurl{#1}}}
\providecommand{\doeprint}[1]{\href{http://ascl.net/#1}{\nolinkurl{http://ascl.net/#1}}}
\providecommand{\doarXiv}[1]{\href{https://arxiv.org/abs/#1}{\nolinkurl{https://arxiv.org/abs/#1}}}

\bibitem[{{Altschuler} \& {Newkirk}(1969)}]{altschulernewkirk1969}
{Altschuler}, M.~D., \& {Newkirk}, G. 1969, \solphys, 9, 131

\bibitem[{{Aly}(1984)}]{aly1984}
{Aly}, J.~J. 1984, \apj, 283, 349, \dodoi{10.1086/162313}

\bibitem[{{Aly}(1991)}]{aly1991}
---. 1991, \apjl, 375, L61, \dodoi{10.1086/186088}

\bibitem[{{Amari} {et~al.}(1996){Amari}, {Luciani}, {Aly}, \&
  {Tagger}}]{amarietal1996}
{Amari}, T., {Luciani}, J.~F., {Aly}, J.~J., \& {Tagger}, M. 1996, \apjl, 466,
  L39, \dodoi{10.1086/310158}

\bibitem[{{Amari} {et~al.}(2000){Amari}, {Luciani}, {Mikic}, \&
  {Linker}}]{amarietal2000}
{Amari}, T., {Luciani}, J.~F., {Mikic}, Z., \& {Linker}, J. 2000, \apjl, 529,
  L49, \dodoi{10.1086/312444}

\bibitem[{{Antiochos} {et~al.}(2007){Antiochos}, {DeVore}, {Karpen}, \&
  {Miki{\'c}}}]{antiochosetal2007}
{Antiochos}, S.~K., {DeVore}, C.~R., {Karpen}, J.~T., \& {Miki{\'c}}, Z. 2007,
  \apj, 671, 936, \dodoi{10.1086/522489}

\bibitem[{{Antiochos} {et~al.}(1999){Antiochos}, {DeVore}, \&
  {Klimchuk}}]{antiochosetal1999}
{Antiochos}, S.~K., {DeVore}, C.~R., \& {Klimchuk}, J.~A. 1999, \apj, 510, 485,
  \dodoi{10.1086/306563}

\bibitem[{{Arge} {et~al.}(2004){Arge}, {Luhmann}, {Odstrcil}, {Schrijver}, \&
  {Li}}]{argeetal2004}
{Arge}, C.~N., {Luhmann}, J.~G., {Odstrcil}, D., {Schrijver}, C.~J., \& {Li},
  Y. 2004, Journal of Atmospheric and Solar-Terrestrial Physics, 66, 1295,
  \dodoi{10.1016/j.jastp.2004.03.018}

\bibitem[{{Arge} {et~al.}(2003){Arge}, {Odstrcil}, {Pizzo}, \&
  {Mayer}}]{argeetal2003}
{Arge}, C.~N., {Odstrcil}, D., {Pizzo}, V.~J., \& {Mayer}, L.~R. 2003, in
  American Institute of Physics Conference Series, Vol. 679, Solar Wind Ten,
  ed. M.~{Velli}, R.~{Bruno}, F.~{Malara}, \& B.~{Bucci}, 190--193,
  \dodoi{10.1063/1.1618574}

\bibitem[{{Barnes} \& {Sturrock}(1972)}]{barnes_sturrock1972}
{Barnes}, C.~W., \& {Sturrock}, P.~A. 1972, \apj, 174, 659,
  \dodoi{10.1086/151527}

\bibitem[{Bell \& Garland(2008)}]{DIACSR}
Bell, N., \& Garland, M. 2008, Efficient sparse matrix-vector multiplication on
  {CUDA}, Tech. rep., Nvidia Technical Report NVR-2008-004, Nvidia Corporation

\bibitem[{Bobra {et~al.}(2014)Bobra, Sun, Hoeksema, Turmon, Liu, Hayashi,
  Barnes, \& Leka}]{sharp}
Bobra, M.~G., Sun, X., Hoeksema, J.~T., {et~al.} 2014, Solar Physics, 289, 3549

\bibitem[{{Caplan} {et~al.}(2016){Caplan}, {Downs}, \& {Linker}}]{caplan16}
{Caplan}, R.~M., {Downs}, C., \& {Linker}, J.~A. 2016, \apj, 823, 53,
  \dodoi{10.3847/0004-637X/823/1/53}

\bibitem[{{Caplan} {et~al.}(2017){Caplan}, {Miki{\'c}}, \& {Linker}}]{POT3DACC}
{Caplan}, R.~M., {Miki{\'c}}, Z., \& {Linker}, J.~A. 2017, ArXiv e-prints.
\newblock \doarXiv{1709.01126}

\bibitem[{Chandrasekaran \& Juckeland(2017)}]{OpenACCBook2}
Chandrasekaran, S., \& Juckeland, G. 2017, OpenACC for Programmers Concepts and
  Strategies (Addison-Wesley Professional)

\bibitem[{Chow \& Saad(1997)}]{ILU_breakdown}
Chow, E., \& Saad, Y. 1997, Journal of Computational and Applied Mathematics,
  86, 387

\bibitem[{{Downs} {et~al.}(2021){Downs}, {Linker}, M., \&
  {Titov}}]{downs21_complexity_in_prep}
{Downs}, C., {Linker}, J.~A., M., C.~R., \& {Titov}, V.~S. 2021, in prep for
  \apjl

\bibitem[{Freeland \& Handy(1998)}]{SSW}
Freeland, S., \& Handy, B. 1998, Solar Physics, 182, 497

\bibitem[{Harvey {et~al.}(1996)Harvey, Hill, Hubbard, Kennedy, Leibacher,
  Pintar, Gilman, Noyes, Toomre, Ulrich, {et~al.}}]{gong}
Harvey, J., Hill, F., Hubbard, R., {et~al.} 1996, Science, 272, 1284

\bibitem[{Hayashi {et~al.}(2016)Hayashi, Yang, \& Deng}]{hayashi2016comparison}
Hayashi, K., Yang, S., \& Deng, Y. 2016, Journal of Geophysical Research: Space
  Physics, 121, 1046

\bibitem[{{Hoeksema} {et~al.}(1983){Hoeksema}, {Wilcox}, \&
  {Scherrer}}]{hoeksemaetal1983}
{Hoeksema}, J.~T., {Wilcox}, J.~M., \& {Scherrer}, P.~H. 1983, \jgr, 88, 9910,
  \dodoi{10.1029/JA088iA12p09910}

\bibitem[{{Hu}(2004)}]{hu2004}
{Hu}, Y.~Q. 2004, \apj, 607, 1032, \dodoi{10.1086/383517}

\bibitem[{Linker {et~al.}(2017)Linker, Caplan, Downs, Riley, Mikic, Lionello,
  Henney, Arge, Liu, Derosa, {et~al.}}]{linker2017open}
Linker, J., Caplan, R., Downs, C., {et~al.} 2017, The Astrophysical Journal,
  848, 70

\bibitem[{{Linker} {et~al.}(2003){Linker}, {Miki\'c}, {Lionello}, {Riley},
  {Amari}, \& {Odstrcil}}]{linkeretal2003}
{Linker}, J.~A., {Miki\'c}, Z., {Lionello}, R., {et~al.} 2003, Phys. of
  Plasmas, 10, 1971

\bibitem[{Linker {et~al.}(1999)Linker, Miki{\'c}, Biesecker, Forsyth, Gibson,
  Lazarus, Lecinski, Riley, Szabo, \& Thompson}]{linker99a}
Linker, J.~A., Miki{\'c}, Z., Biesecker, D.~A., {et~al.} 1999, Journal of
  Geophysical Research: Space Physics, 104, 9809

\bibitem[{Linker {et~al.}(2016)Linker, Caplan, Downs, Lionello, Riley, Mikic,
  Henney, Arge, Kim, \& Pogorelov}]{linker2016empirically}
Linker, J.~A., Caplan, R.~M., Downs, C., {et~al.} 2016, in Journal of Physics:
  Conference Series, Vol. 719, IOP Publishing, 12012--12023

\bibitem[{{McGregor} {et~al.}(2008){McGregor}, {Hughes}, {Arge}, \&
  {Owens}}]{mcgregoretal2008}
{McGregor}, S.~L., {Hughes}, W.~J., {Arge}, C.~N., \& {Owens}, M.~J. 2008,
  Journal of Geophysical Research (Space Physics), 113, A08112,
  \dodoi{10.1029/2007JA012330}

\bibitem[{Miki\'c \& Linker(1994)}]{mikiclinker1994}
Miki\'c, Z., \& Linker, J.~A. 1994, Ap. J., 430, 898

\bibitem[{{Miki{\'c}} {et~al.}(2018){Miki{\'c}}, {Downs}, {Linker}, {Caplan},
  {Mackay}, {Upton}, {Riley}, {Lionello}, {T{\"o}r{\"o}k}, {Titov}, {Wijaya},
  {Druckm{\"u}ller}, {Pasachoff}, \& {Carlos}}]{mikic18}
{Miki{\'c}}, Z., {Downs}, C., {Linker}, J.~A., {et~al.} 2018, Nature Astronomy,
  2, 913, \dodoi{10.1038/s41550-018-0562-5}

\bibitem[{{Nitta} {et~al.}(2006){Nitta}, {Reames}, {De Rosa}, {Liu}, {Yashiro},
  \& {Gopalswamy}}]{nittaetal2006}
{Nitta}, N.~V., {Reames}, D.~V., {De Rosa}, M.~L., {et~al.} 2006, \apj, 650,
  438, \dodoi{10.1086/507442}

\bibitem[{{Pizzo} {et~al.}(2011){Pizzo}, {Millward}, {Parsons}, {Biesecker},
  {Hill}, \& {Odstrcil}}]{pizzoetal2011}
{Pizzo}, V., {Millward}, G., {Parsons}, A., {et~al.} 2011, Space Weather, 9,
  3004, \dodoi{10.1029/2011SW000663}

\bibitem[{{Riley} {et~al.}(2015){Riley}, {Linker}, \& {Arge}}]{rileyetal2015}
{Riley}, P., {Linker}, J.~A., \& {Arge}, C.~N. 2015, Space Weather, 13, 154,
  \dodoi{10.1002/2014SW001144}

\bibitem[{Riley {et~al.}(2001)Riley, Linker, \& Miki\'c}]{rileyetal2001}
Riley, P., Linker, J.~A., \& Miki\'c, Z. 2001, J. Geophys. Res., 106, 15889

\bibitem[{{Riley} {et~al.}(2014){Riley}, {Ben-Nun}, {Linker}, {Mikic},
  {Svalgaard}, {Harvey}, {Bertello}, {Hoeksema}, {Liu}, \& {Ulrich}}]{riley14}
{Riley}, P., {Ben-Nun}, M., {Linker}, J.~A., {et~al.} 2014, \solphys, 289, 769,
  \dodoi{10.1007/s11207-013-0353-1}

\bibitem[{Saad(2003)}]{IterativeMethods_SAAD_Book}
Saad, Y. 2003, Iterative methods for sparse linear systems (Siam)

\bibitem[{{Schatten}(1971)}]{schatten1971}
{Schatten}, K.~H. 1971, Cosmic Electrodynamics, 2, 232

\bibitem[{{Schatten} {et~al.}(1969){Schatten}, {Wilcox}, \&
  {Ness}}]{schattenetal1969}
{Schatten}, K.~H., {Wilcox}, J.~M., \& {Ness}, N.~F. 1969, \solphys, 6, 442

\bibitem[{Scherrer {et~al.}(2012)Scherrer, Schou, Bush, Kosovichev, Bogart,
  Hoeksema, Liu, Duvall, Zhao, Schrijver, {et~al.}}]{hmi}
Scherrer, P.~H., Schou, J., Bush, R., {et~al.} 2012, Solar Physics, 275, 207

\bibitem[{Schrijver \& DeRosa(2003)}]{schrijver2003photospheric}
Schrijver, C.~J., \& DeRosa, M.~L. 2003, Solar Physics, 212, 165

\bibitem[{{Schrijver} {et~al.}(2004){Schrijver}, {Sandman}, {Aschwanden}, \&
  {De Rosa}}]{schrijveretal2004}
{Schrijver}, C.~J., {Sandman}, A.~W., {Aschwanden}, M.~J., \& {De Rosa}, M.~L.
  2004, \apj, 615, 512, \dodoi{10.1086/424028}

\bibitem[{Smith \& Zhang(2011)}]{CSRopt}
Smith, B., \& Zhang, H. 2011, International Journal of High Performance
  Computing Applications, 25, 386

\bibitem[{Smith \& Balogh(1995)}]{smith_balogh1995}
Smith, E.~J., \& Balogh, A. 1995, Geophysical research letters, 22, 3317

\bibitem[{Smith \& Balogh(2008)}]{smith_balogh2008}
---. 2008, Geophysical research letters, 35

\bibitem[{Stansby {et~al.}(2020)Stansby, Yeates, \& Badman}]{pfsspy}
Stansby, D., Yeates, A., \& Badman, S. 2020, Journal of Open Source Software, 5

\bibitem[{{Sturrock}(1991)}]{sturrock1991}
{Sturrock}, P.~A. 1991, \apj, 380, 655, \dodoi{10.1086/170620}

\bibitem[{Sun(2018)}]{polarHMI}
Sun, X. 2018, arXiv preprint arXiv:1801.04265

\bibitem[{Titov {et~al.}(2012)Titov, Mikic, T{\"o}r{\"o}k, Linker, \&
  Panasenco}]{titov20122010}
Titov, V., Mikic, Z., T{\"o}r{\"o}k, T., Linker, J., \& Panasenco, O. 2012, The
  Astrophysical Journal, 759, 70

\bibitem[{Titov(2007)}]{titov2007generalized}
Titov, V.~S. 2007, The Astrophysical Journal, 660, 863

\bibitem[{{Titov} {et~al.}(2011){Titov}, {Miki{\'c}}, {Linker}, {Lionello}, \&
  {Antiochos}}]{titovetal2011}
{Titov}, V.~S., {Miki{\'c}}, Z., {Linker}, J.~A., {Lionello}, R., \&
  {Antiochos}, S.~K. 2011, \apj, 731, 111, \dodoi{10.1088/0004-637X/731/2/111}

\bibitem[{T{\'o}th {et~al.}(2011)T{\'o}th, Van~der Holst, \&
  Huang}]{toth2011obtaining}
T{\'o}th, G., Van~der Holst, B., \& Huang, Z. 2011, The Astrophysical Journal,
  732, 102

\bibitem[{Veldman \& Rinzema(1992)}]{NUG_PLAY_1992}
Veldman, A., \& Rinzema, K. 1992, Journal of engineering mathematics, 26, 119

\bibitem[{{Wang} {et~al.}(1996){Wang}, {Hawley}, \& {Sheeley}}]{wangetal1996}
{Wang}, Y.-M., {Hawley}, S.~H., \& {Sheeley}, Neil~R., J. 1996, Science, 271,
  464, \dodoi{10.1126/science.271.5248.464}

\bibitem[{{Wang} \& {Sheeley}(1990)}]{wangsheeley1990}
{Wang}, Y.~M., \& {Sheeley}, N.~R., J. 1990, \apj, 355, 726,
  \dodoi{10.1086/168805}

\bibitem[{{Wang} \& {Sheeley}(1994)}]{wangsheeley1994}
---. 1994, \jgr, 99, 6597, \dodoi{10.1029/93JA02105}

\bibitem[{{Wang} \& {Sheeley}(2004)}]{wangsheeley2004}
---. 2004, \apj, 612, 1196, \dodoi{10.1086/422711}

\bibitem[{{Wang} \& {Sheeley}(1992)}]{wang_sheeley1992}
{Wang}, Y.-M., \& {Sheeley}, Jr., N.~R. 1992, \apj, 392, 310,
  \dodoi{10.1086/171430}

\end{thebibliography}

\end{document}